\documentclass[twocolumn]{aastex631}
\usepackage{amsmath}
\usepackage{txfonts}
\usepackage{multirow}
\usepackage{xcolor,colortbl}
\usepackage[utf8]{inputenc} 

\definecolor{green}{rgb}{0.1,0.1,0.1}
\newcommand{\new}{\mathrm{new}}

\newcommand{\vkep}{v_{\mathrm{kep}}}
\newcommand{\eff}{\mathrm{eff}}

\newcommand{\SMBH}{\mathrm{SMBH}}

\newcommand{\Msun}{M_{\odot}}
\newcommand{\Edd}{\mathrm{Edd}}

\newcommand{\back}{\mathrm{back}}
\newcommand{\DBH}{\mathrm{DBH}}
\newcommand{\Ds}{\mathrm{Ds}}

\newcommand{\vrel}{v_{\mathrm{rel}}}
\newcommand{\rHill}{r_{\mathrm{Hill}}}
\newcommand{\erf}{\mathrm{erf}}

\newcommand{\BH}{\mathrm{BH}}
\newcommand{\ini}{\mathrm{ini}}
\newcommand{\rmin}{\mathrm{in}}
\newcommand{\out}{\mathrm{out}}
\newcommand{\disk}{\mathrm{disk}}
\newcommand{\IMF}{\mathrm{IMF}}
\newcommand{\cre}{\mathrm{cre}}
\newcommand{\change}{\mathrm{change}}
\newcommand{\gas}{\mathrm{gas}}

\newcommand{\AGN}{\mathrm{AGN}}

\newcommand{\GDF}{\mathrm{GDF}}
\newcommand{\acc}{\mathrm{acc}}
\newcommand{\mig}{\mathrm{mig}}
\newcommand{\CBD}{\mathrm{CBD}}

\newcommand{\BS}{\mathrm{BS}}
\newcommand{\WS}{\mathrm{WS}}
\newcommand{\rel}{\mathrm{rel}}

\newcommand{\capt}{\mathrm{cap}}

\newcommand{\cell}{\mathrm{cell}}
\newcommand{\diff}{\mathrm{diffusion}}
\newcommand{\DF}{\mathrm{DF}}
\newcommand{\preexist}{\mathrm{pre}}

\newcommand{\uni}{\mathrm{uni}}

\begin{document}

\title{What Determines the Maximum Mass of AGN-assisted Black Hole Mergers?}

\author{LingQin Xue}
\affiliation{Department of Physics, University of Florida, P.O. Box 118440, Gainesville, Florida 32611-8440, USA}

\author{Hiromichi Tagawa}
\affiliation{Shanghai Astronomical Observatory, Shanghai, 200030, People's Republic of China}

\author{Zolt\'an Haiman}
\affiliation{Institute of Science and Technology Austria, Am Campus 1, Klosterneuburg 3400 Austria}
\affiliation{Department of Astronomy, Columbia University, MC 5246, 538 West 120th Street, New York, New York 10027, USA}
\affiliation{Department of Physics, Columbia University, MC 5255, 538 West 120th Street, New York, New York 10027, USA}

\author{Imre Bartos}
\affiliation{Department of Physics, University of Florida, P.O. Box 118440, Gainesville, Florida 32611-8440, USA}
\email{imrebartos@ufl.edu}
 
\begin{abstract}
The origin of merging binary black holes detected through gravitational waves remains a fundamental question in astrophysics. While stellar evolution imposes an upper mass limit of $\sim50$M$\odot$ for black holes, some observed mergers—most notably GW190521—involve significantly more massive components, suggesting alternative formation channels. Here we investigate the maximum masses attainable by black hole mergers within active galactic nucleus (AGN) disks. Using a comprehensive semianalytic model incorporating 27 binary and environmental parameters, we explore the role of AGN disk conditions in shaping the upper end of the black hole mass spectrum. We find that the AGN disk lifetime is the dominant factor, with high-mass mergers ($\gtrsim200$M$\odot$) only possible if disks persist for $\gtrsim40$\,Myr. The joint electromagnetic observation of an AGN-assisted merger could therefore lead to a direct measurement of the age of an AGN disk.
\end{abstract}

%\keywords{Classical Novae (251) --- Ultraviolet astronomy(1736) --- History of astronomy(1868) --- Interdisciplinary astronomy(804)}
\section{Introduction} \label{sec:intro}

Binary black hole mergers are now regularly being observed through gravitational wave emission. The LIGO 
\citep{2015CQGra..32g4001L}, Virgo \citep{VIRGO:2014yos} and KAGRA \citep{KAGRA:2018plz} collaborations have reported around 90 previous discoveries, while their fourth observing run, O4, is currently ongoing.

The origin of the observed black hole mergers, and the astrophysical processes that lead to the mergers, are some of the main questions of interest. These questions are not settled at present, and multiple possible astrophysical scenarios are being investigated. These include formation from isolated stellar binaries (e.g., \citealt{Postnov2014,van_den_Heuvel_2017,Mandel2016}), dynamical formation in dense clusters (e.g., \citealt{Benacquista2013,2016PhRvD..93h4029R,2022hgwa.bookE..15K}), or gas capture in the disks of active galactic nuclei (AGNs; e.g., \citealt{2017ApJ...835..165B,2018ApJ...866...66M,2019PhRvL.123r1101Y,2020ApJ...899...26T,2025arXiv250311721D}). 

Differentiating between these formation mechanisms for either a single detection or the overall set of detections is made difficult by both the currently large model uncertainties and the expected substantial overlap between the binary properties these mechanisms are expected to produce. A promising direction to avoid this overlap is to specifically target outliers, or in general parts of the parameter space not expected to be covered by all models. 

The high end of the observed black hole mass distribution is a particularly promising part of the parameter space. Generally, heavier black holes are expected to be rarer, making processes that preferentially select heavier black holes for mergers stand out. Even more importantly, black hole formation from stellar evolution is expected to reach only up to a mass of around $50\,
$M$_\odot$ \citep{2001ApJ...550..372F}. Massive stars above a critical mass are expected to experience pair instability in their center, resulting in their early explosion before they could form a sufficiently massive core. These stars are expected to leave no compact remnant behind, effectively putting an upper limit on how massive black holes can be formed from stars.

Black holes have nonetheless been observed with over $50
\,$M$_\odot$ mass. In particular, merger GW190521 was recorded to have been a binary of two black holes with masses estimated to be $95.3^{+28.7}_{-18.9}$M$_\odot$ and $69.2^{+17.0}_{-10.6}$\,M$_{\odot}$  \citep{2020PhRvL.125j1102A,2021PhRvX..11b1053A}. Such high masses are difficult to explain with stellar evolution (although see \citealt{2020ApJ...905L..15B}). In addition, GW190521 featured other outlying properties, including its apparent high precessing spin, and possibly highly eccentric orbit prior to merger \citep{Gayathri2022,2020ApJ...903L...5R,2023NatAs...7...11G}, pointing toward a dynamical or AGN origin \citep{2022Natur.603..237S}.

A promising mechanism to explain the creation of black hole masses beyond stellar evolution limits is the consecutive merger of black holes, or so-called {\it hierarchical mergers} \citep{2017PhRvD..95l4046G,2017ApJ...840L..24F,2019PhRvL.123r1101Y,tagawa2020formation,tagawa2021mass,2021ApJ...918L..31M,2024ApJ...975..117M,2025PhRvD.111b3013M}. The remnant of the merger of two black holes can naturally exceed the maximum mass allowed by stellar evolution; therefore, if such remnants can participate in further mergers, possibly much higher black hole masses can be reached. 

How high the masses can get depends on the physical processes and environmental conditions that produce the hierarchical mergers, and possibly accretion.

In this paper we explore this connection between maximum mass and environment. The goal of this investigation is to understand how the observation of the heaviest black holes can be converted to understanding of the environment of the black holes, or astrophysical processes more generally.
While AGN disks are the primary focus of this work, it is important to note that other environments may also facilitate the formation of intermediate-mass black holes (IMBHs). For example, globular clusters and nuclear star clusters \citep{chattopadhyay2023double,barber2025formation,purohit2024binary,atallah2023growing,askar2017mocca,gieles2025globular}, and possibly gas-rich dense stellar systems can support repeated mergers or gas accretion. 

We focus in particular on black hole mergers in AGN disks. AGNs are promising black hole assembly lines that are expected to promote hierarchical mergers \citep{2019PhRvL.123r1101Y}, and also may grow black hole masses through accretion \citep{2020ApJ...901L..34Y}. 

AGN-assisted mergers represent possibly the most complex binary evolution process that features not only the properties of populations of compact objects and stars but also accretion and interaction with the AGN disk. This complexity means that there are a large number of binary and environmental parameters, many of which are not well understood. We will include 27 such parameters in our model, which are described in Table \ref{table:par}. Due to the difficulty of exploring a 27-dimensional parameter space, our strategy is to consider a fiducial value for each parameter, and explore the change of the maximum mass compared to the fiducial value by changing only a single parameter at a time. This gives an informative picture of the dependence of the maximum mass on each of the 27 parameters. 

As there is no hard cutoff for black hole masses but rather a continuous, if steep, distribution, we will define maximum mass for the purposes of our analysis as the average mass of the heaviest 1\% of the black holes that undergo mergers in our simulations for each set of parameters. This mass is informative of what is expected to be observable as the highest mass through gravitational waves from the AGN channel, and therefore we consider it a meaningful basis for observational model testing.

The paper is organized as follows. Sec.  \ref{sec:Method} presents our AGN model and binary gas capture, evolution and merger simulation framework with additional technical details provided in the Appendices \ref{sec:pre-exist}-\ref{sec:time_step} . We present our results in Sec.  \ref{sec:results}. We discuss some of the limitations of our model and future work in \ref{sec:discussion}, and conclude in Sec. \ref{sec:conclusion}.

\section{Method}
\label{sec:Method}

\begin{figure*}[!t]
    \centering
    \includegraphics[width=1.0\linewidth]{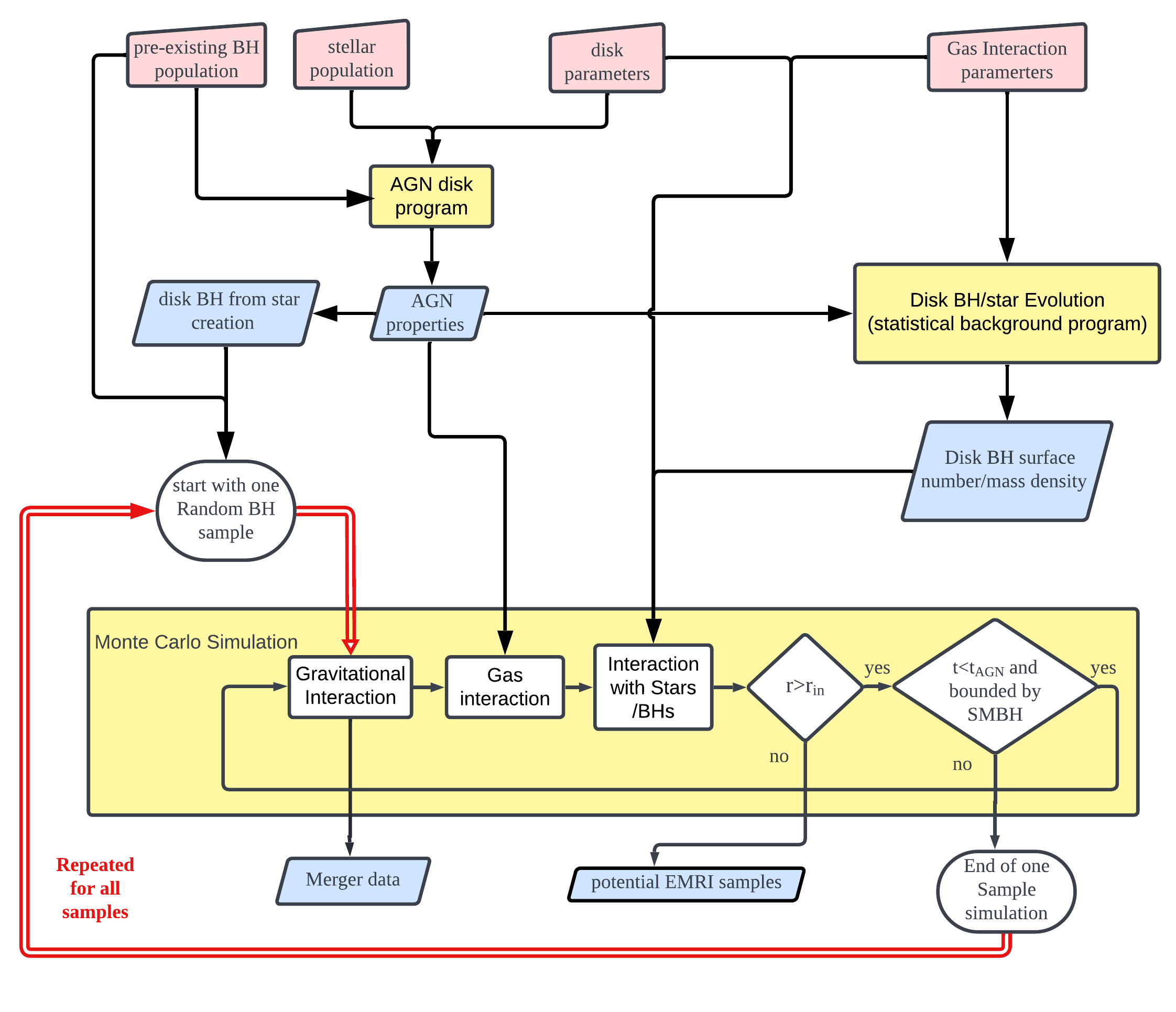}
    \caption{Schematic flow chart of the simulation framework. Pink blocks represent configurable models and parameters, including AGN disk properties, initial BH/stellar populations, and gas interaction models. Yellow blocks correspond to different computational processes, such as AGN disk property calculations, statistical background estimations, and Monte Carlo simulations for BH evolution. Blue blocks indicate simulation outputs, such as merger rates, mass distributions, and population evolution trends. The Monte Carlo simulations are repeated (red double arrows) to track BH status throughout the AGN lifetime, with iterations terminated when either 10,000 mergers are recorded or 10 AGNs are fully simulated.}
    \label{fig:process}
\end{figure*}

\begin{table*}[h]
\centering
\begin{tabular}{ |p{3cm}|p{1.5cm}|p{4cm}|p{3cm}|p{3cm}| }
 \hline
 Type& Symbol &Description& Fiducial value& Plot range\\
 \hline
  \hline
 Simulation   & $N_{\rm{cell}}$    &number of radial cells in simulation &   120& {convergence test}\\
 \hline
  Simulation   & $N_{\rm{mass}}$    &number of mass cells in background calculation &   100&  {convergence test}\\
 \hline
  Simulation   & $\eta_t$    &time step & 0.1 &   {Convergence test}\\
 \hline
 Simulation   & $r_{\rmin}$    & inner bound of simulation&   $r_{\disk,\rmin}$&  {convergence test}\\
 \hline
 Simulation   & $r_{\out}$    & outer bound of simulation&   $r_{\disk,\out}$&  {convergence test}\\
 \hline
 \hline
 SMBH   & $M_{\SMBH}$    & SMBH mass&   $4\times10^6\;M_{\odot}$&$10^5-10^9\;\Msun$\\
   \hline
  \hline
   AGN disk    & $r_{\disk,\rmin}$    &inner bound of AGN disk & $10^{-4}$\;pc, following Eq.  \ref{eq:disk_in}& $0.5\times10^{-4}$-$1.5\times10^{-4}$ pc\\
 \hline
  AGN disk   & $r_{\disk,\out}$    &outer bound of AGN disk & $5$\;pc, following Eq. \ref{eq:disk_out}& 5-10 pc \\
 \hline
 AGN disk   & $\dot{M}_{\out}$    &accretion rate at the outer bound of AGN disk &   $0.1\dot{M}_{\Edd}$&$0.025\dot{M}_{\Edd}$-$0.20\dot{M}_{\Edd}$\\
 \hline
 AGN disk    & $\xi$    &pressure ratio &   1 & no change due to instability\\
 \hline
 AGN disk   & $m_{AM}$    &angular momentum transport rate &   0.1& 0.05-0.3\\
 \hline
 AGN disk   & $\alpha_{\mathrm{SS}}$    & viscosity $\alpha$ parameter&   0.1& 0.05-0.25\\
 \hline
  AGN disk   & $\epsilon$    & conversion efficiency &  $10^{-4}$ & $10^{-4}$-$10^{-3}$\\
 \hline
 AGN disk   & $t_{\AGN}$    & AGN life time&   10 Myr & 10-100 Myr\\
 \hline
 AGN disk   & $\beta_{\IMF,\cre}$    &star creation IMF index &   2.35& 1.7-2.35 \\
 \hline
 \hline
Mass distribution   & $\beta_{\IMF}$    & IMF index of preexisting BH&   2.35& 1.7-2.35\\
 \hline
Mass distribution   & $M_{\mathrm{\BH,\ini,\min}}$   & minimum single BH mass of pre-existing BH &   $5\Msun$& no change due to observational fact\\
 \hline
Mass distribution   & $M_{\mathrm{\BH,\ini,\max}}$   & maximum single BH mass of pre-existing BH&   15 $\Msun$&15-30 $\Msun$\\
 \hline
Radial distribution   & $\gamma_{\rho}$    & index of radial distribution&   0.0&-0.4-0.4\\
 \hline
Radial distribution   & $N_{\mathrm{\BH,\ini}}$    & initial BH number in the AGN (within $r_{BH,out}$)&   20000, following Eq. \ref{eq:N_BH_ini}& 20000-100000\\
 \hline
Radial distribution   & $r_{\BH,\out}$    & maximum BH radial distance to SMBH of preexisting BH&  following Eq. \ref{eq:r_BH,out}& 2.5-3.5 pc\\
 \hline
Velocity(inclination distribution)& $\beta_v$    & velocity dispersion parameter (Gaussian distribution only) &   0.2 & 0.2-1.0 \\
 \hline
Binary& $f_{\rm{pre}}$    &preexisting binary BH fraction &   0.15& 0-0.2 \\
 \hline
Binary& $R_{\max}$    &maximum separation of preexisting binary BH &   $10^5R_{\odot}$& $10^4\;-\;10^5\,R_{\odot}$ \\
 \hline
 \hline
Gas interaction& $f_{\mig}$    & migration factor &   2.0 & 1.0-3.0\\
 \hline
Gas interaction& $\ln\Lambda_{\gas}$    & Coulomb logarithm of gas dynamical friction&   3.1& 1.0-5.0\\
 \hline
Gas interaction& $\Gamma_{\Edd}$    &max accretion rate with respect to Eddington ratio &   1 & 0.5-2\\
 \hline
Gas interaction&  $\eta_c$    & radiative efficiency&   0.1 & considered together using $\Gamma_{\Edd}/(\eta_c/0.1)$\\
 \hline \hline
\end{tabular}
\caption{Parameter List for all variables changed in the simulation. The fiducial values of the parameters except for the parameters plotted are applied to all simulations. }
\label{table:par}
\end{table*}

Our model largely builds on the framework established by \cite{tagawa2020formation}. Our main modification compared to this model aims to enable faster simulations to be able to cover a larger part of the parameter space. Instead of employing N-body simulations, we adopt a statistical approach to efficiently calculate background properties, significantly reducing computational costs. Especially, this approach enables us to simulate systems with high-mass SMBHs and large populations of stellar-mass BHs. Comparable studies using statistical methods have been conducted, focusing on dynamical interactions \citep{gondan2018eccentric} and gas interactions \citep{mckernan2012intermediate,vaccaro2024impact, gayathri2021black, yang2019agn} independently. Additionally, Fokker-Planck numerical methods have been used to investigate extreme mass ratio inspirals near SMBHs \citep{pan2021formation}. Many studies also explore spin evolution due to these mechanisms \citep{gondan2018eccentric, vaccaro2024impact, tagawa2020spin, gayathri2021black}. In the present paper, we consider both the gas hardening process and star interactions, while spin evolution will be explored in a future paper.

Our simulations focus on a system consisting of multiple stellar-mass black holes (BHs) and binary BHs near a central supermassive black hole (SMBH) of mass $M_{\mathrm{SMBH}}$. The system also includes a gaseous AGN disk, a background stellar distribution, and an evolving population of BHs. A flowchart summarizing the simulation framework is presented in Fig. \ref{fig:process}, and Table \ref{table:par} summarizes all free parameters and the range of values explored in our simulations.

The simulations are constrained within radial distances between $r_{\mathrm{in}}$ and $r_{\mathrm{out}}$ from the central SMBH. This radial range is divided into $N_{\cell}=120$ cells with log-uniform distribution. The inner radius, $r_{\mathrm{in}}$, is set to be the inner boundary of the AGN disk, $r_{\mathrm{disk,in}}$, while the outer radius, $r_{\mathrm{out}}$, is chosen to be the minimum value of 5pc and the outer boundary of AGN disk $r_{\mathrm{disk,out}}$ to optimize computational efficiency. We provide a convergence test in Sec.~\ref{sec:convergence}.

The BH and stellar populations are described in Appendix \ref{sec:pre-exist}, which provides their spatial and mass distribution. The Keplerian speed is slightly modified by the presence of the spatially extended stellar populations (Eq.~\ref{eq:kep}). 

The properties of the AGN disk are calculated using the model described in Appendix \ref{sec:AGN}. The inner ($r_{\mathrm{disk,in}}$) and outer ($r_{\mathrm{disk,out}}$) boundaries of the AGN disk are treated as free parameters to estimate their effect, though it could be set by the innermost stable circular orbit (ISCO) of the SMBH. Details are discussed in Appendix \ref{sec:AGN} and \ref{sec:mig}. The simulation begins by determining the accretion rate at the outer boundary of the AGN disk, $\dot{M}_{\mathrm{out}}$, and uses this to compute the disk's properties from the outer radius inward. It is assumed that stars and BHs are formed in the outer region of the AGN disk, which helps to stabilize the disk. Star formation ceases when the accretion heat at the outer boundary is sufficient to prevent gravitational collapse. This typically occurs in systems with either a low SMBH mass or a low accretion rate at the outer boundary. Given the nonzero star formation rate, disk properties are calculated on a denser radial grid, as suggested in Appendix \ref{sec:AGN}, and then translated to the standard values at grid  boundaries and center of $N_{\cell}=120$.

The distribution of BHs within the AGN disk is then estimated as outlined in Appendix \ref{sec:disk_com}. We apply a mass step on a log-uniform scale between the minimum BH mass $M_{\BH,\min,\ini}$ and 3 times the maximum single BH mass, $3M_{\BH,\max,\ini}$. 
The joint surface number density and mass density function $\Sigma(r,M)$ evolves over time,  where the total surface number density of disk BHs is given by $\Sigma_{\DBH}=\int dM\,\Sigma(r,M)$. The background calculations’ time step is discussed in Appendix \ref{sec:disk_com}. The time-evolving disk BH number density $\Sigma_{\DBH}(r,t)$ and mass density $M_{\mathrm{tot},\DBH}(r,t)$ are then established at the radial grid boundaries and subsequently converted to values at the cell centers for individual BH sample simulations.
 
With the background fully established, we begin by considering single or binary BHs, characterized by BH mass $M_{\mathrm{BH}}$, radial position $r$ from the SMBH, velocity components $v_{x,y,z}$ relative to the disk velocity, and, for binaries, the mass ratio $q$ and separation $s$. We examine the evolution of these quantities as the BHs interact gravitationally (Appendix \ref{sec:GW}), with the surrounding gas (Appendix \ref{sec:Gas}) and with the surrounding stellar objects as an evolving background (Appendix \ref{sec:int}).  

\begin{figure*}[!t]
    \centering
    \includegraphics[width=1.0\linewidth]{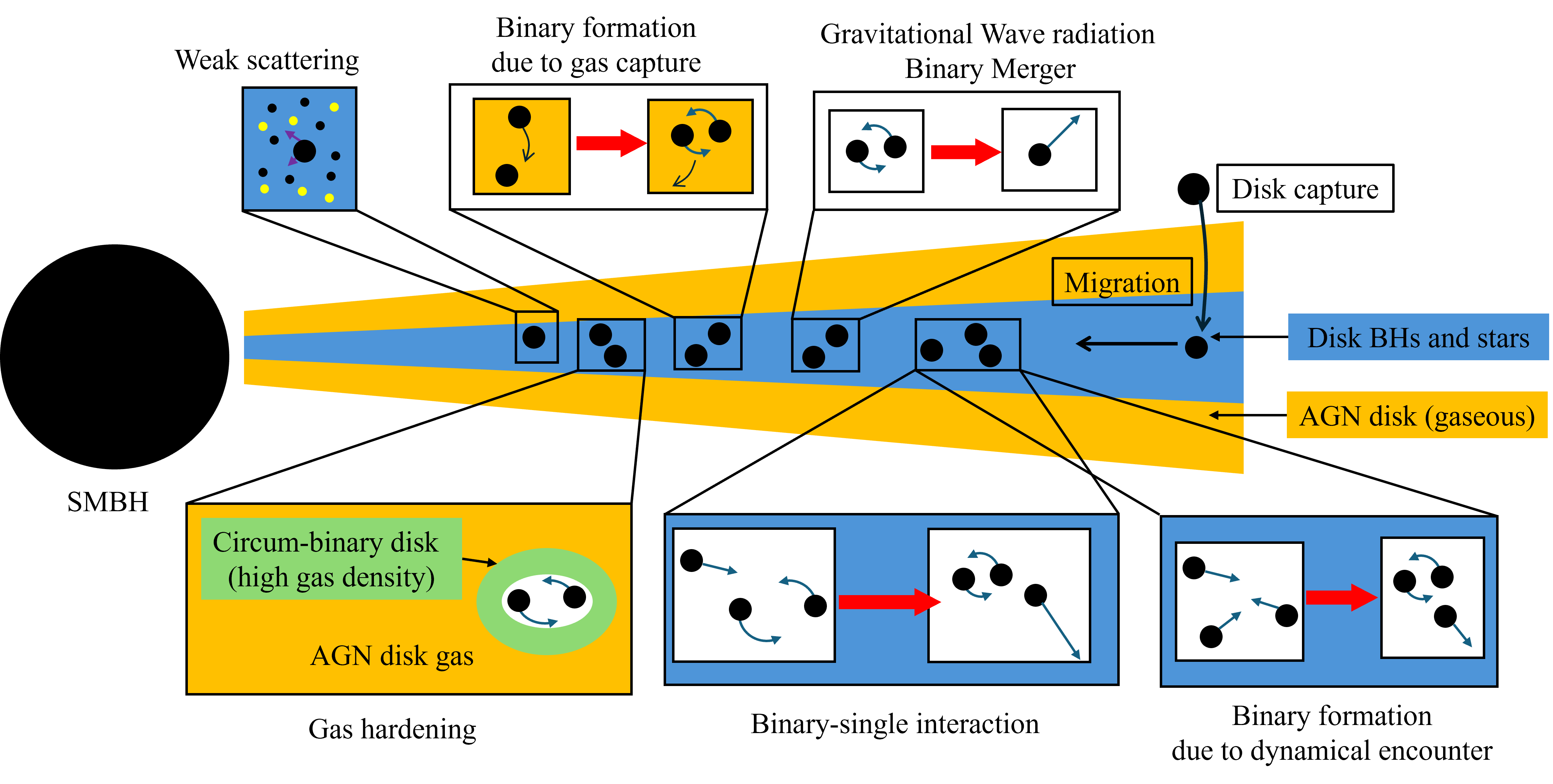}
    \caption{Schematic diagram illustrating the mechanisms incorporated in the program. The orange region represents the AGN disk, while the blue region denotes disk components (including both stars and BHs), which have a much smaller thickness compared to the AGN disk. All interaction types are depicted, though their positions are not drawn to scale.}
    \label{fig:Interactions}
\end{figure*}

Fig.~\ref{fig:Interactions} illustrates all interaction types considered in the program. In the gravitational interaction, we discuss binary hardening due to gravitational wave emission (Appendix \ref{sec:GW_radiatio}) and, eventually, the merger into a new single BH (Appendix \ref{sec:Merger}). The interaction between the BH and the surrounding gas affects mass, position, and velocity through gas accretion (Appendix \ref{sec:acc}), gas dynamical friction (Appendix \ref{sec:GDF}), and type I/II migration (Appendix \ref{sec:mig}). Gas dynamical friction may also assist in binary formation (Appendix \ref{sec:gas_cap}) and the binary hardening process (Appendix \ref{sec:gas_hard}). The interaction between the BH or BH binary and the evolving background includes both weak encounters (weak scattering, Appendix \ref{sec:WS}) and strong encounters, such as binary-single interactions (Appendix \ref{sec:BS_int}) and binary formation through three-body interactions (Appendix \ref{sec:BBF}). These effects renew the BH sample's properties iteratively with time step given in Appendix \ref{sec:time_step} until the AGN disk dissipates.

We record all merger events that occur during the simulation, aiming for at least 10,000 mergers.  However, since the merger rate varies significantly, strictly requiring 10,000 mergers may lead to excessive computational time. To balance efficiency and accuracy, we maintain a minimum of 10 AGNs if the merger rate is low, where the initial number of BHs $N_{\BH,\ini}$ is set by default according to Eq. \ref{eq:N_BH_ini}.  In the individual sample simulation, the radial distance $r$, Keplerian speed $\vkep$ and the disk properties (e.g. gas density $\rho$) are represented by the cell center values to reduce computational cost.

Our program is written in C++ to optimize computational performance. The AGN disk module takes approximately 7 seconds per AGN and shows minimal variation across parameter choices, as the runtime is primarily dominated by data input/output operations. The statistical background calculation for the fiducial models (both Gaussian and isotropic) requires roughly 600 seconds (CPU time) and scales approximately linearly with the AGN lifetime and the number of time intervals specified in Eq.~\ref{eq:time_step_back}. This computational cost also decreases with increasing SMBH mass.

The runtime of the Monte Carlo simulation depends on the chosen model and parameter settings. For instance, the fiducial Gaussian model with $t_{\AGN}=10$ Myr takes 7,825 seconds (CPU time) to simulate 2 AGNs and produce 10,000 merger events. In comparison, the fiducial isotropic model with the same AGN lifetime takes 5,204 seconds (CPU time) to simulate 6.5 AGNs and yield over 10,000 merger events.The runtime does not increase linearly with AGN lifetime but instead grows more slowly for longer AGN lifetime. For instance, with $t=100$ Myr, the Gaussian model requires approximately 129,000 seconds (CPU time) to simulate a single AGN, while the isotropic model takes about 50,000 seconds. Importantly, the Monte Carlo simulations are straightforward to parallelize using the \texttt{OpenMP} package, enabling substantial speed-up when running on multiple CPU cores or computing nodes.

%\section{Method}

\section{Results} \label{sec:results}

\begin{figure*}[!t]
    %\centering
    \includegraphics[width=1.0\linewidth]{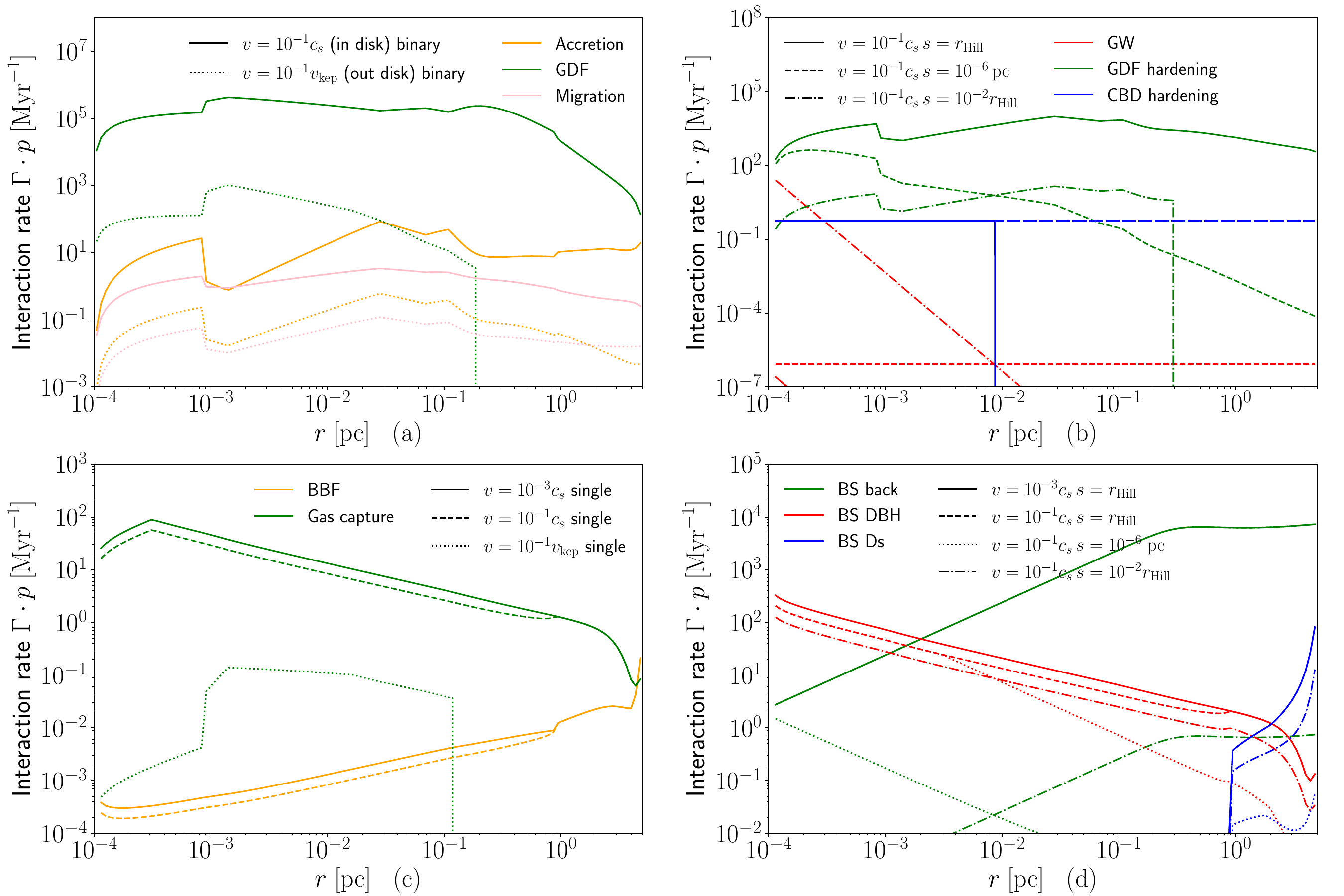}
    \caption{Interaction rate (in $\mathrm{Myr}^{-1}$) of a $10\Msun +10\Msun$ binary BH and single BH $10\Msun$ using the fiducial model with default SMBH mass ($4\times10^6\Msun$) and Gaussian ($\beta_V=0.2$) inclination distribution. The relative velocity to disk and the separation of the binary BH is labeled in the figure. Panel (a) shows how the accretion, migration and the gas dynamical friction rate of different relative velocity at different radius. Panel (b) shows how the hardening rate of a $10\Msun +10\Msun$ binary BH changes due to radius and separation. Panel (c) shows how the binary formation interaction rate of a $10\Msun$ single BH changes at different radius. Panel (d) shows how the binary-single interaction rate with different surrounding components of a $10\Msun +10\Msun$ binary BH changes due to radius and separation. }
    \label{fig:timecale}
\end{figure*}

\subsection{Life of black hole binaries}

We illustrate the interaction rate of binary or single BHs in Fig.~\ref{fig:timecale}. For this analysis, we use SMBH mass $4\times10^6\Msun$, with the background disk parameters evaluated at $t=3$ Myr, adopting a Gaussian inclination distribution with $\beta_v=0.2$. The single BH mass is set to $10\Msun$, while the BH binary mass of two BHs is set to $10\Msun+10\Msun$. The single and binary BHs are assumed to move at relative velocities $v_{x,y,z}=10^{-3}c_s/\sqrt{3}$ (inside the disk components), $v_{x,y,z}=10^{-1}c_s/\sqrt{3}$ (inside the disk but outside the disk components) or $v_{x,y,z}=10^{-1}\vkep/\sqrt{3}$  (outside the disk), where $c_s=\vkep (h/r)$ represents the local sound speed of the AGN disk. The choice of outer-disk BH velocity indicates the inclination angle $i=0.058\,\mathrm{rad}=3.3^{\circ}$. The binary’s separation is taken as the Hill radius $\rHill(r,M=20\Msun)$, which is the most newly formed binaries' separation, $10^{-2}\rHill$,  and a fixed separation $10^{-6}$ pc. In Fig. \ref{fig:timecale}, we account for the fraction of time the BHs spend within the AGN disk $p_{\disk}$ or within the disk components $p_{c}$. Additionally, we set $p_{\rm{uni}}=0.5$ in the shear velocity equation (Eq. \ref{eq:shear}) to mitigate randomness.

In Fig.~\ref{fig:timecale} (a), the interaction rates for a BH in the AGN disk reveal that gas dynamical friction (green) generally dominates over other interactions. The accretion rate (orange) and migration rate (pink) are typically around 1000 times smaller than the gas dynamical friction rate. For BHs located outside the disk (dotted line in Fig.~\ref{fig:timecale} (a) and (c)), the interaction rates are lower due to the factor $p_{\disk}$, particularly for binary formation due to gas capture, which is more likely within the disk environment. Notably, gas dynamical friction does not operate effectively at certain velocities due to the feedback condition Eq. \ref{eq:feedback}, leading to a disappearance in the green line at larger radii, observed in both panels (a) and (c). 

In Fig.~\ref{fig:timecale} (b), we plot the hardening rate for BH binaries within the disk, as gas hardening occurs only in the disk. When a binary forms at a separation of $s \sim \rHill$, gas dynamical friction plays a dominant role in hardening the system. As the separation decreases quickly to a smaller value $10^{-2}\rHill$, the effectiveness of dynamical friction diminishes, while interactions with the circum-binary disk become increasingly significant. These two processes continue to shrink the binary orbit until gravitational wave emission eventually takes over, driving a rapid merger. When the BH binary migrates to smaller radii, the Hill radius $\rHill$ decreases, resulting in a stronger gas dynamical friction rate. Consequently, the transition from gas-hardening dominance to gravitational-wave dominance occurs over a shorter timescale, effectively accelerating the hardening process (e.g. dash-dot line at small radii) . This progression aligns well with our previous understanding of binary mergers in AGN disks.

In Fig.~\ref{fig:timecale} (c), binaries form at a very high rate within the AGN disk, primarily due to gas capture. The time fraction spent in the disk components, $p_{c}$, has only a minor effect on the interaction rate (solid and dashed lines) when the BH is inside the disk. This process provides an abundant supply of binaries within the AGN disk, facilitating the formation of higher-generation BH binaries from remnant single BHs.

During the migration and hardening process, binary-single interactions with disk stars play a crucial role in shaping binary dynamics. As shown in Fig.~\ref{fig:timecale} (d), interactions with background spherical stars (green line)  regularly soften newly formed binaries ($s\sim\rHill$) at large radii. However, as the binary separation decreases, the impact of these interactions diminishes significantly. The interaction rate with disk BHs (red line) and disk stars (blue line) remains relatively unaffected by binary separation and the relative velocity $v$. Within the AGN disk, binaries frequently undergo binary-single interactions with disk BHs, especially at smaller radii. These encounters rapidly harden the binary but can also eject it from the disk, effectively halting gas-driven hardening and slowing its inward migration toward the center.

We can then construct a binary evolution pathway similar to that presented in Fig.~5 of \cite{tagawa2020formation}. In AGN disks, BH binaries can form at larger radii, migrate inward, while rapidly hardening through gas dynamical friction. As the binary separation decreases, gas dynamical friction becomes less effective, and circum-binary disk interactions take over as the dominant hardening mechanism. At very small separations, gravitational wave emission ultimately governs the final stage of hardening, leading to the binary’s merger into a single BH. Binaries at smaller radii typically transition more quickly from the gas dynamical friction hardening phase to the gravitational wave-driven hardening phase. However, during the inward migration, binaries may undergo multiple binary-single interactions with other disk objects, which can further harden the system but eject the binary from the AGN disk. Ejected binaries can often reenter the disk relatively quickly, resuming their inward migration and continuing to harden through gas interactions. This dynamic interplay of migration, hardening, and interactions shapes the overall evolution and merger likelihood of BH binaries in AGN disks. Our results align well with those of \cite{tagawa2020formation}.

\begin{figure*}[!t]
    \centering
    \includegraphics[width=1.0\linewidth]{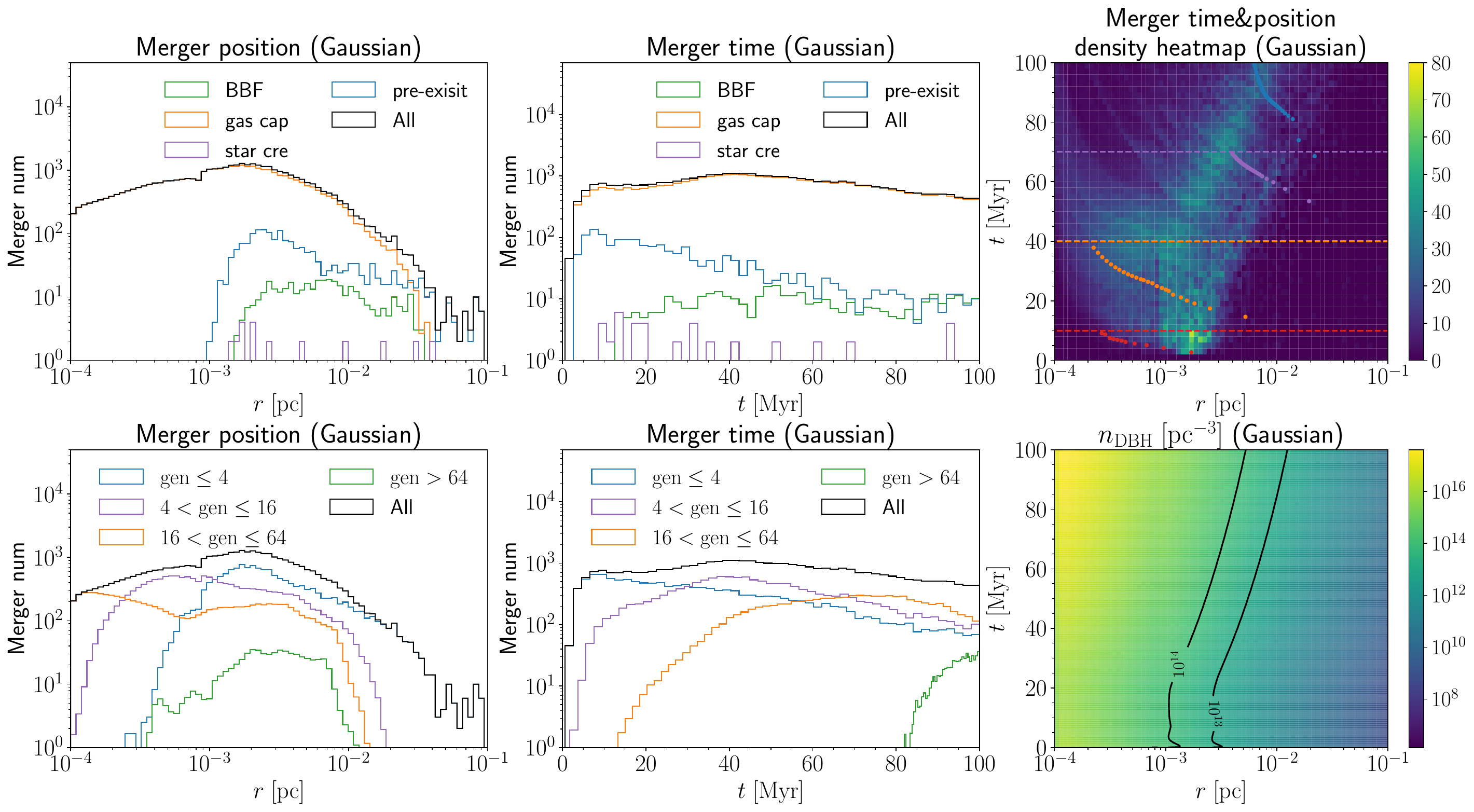}
    \includegraphics[width=1.0\linewidth]{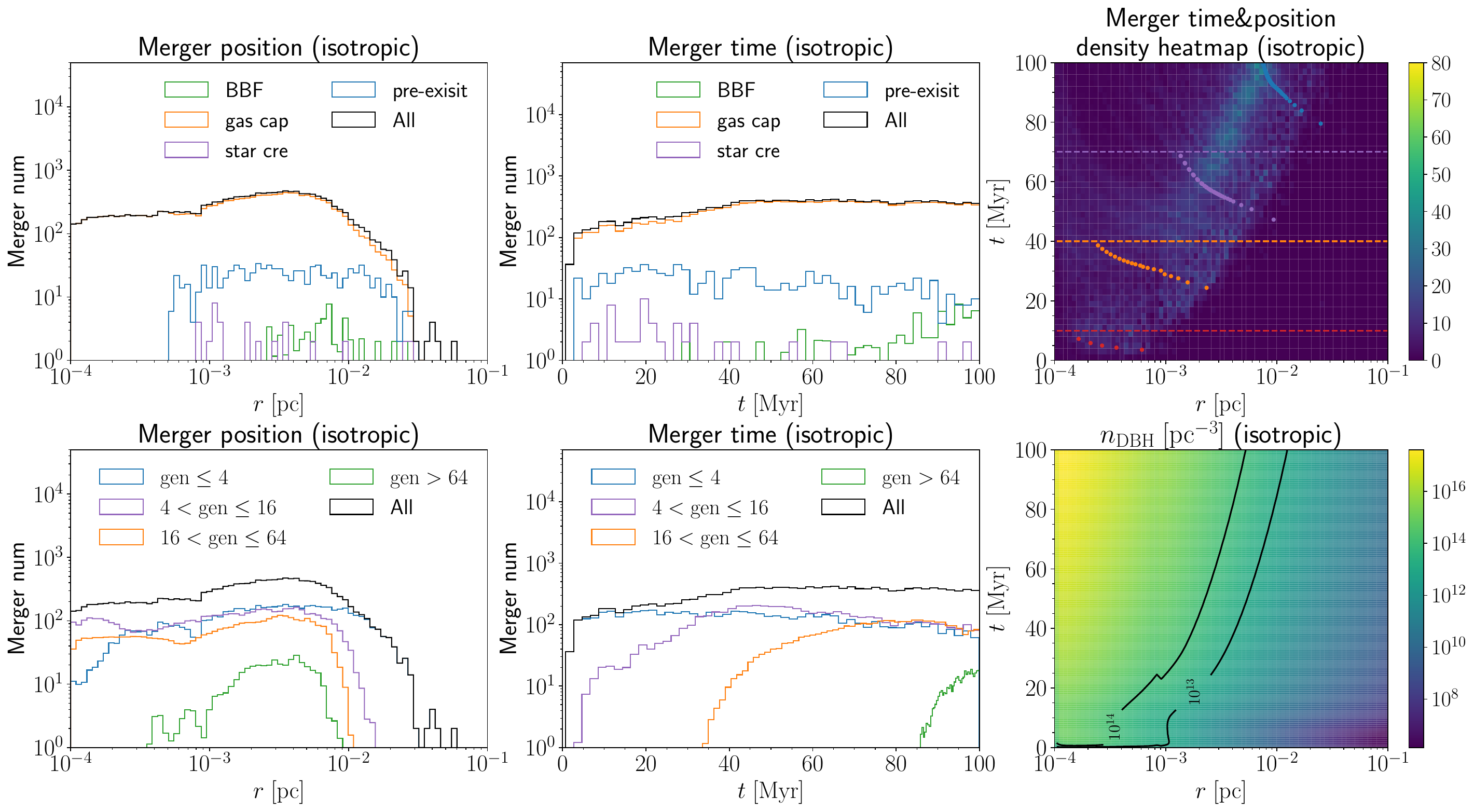}
    \caption{Merger positions and times over 100 Myr (fiducial AGN lifetime: 10 Myr) for the Gaussian model (top two rows) and the isotropic model (bottom two rows). The first column shows the spatial distribution of mergers, while the second column presents the time distribution of mergers, reflecting the shape of the instantaneous merger rate. The black lines represent all mergers, while colored lines distinguish different binary formation types or generations of merger remnants. The third column displays density heatmaps of merger position versus time for the Gaussian model (first row) and isotropic model (third row), with corresponding disk BH number density evolution shown in the second and fourth rows, respectively. The heatmaps also trace the merger history of the most massive merger before 10, 40, 70, and 100 Myr, marked by red, orange, purple, and green dots, respectively. In the number density plots, contours at $10^{13}\;\rm{pc^{-3}}$ and $10^{14}\;\rm{pc^{-3}}$ are overlaid, corresponding well to the bright, high-merger-density regions in the heatmaps.}    
    \label{fig:merger_feature}
\end{figure*}

\subsection{Merger features}
\label{sec:merger_feature}

In the binary formation channel, Eqs.~\ref{eq:BBF} and \ref{eq:cap} introduce a double-counting effect. To correct for this, we assign a weighting factor to each merger, given by the inverse of the remnant BH’s generation $w=1/\rm{gen}$. The classification of BH generations is defined as follows:  preexisting binaries and binaries formed within the AGN disk are designated as the zeroth generation. After a merger, their remnant single BHs become first generation BHs.  Additionally, preexisting single BHs and those originating from the AGN disk are also classified as first generation. When an $n$th generation BH forms a binary and subsequently merges, the remnant single BH is classified as $(n+1)$th generation. 

\begin{figure}
    \centering
    \includegraphics[width=1.0\linewidth]{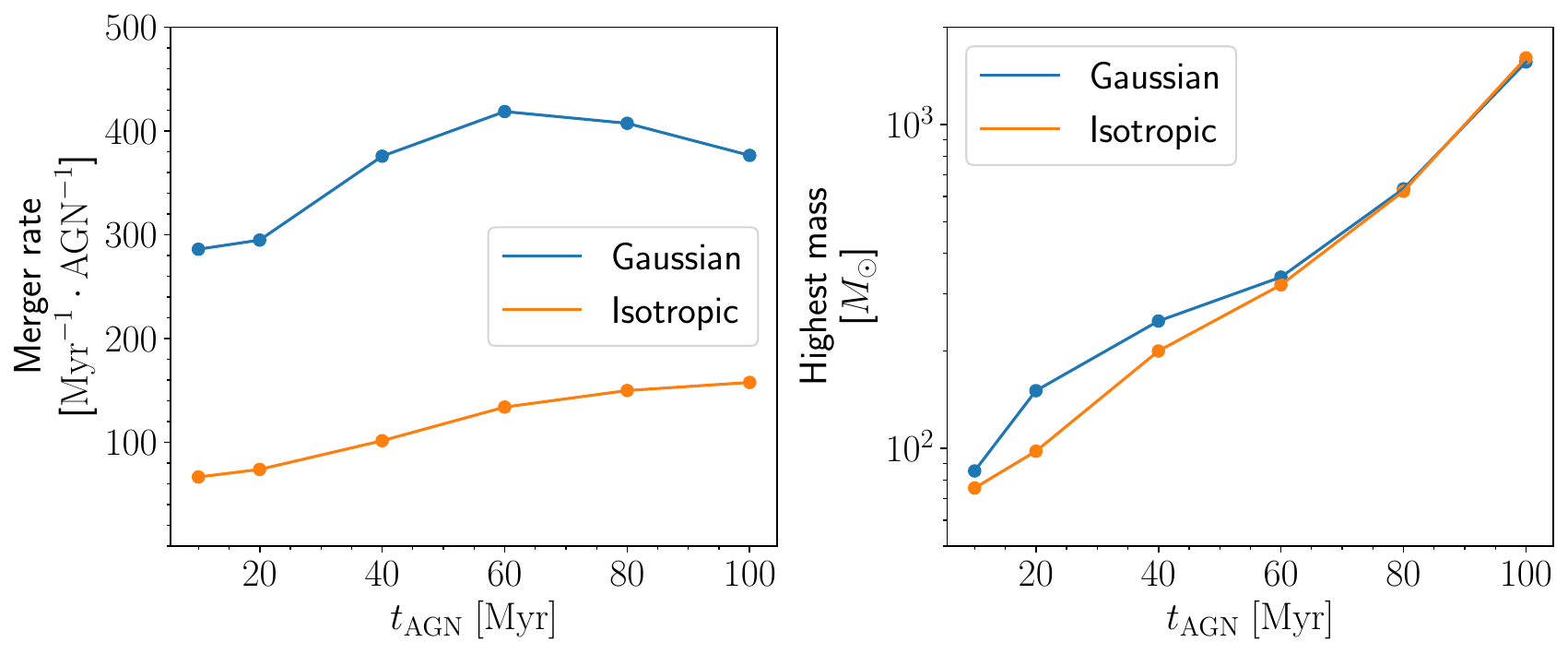}
    \caption{The averaged merger rate and highest 1\% merger mass as a function AGN lifetime $t_{\AGN}$. The blue line represents the merger rate for a Gaussian distribution inclination, and the orange line is for an isotropic distribution inclination. The merger rate figure is the averaged merger rate of Fig.~\ref{fig:merger_feature}.}
    \label{fig:t_AGN}
\end{figure}

We primarily consider two inclination (velocity) models: an isotropic distribution and an anisotropic Gaussian distribution with $\beta_v=0.2$. These two models behave very differently in the number of BHs embedded in the disk due to the inclination difference, as shown in Fig.~\ref{fig:inclination_dis}. Fig.~\ref{fig:merger_feature} illustrates the radial and time distribution of mergers for both the Gaussian model (top two rows) and the isotropic model (bottom two rows), categorized by binary formation channel or remnant generation. We assume an AGN lifetime of 100 Myr; however, in our model, the AGN lifetime solely determines the simulation duration and does not affect the AGN disk structure or BH distribution properties. Consequently, the characteristics of mergers in a shorter AGN lifetime scenario can also be inferred from the figure. The averaged merger rate and top 1\% merger mass is plotted in Fig.~\ref{fig:t_AGN}.

It is evident that mergers originating from gas-captured binaries dominate the merger events, occurring at rates approximately $\mathcal{O}(10)$ times higher than those of preexisting binary mergers, both with a slight decline over time. This decrease occurs as disk BHs gradually deplete, while mergers from the dynamical formation and star formation channels continue at a steady rate. Overall, the merger rate across all binary types remains roughly constant, aligning well with Fig.~7 in \cite{tagawa2020formation}.

At very early time, we observe a rapid rise in the merger rate, which stabilizes around $t\approx5$ Myr. The rate peaks at $t\approx50$ Myr and then slightly decreases for the Gaussian model, while remaining approximately constant for the isotropic model. This divergence stems from the difference in initial inclinations: the Gaussian model contains fewer BHs with large initial inclinations. As a result, the averaged merger rate shows only weak dependence on AGN lifetime, as illustrated in the left panel of Fig.~\ref{fig:t_AGN}.

The merger events are clustered at approximately $10^{-3}$ pc at early time, with their positions being closer to the SMBH in the isotropic model. These regions of frequent mergers generally correspond to disk BH number densities of $n\approx10^{13}-10^{14}\;\rm{pc^{-3}}$. Such high number densities (calculated by Eq. \ref{eq:DBH_density}) arise from the extremely low scale height of the disk BH component, as discussed in Appendix~\ref{sec:v_disp}. As time progresses, the location of the critical density shifts outward as shown in the number density map, causing the gathered merger positions to increase correspondingly in the density heatmap. A small bump is observed in the merger position figure due to the sudden slow down of migration speed at $r\sim 10^{-3}$ pc (a discontinuous migration trap), which is also visible in the density heatmap of merger time and mass.

Merger remnant generations can reach extremely high values, with a maximum of approximately 100 observed at late times. The highest-generation mergers closely follow their merger time and location, as these BHs require extended time in dense regions of the disk BHs to undergo successive mergers. Mergers of generation greater than 64 are concentrated in the region $10^{-3}-10^{-2}$ pc, where the high disk BH density facilitates frequent binary-single interactions, providing a continuous supply of binary seeds. Moreover, the high gas density in this region enables BHs to rapidly reenter the disk, accelerating the hardening and merger process. These very high-generation BHs are typically formed at late times ($t\gtrsim 80$ Myr), with their merger masses growing significantly through repeated mergers.

\subsection{Merger rate and highest mass}
\label{sec:Par_rel}
\begin{table*}[t]
\begin{tabular}{ |p{2.7cm}|p{2.25cm}|p{2.0cm}|p{2.0cm}|p{2.0cm}|p{2.0cm}|}
 \hline
   \multicolumn{6}{|c|}{Parameter relevance $R$}\\
 \hline
     Type& Symbol& Merger rate (Gaussian)&Highest mass (Gaussian) & Merger rate (Isotropic)&Highest mass (Isotropic)\\
     \hline
     \hline
 SMBH&$\log_{10}(M_{\SMBH}/\Msun)$&{3.271664742}	& 2.283997134		
	& 3.227080137	& 1.481548969
\\
   \hline 
   \hline

AGN disk&$r_{\disk,\out}$&0.150934513	&{0.479450428}
&0.325568885	&0.254059658
\\
 \hline
 AGN disk&$r_{\disk,\rmin}$&0.367946993	& 0.470898416	&0.227961833	&0.048850026
\\
 \hline
AGN disk&$\dot{M}_{\out}/(0.1\dot{M}_{\Edd})$&0.128298205	&0.149516024	&0.123283759	&0.01802458
\\
 \hline
AGN disk&$m_{\mathrm{AM}}$& 0.975586098	& 0.568123545	& 0.869958957	&0.213075973
\\
 \hline
 
AGN disk&$\alpha_{\mathrm{SS}}$&0.26617256	&0.291467832	&0.29347429	 &0.234098288
    \\
     \hline
    AGN disk&$\epsilon$& 0.476165094	&0.214333161	&0.148482635	&0.048634006
\\
 \hline
 AGN disk& $t_{\AGN}$ &0.241828557			
	& 1.094862877	& 0.51226985	
 & 1.115050771
\\
 \hline
AGN disk&$\beta_{\cre}$&0.079517332	&0.113970213	&0.315327243	&0.060308176
\\
\hline
\hline
Mass distribution& $\beta_{\IMF}$&0.085197411	&0.222676003	&0.219689462	&0.209508684
\\
 \hline
 Mass distribution&$M_{\mathrm{BH,ini,max}}$&0.051343959	&0.199015371	&0.150541924	& 0.609525054
\\
 \hline
 Radial distribution&$1+\gamma_{\rho}$& 0.71035059	&0.174521094	& 1.519590275	& 0.440812756
\\
 \hline
 Radial distribution&$N_{\BH,\mathrm{ini}}$&  0.995658439	&0.273836794	& 1.043597729	&0.168824286
\\
 \hline
 Radial distribution&$r_{\BH,\out}$&0.911899634	&0.20054724	&1.16909195	&0.143582742
\\
   \hline
 Velocity distribution& $\beta_v$& 1.201186702	&0.086831942 &NA	&NA
\\
 \hline
 Binary&$f_{\preexist}$&0.076102625	&0.023543428	&0.114463463	&0.132984591
\\
 \hline
 Binary&$R_{\max}$&0.064121306	&0.069069019	&0.071751937	&0.026716585
\\
 \hline
 \hline
 Gas interaction&$f_{\mig}$& 0.966145685	&0.463338762	&1.06062715	&0.17690834
\\
 \hline
 Gas interaction&$\ln\Lambda_{\gas}$&0.047618704    &0.08551462	&0.179980313	&0.080284549
\\
 \hline
 Gas interaction&$0.1\Gamma_{\Edd}/\eta_c$&0.147286371	&0.159534718	&0.055696851	&0.113834999
\\
\hline
 Gas interaction&$\alpha_{\CBD}$&0.016120167	&0.023155906	&0.030935634	&0.006721593
\\
 
\hline
\end{tabular}

\caption{The parametric relevance $R$ (defined in Eq. \ref{eq:relevance}) for preexisting BHs (including the star forming "preexisting" binaries) during the simulation under isotropic and Gaussian distributions of inclination. The plot range of each parameter is listed in Table \ref{table:par}. $R>0.75$ indicates that the parameter plays a dominant role, $0.4<R<0.75$ suggests that the parameter is relevant but less influential in determining the merger rate or maximum BH mass for the Gaussian and isotropic distributions. We estimate the relevance of $(1+\gamma_{\rho})$ instead of $\gamma_{\rho}$ to remove the singularity in Eq. \ref{eq:relevance}.}
\label{table:rel}
\end{table*}

\begin{figure*}[!t]
    \centering
    \includegraphics[width=1.0\linewidth]{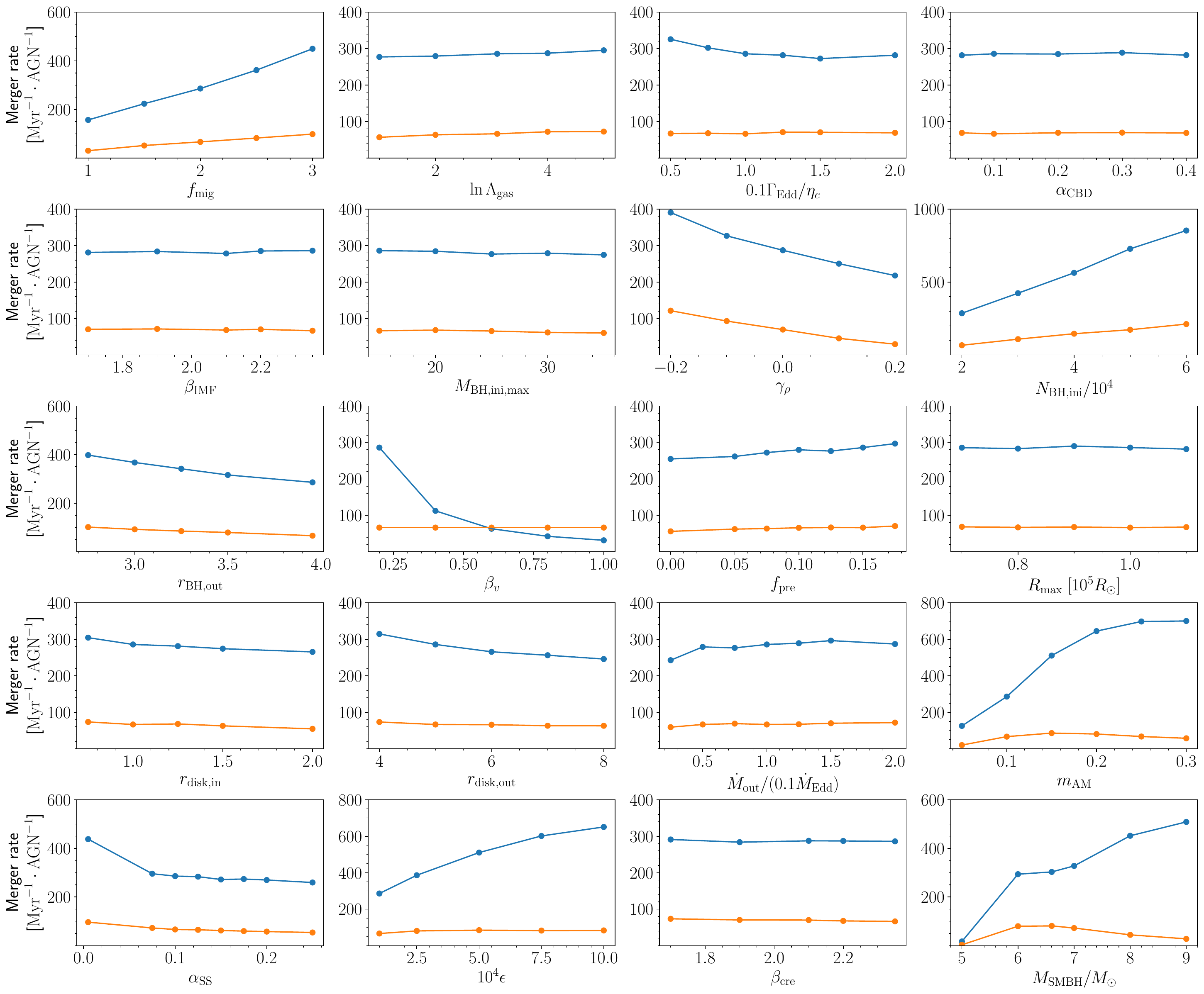}
    \caption{The averaged merger rate as a function of different parameters listed in Table \ref{table:par} (except for $t_{\AGN}$). The blue line represents the merger rate for a Gaussian distribution inclination, and the orange line is for an isotropic distribution inclination.}
    \label{fig:merger_rate}
\end{figure*}

\begin{figure*}[t]
    \centering
    \includegraphics[width=1.0\linewidth]{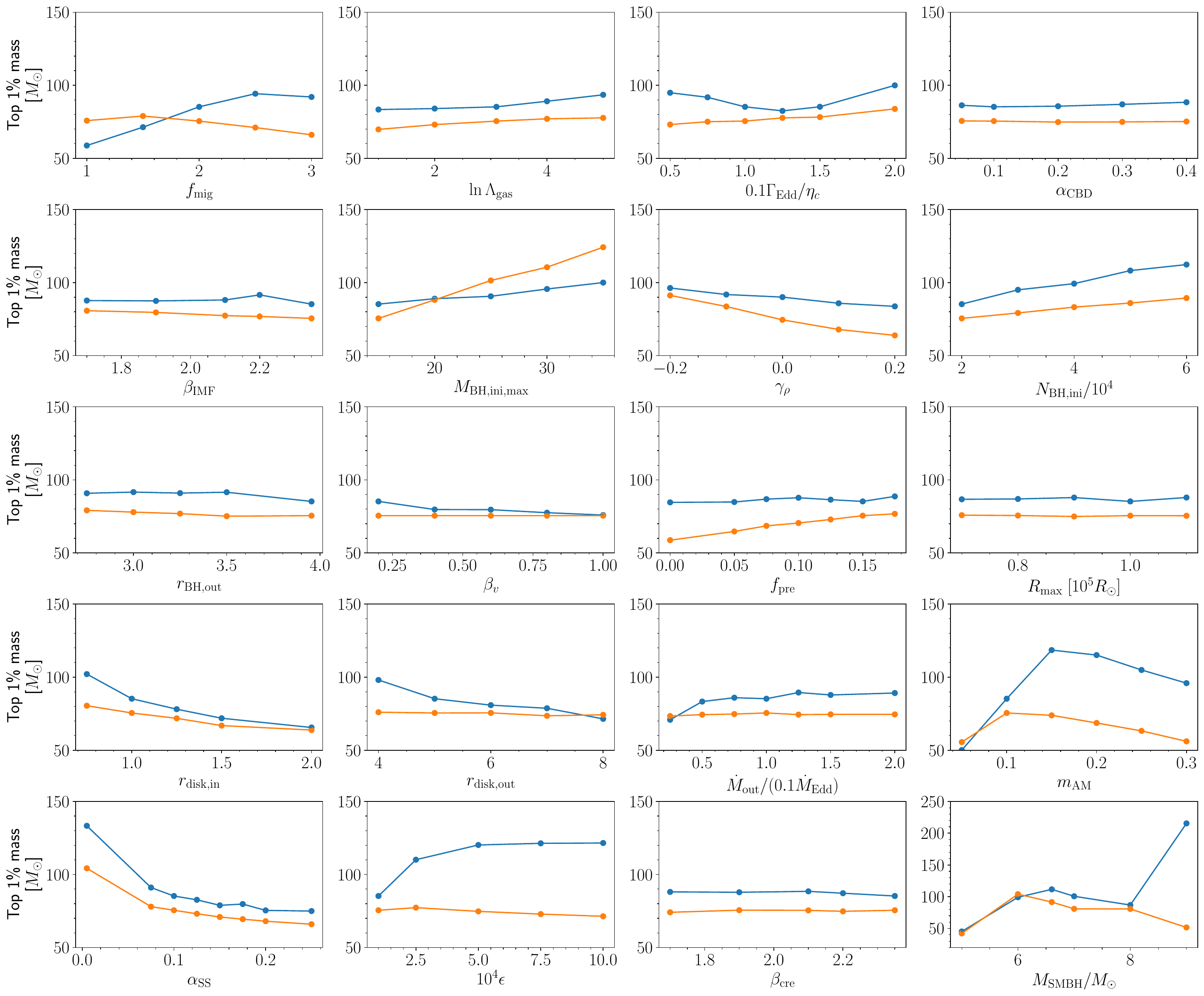}
    \caption{Highest 1\% merger mass as a function of different parameters listed in Table \ref{table:par} (except for $t_{\AGN}$). The blue line represent the merger rate for a Gaussian distribution inclination, and the orange line is for an isotropic distribution inclination. }
    \label{fig:merger_mass}
\end{figure*}
We then estimate the average merger rate during the AGN disk lifetime (in $\mathrm{Myr}^{-1}\cdot\mathrm{AGN^{-1}}$) and average mass of the top 1\% merger mass ($M_{\BH}=M_1+M_2$ in $\Msun$) according to their merger generations across different parameter configurations listed in Table \ref{table:par}. The average merger rate is given by        
\begin{equation}
    \Gamma_{\mathrm{merger}}=\frac{1}{t_{\AGN}}\sum\frac{1}{\mathrm{gen}},\label{eq:weight_rate}
\end{equation}
  while the average mass of the top 1\% of mergers is computed as
 \begin{equation}
     M_{\mathrm{top}}=\left(\sum\frac{M_{\BH}}{\mathrm{gen}}\right)/\left(\sum\frac{1}{\mathrm{gen}}\right)\;\;\text{until }\sum\frac{1}{\mathrm{gen}}\ge1\%.\label{eq:weight_mass}
 \end{equation}
 Here, the summation $\sum$ is performed in descending order of BH merger mass. We plot the merger rate and highest mass varied with the parameter (Table~\ref{table:par}) in Fig.~\ref{fig:merger_rate} and Fig.~\ref{fig:merger_mass}.

 To assess the impact of various parameters, we define the parametric relevance $R$ as:
\begin{equation}
    R=\left(\frac{A_{\max}-A_{\min}}{A_{\max}+A_{\min}}\right)/\left(\frac{f_{\max}-f_{\min}}{f_{\max}+f_{\min}}\right),\label{eq:relevance}
\end{equation}
where $A$ is the merger rate $\Gamma_{\mathrm{merger}}$ or highest mass $M_{\mathrm{top}}$, and $f$ denotes the parameters varied in the simulation. A parameter is considered \textit{significant} if $R\gtrsim 1$, whereas \textit{irrelevant} parameters generally correspond to $R\sim \mathcal{O}(0.1)$. Parameters with \textit{moderate relevance} $R\approx 0.5$. The corresponding results are listed in Table \ref{table:rel}. 

\subsubsection{Impact of AGN lifetime}

The AGN lifetime $t_{\AGN}$ is a unique disk parameter in that it only extends the simulation duration without modify any interaction strength or disk properties directly. The maximum merger mass increases exponentially with AGN lifetime, without affecting the merger rate a lot, as shown in the Fig.~\ref{fig:t_AGN}. In the density heatmap of Fig.~\ref{fig:merger_feature}, we trace the merger history of the most massive mergers before 10, 40, 70, and 100 Myr. The merger positions of 1st generation and the inward migration speeds of these BHs progressively slow down due to increasing disk BH number densities over time. At early times $t\lesssim 40$ Myr, the maximum BH masses differ slightly between the Gaussian and isotropic models. However, these differences diminish at longer AGN lifetimes, primarily reflecting disparities in BH number density during the initial stages of disk evolution.

Directly determining the ages of AGNs remains observationally challenging. Nevertheless, both theoretical frameworks and indirect observational approaches have been developed to estimate AGN lifetimes across cosmic time \citep{enoki2014anti, schawinski2015active, adelberger2005possible, miller2003environment}. Given the strong sensitivity of the maximum BH merger mass to AGN lifetime observed in our simulations, this relationship may offer a novel method for inferring the current age of AGN disks—measured from the onset of disk formation—based on the most massive binary mergers. This approach provides a complementary alternative to traditional optical diagnostics.

\subsubsection{Impact of gas interaction}
\label{sec:gas_int_impact}
\begin{figure*}
    \centering
    \includegraphics[width=1.0\linewidth]{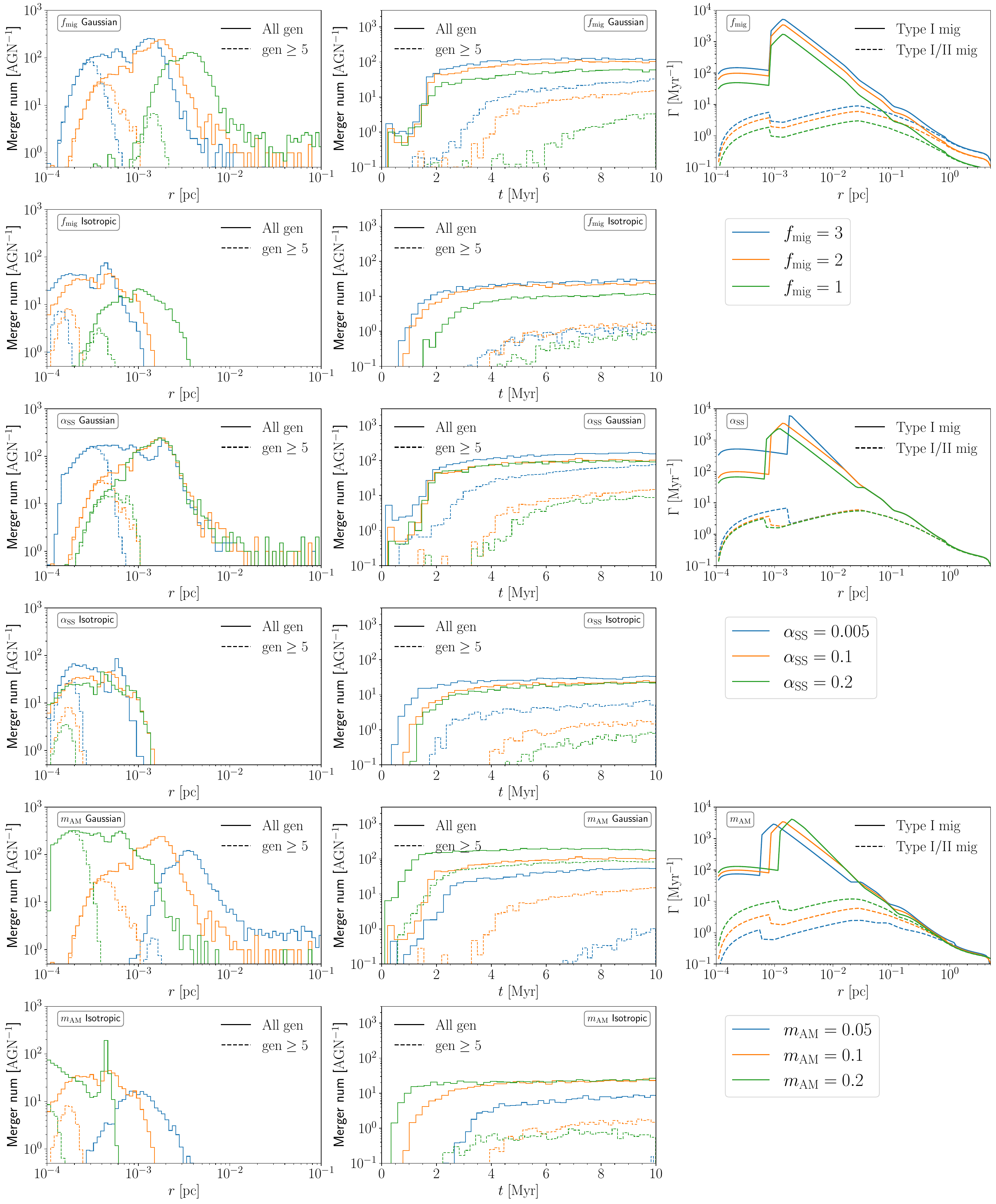}
    \caption{Correlation between merger time, position, and migration rate for isotropic and Gaussian distributions due to different parameter variations. The plotted merger radii range from $10^{-4}\sim10^{-1}$ pc, covering the region where most mergers occur. The dashed line represents the distribution of remnants with generation $\ge 5$ where most high-mass mergers are found. The three rightmost panels show the migration rate for a $10\Msun$ BH comparing type I migration (solid line) and type I/II migration (dashed line). This plot includes variations in the migration factor $f_{\mig}$, vicious $\alpha$ parameter $\alpha_{\rm{SS}}$, and angular momentum transport rate $m_{\rm{AM}}$.}
    \label{fig:merger_par_1}
\end{figure*}

\begin{figure*}
    \centering
    \includegraphics[width=1.0\linewidth]{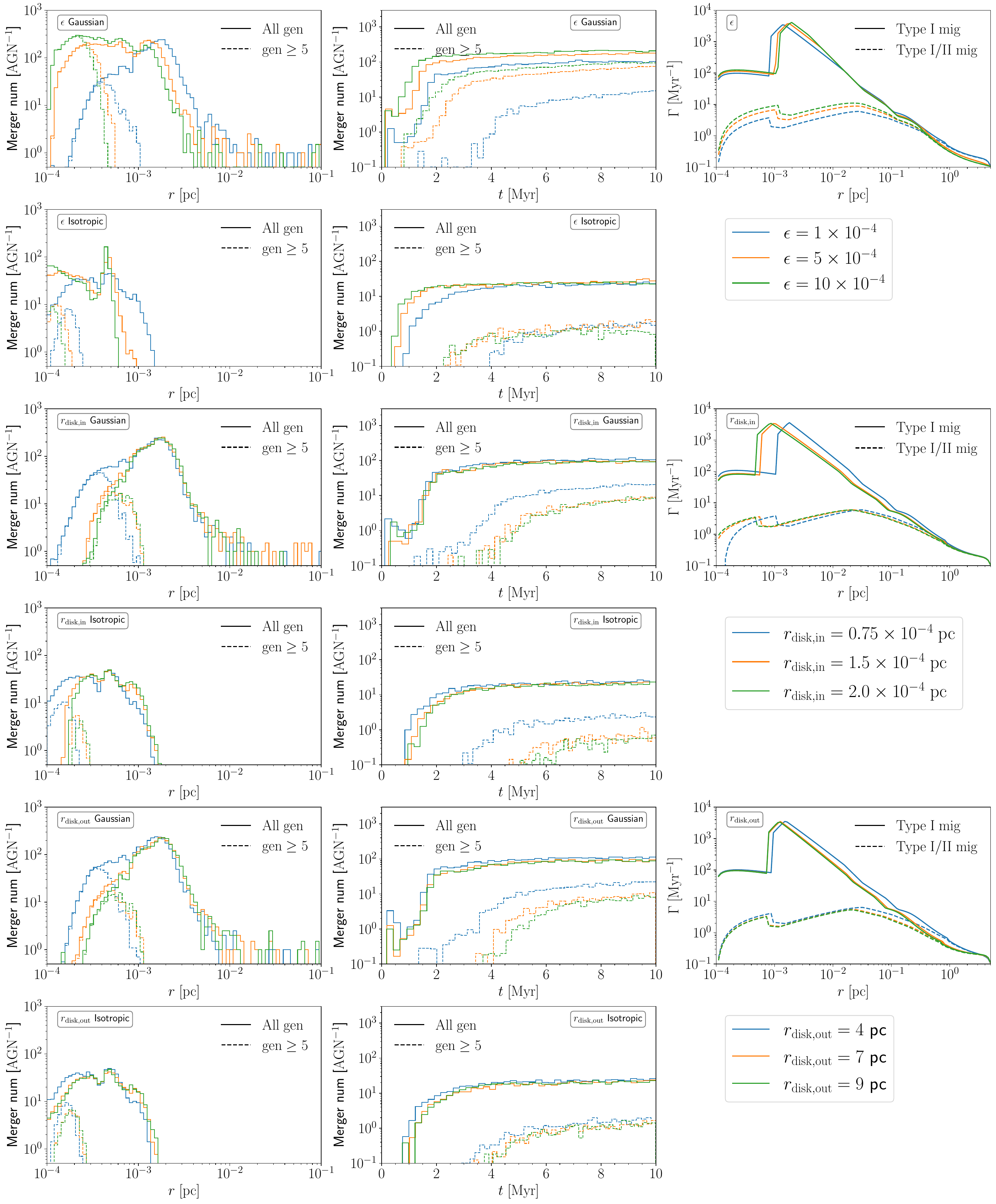}
    \caption{Same as Fig.~\ref{fig:merger_par_1}. This plot includes variations in the conversion efficiency $\epsilon$, AGN disk inner size $r_{\disk,\rmin}$, and AGN disk outer size $r_{\disk,\out}$. }
    \label{fig:merger_par_2}
\end{figure*}

\begin{figure*}
    \centering
    \includegraphics[width=1.0\linewidth]{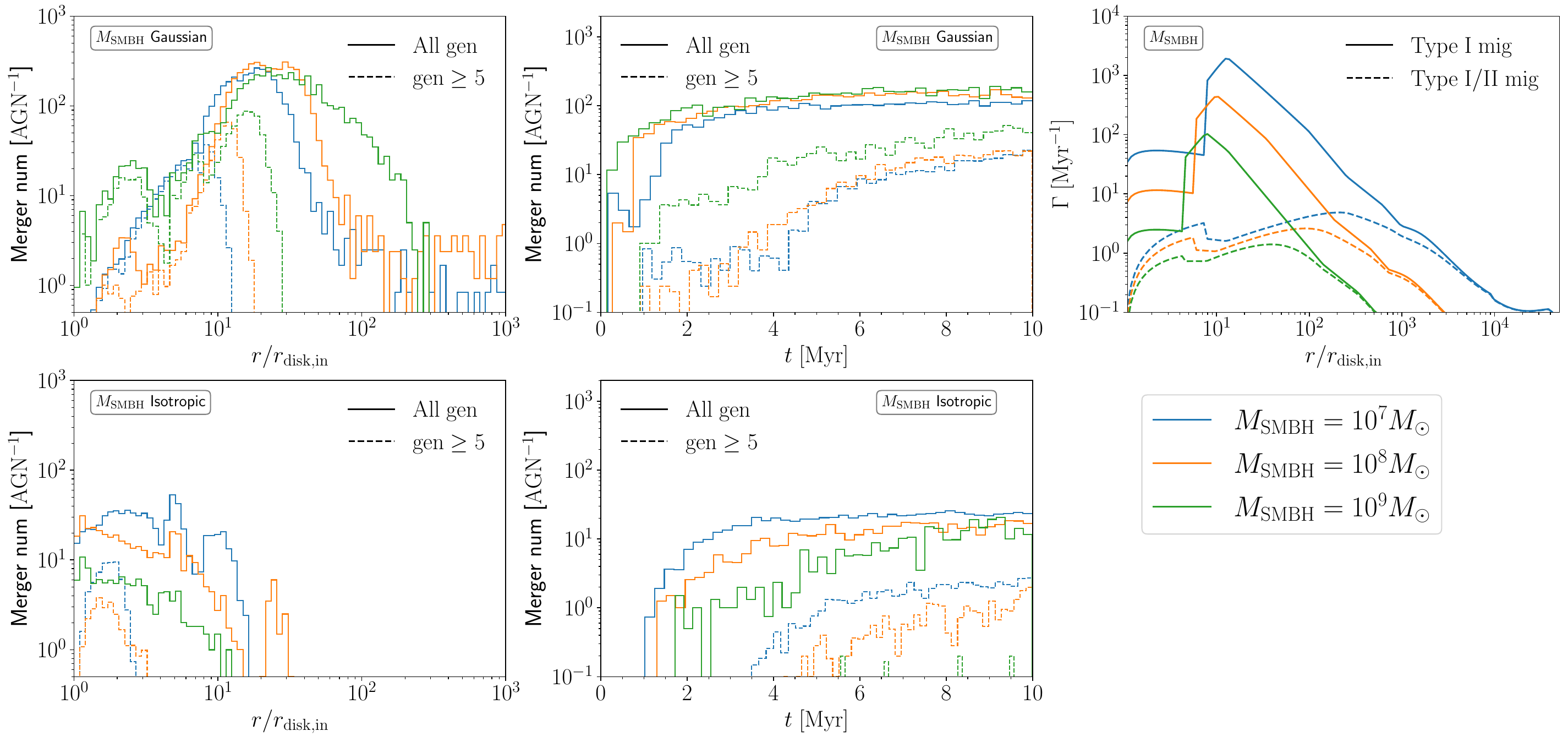}
    \caption{Same as Fig.~\ref{fig:merger_par_1}. This plot includes variations in the SMBH mass $M_{\SMBH}=10^7-10^9\Msun$. The radius is rescaled to $r/r_{\disk,\rmin}$.}
    \label{fig:merger_par_SMBH}
\end{figure*}
The gas interaction factors considered are $f_{\mig}$, $\Gamma_{\Edd}$, $\ln\Lambda_{\gas}$, and the circum-binary disk viscous parameter $\alpha_{\CBD}$, representing the interaction strength of migration, maximum accretion, gas dynamical friction (both for velocity and gas hardening), and gas hardening due to circum-binary disks. Only the migration factor $f_{\mig}$ is significant affecting the merger rate and on the merger mass. 

 Faster migration speeds (larger $f_{\mig}$) mean that BHs spend less time in low gas density regions, thereby accelerating their mergers. Additionally, BHs concentrated in high-density gas regions experience more frequent binary-single interactions, which enhance binary hardening but can also reduce migration speed by ejecting BHs. As illustrated in the upper two lines of  Fig.~\ref{fig:merger_par_1}, the merger time decreases and the merger position gets closer to the center as the migration speeds up. 

Interestingly, $f_{\mig}$ appears to have little effect on the highest merger mass in the isotropic distribution. While increasing migration speed slightly increases the maximum merger mass in the Gaussian model, the isotropic model reduces the maximum mass slightly. As seen in Fig.~\ref{fig:merger_par_1}, high-generation mergers of isotropic model occur at smaller radii, very close to $r_{\rmin}$, and exhibit little dependence on $f_{\mig}$ when $f_{\mig}>2$. Additionally, in the Gaussian model, no BHs are observed migrating into the $r<r_{\rmin}$ region, whereas in the isotropic model, the BHs migrating into $r<r_{\rmin}$ increases as $f_{\mig}$ increases. This difference arises due to the larger disk BH number density in the Gaussian model, leading to more frequent binary single interactions and the increased potential for higher-generation mergers. Notably, under realistic conditions, $f_{\mig}$ varies with radial distance on a small scale but remains around $f_{\mig}\sim2.0$, as discussed in Sec.~\ref{sec:mig_trap}.

The maximum accretion rate $\Gamma_{\Edd}$ which typically increases BH masses, has a low relevance $R\sim0.2$ for the merger mass, making it nearly insignificant.  In Figure~15 of \cite{tagawa2020formation}, the accreted mass of BH binaries during mergers is generally $\mathcal{O}(0.1)$ of their initial mass, and our results align with this conclusion.

The gas dynamical friction factor $\ln\Lambda_{\gas}$, which affects the rate of BHs migrating into the disk and %ffects 
the scale height of disk components on an $\mathcal{O}(1)$ scale, is largely irrelevant. Its impact is limited to scenarios where BHs are near/in the AGN disk, and it %s impact 
diminishes logarithmically outside the AGN disk, as shown in Figure~\ref{fig:GDF}. 

The circum-binary disk viscous parameter $\alpha_{\CBD}$ is of minor importance, with $R_{\CBD}\sim\mathcal{O}(0.1)R_{\GDF}$, indicating that gas hardening due to the circum-binary disk is a secondary effect compared to gas dynamical friction hardening. This conclusion is supported by hydrodynamical simulations \citep{baruteau2010binaries,li2022long,rowan2023black}.

\subsubsection{Impact of BH populations}
The BH mass parameters naturally play a significant role in determining merger masses. As expected, increasing the maximum mass of preexisting single BHs leads to a higher upper limit on merger mass, while the overall merger rate remains relatively unchanged. However, we observe that the initial mass function index $\beta_{\IMF}$ and the initial maximum single mass $M_{\mathrm{BH,ini,max}}$ exhibit little relevance to either the merger rate or the highest merger mass, except for the highest mass of the isotropic distribution. This suggests that increases in individual BH masses are diluted by frequent binary-single interactions and binary formation, which promote hierarchical mergers across multiple generations. In contrast, the isotropic distribution, which has fewer disk BHs, relies more heavily on the initial single BH mass to determine the highest merger mass.

Parameters related to the spatial distribution significantly affect the merger rate but have minimal impact on merger masses. These parameters,  $\gamma_{\rho}$, $N_{\BH,\ini}$ and $r_{\BH,\out}$ determine the BH number density near the SMBH (Eq.\ref{eq:n_BH_gamma_rho}), influencing the merger rate without substantially altering the highest merger mass. The inclination distribution, particularly $\beta_v$ in the Gaussian model, governs the number of BHs that settle into the disk, serving as potential seeds for binary formation and binary-single interactions. All these parameters are at least relevant to the merger rate, and changing $\gamma_{\rho}$ becomes a more effective way to increase the number density of disk BHs for the isotropic model.

The preexisting binary fraction $f_{\mathrm{pre}}$ and the initial maximum separation $R_{\max}$ both have negligible influence on both the merger rate and the highest BH mass. These two parameters mainly determine the gen=1 BH remnants, which is a secondary effect compared to the higher generation binaries both to merger rate and highest mass. What's more, the separation rapidly decreases due to gas dynamical friction hardening as BHs migrate into the disk, making the initial separation less important.

\subsubsection{Impact of disk parameters}
In our analysis of disk parameter effects, we examine the influence of the AGN disk size ($r_{\disk,\rmin},r_{\disk,\out}$), accretion rate at the outer disk boundary $\dot{M}_{\out}$, the angular momentum transport rate $m_{\rm{AM}}$, Shakura-Sunyaev $\alpha$-parameter $\alpha_{\mathrm{SS}}$, the conversion efficiency $\epsilon$, and the star-formation initial mass function index $\beta_{\cre}$.  Some disk variables, like $m_{\rm{AM}}$, are analyzed beyond their typical range ($m_{\rm{AM}}$ is normally 0.1$-$0.2) to assess their influence. 

Among these disk parameters, the angular momentum transport rate $m_{\rm{AM}}$ has a significant impact on the merger rate in both the isotropic and Gaussian distributions, whereas the conversion efficiency $\epsilon$ plays a comparatively minor role, primarily affecting the merger rate of Gaussian distribution. Regarding the maximum merger mass, parameters such as the inner and outer AGN disk radii ($r_{\disk,\rmin}$, $r_{\disk,\out}$), and $m_{\rm{AM}}$ primarily influence the Gaussian distribution, while their effect on the isotropic distribution remains relatively small.

While all disk parameters influence disk properties, minor variations in these parameters may result in negligible changes in scale height or gas density. However, certain variations can significantly impact the merger rate, primarily due to their effect on type I/II migration of BHs within the disk. Changes in disk properties can lead to variations in migration speed by a factor of $\sim$ 2, thereby influencing the merger rate. Notably, parameters such as $m_{\rm{AM}}$ and $\epsilon$ induce by a factor of $\sim 2$ changes in migration speed at $r\sim10^{-4}-10^{-1}$ pc. In contrast, the Shakura-Sunyaev viscosity parameter ($\alpha_{\rm{SS}}$) affects migration speed only at $r\lesssim10^{-3}$ even when varied to an extremely small value $\alpha_{\rm{SS}}=0.005$, resulting in minimal impact on the merger rate, as illustrated in Fig. \ref{fig:merger_par_1} and \ref{fig:merger_par_2}.

Modifying both the inner and outer sizes of the AGN disk significantly affects the maximum merger mass in the Gaussian distribution but generally has a minor impact on the isotropic distribution. With a migration speed variation of at most a factor of $<2$, we observe a noticeable decrease in the number of high-generation mergers when increasing either $r_{\disk,\out}$ or $r_{\disk,\rmin}$. This decline may be attributed to the substantial difference in number density at $r\sim2\times10^{-4}$ pc. However, this effect is prominent only in the Gaussian distribution, as the number of disk BHs in the isotropic model is insufficient to reveal a similar trend. Notably, when $\alpha_{\rm{SS}}$ takes very small values, it induces a similar trend in the maximum merger mass, it induces a similar trend in the maximum merger mass, but its effect is not significant enough to be considered a relevant parameter.

Surprisingly, direct changes to the accretion rate at outer radius $\dot{M}_{\out}$ have a less pronounced effect on the merger rate compared to variations in $m_{AM}$ and $\epsilon$, while increasing $\dot{M}_{\out}$ generally enhances the star-formation rate, leading to minimal influence on the inner regions of the AGN disk. As a result,  as long as $\dot{M}_{\out}$ varies within its typical range, both the number density and migration speed remain largely unchanged, making its impact relatively insignificant. However, significant changes in $r_{\disk,\out}$ could, in turn, influence the accretion rate at $r_{\out}=5\;\rm{pc}$, potentially altering the merger dynamics affecting the merger rate and mass.

\subsubsection{Impact of SMBH mass}

Changes in SMBH mass can lead to complex effects on both the merger rate and the maximum merger mass. As shown in Figs. \ref{fig:merger_par_1} and \ref{fig:merger_par_2}, when the SMBH mass is below $10^7\Msun$, the merger rate and maximum BH mass remain very low, primarily constrained by the initial number of BHs in the disk. As the SMBH mass increases, both the merger rate and maximum mass stabilize around $10^6-10^7\Msun$.

However, for $M_{\SMBH} \gtrsim 10^{7}\Msun$, the merger rate and maximum mass exhibit divergent trends between the isotropic and Gaussian distributions. In the isotropic distribution, fewer mergers and hierarchical mergers occur due to the limited number of disk BHs, as shown in Fig.~\ref{fig:merger_par_SMBH}. In contrast, for the Gaussian distribution, the merger rate increases significantly, while higher-generation mergers occur at smaller radii, leading to a substantial increase in the maximum merger mass.

We can roughly estimate the two major factors influence the merger rate and mass. First, the type I/II migration speed at small radii remains roughly constant at the same normalized radius, $r_0=r/r_{\disk,\rmin}$, regardless of SMBH mass (Fig.~\ref{fig:mig}), However, at larger radii, the migration speed slows down as SMBH mass increases. This should reduce the merger rate while it takes more time to migrate to dense gas region.

Second, we estimate the disk BH number density at the same $r_0$ as: 
\begin{equation} 
n(r_0)\propto\frac{N_{\BH,\ini}}{r_0^2r_{\disk,\rmin}^3h_{\DBH}(r_0)}\propto\frac{M_{\SMBH}}{r_0^3r_{\disk,\rmin}^3}\frac{\vkep}{\sigma_{\DBH}}\propto M_{\SMBH}^{-1/4}, 
\end{equation} 
where we use the velocity dispersion estimate from Eq.\ref{eq:v_dispersion}, gas dynamical friction interactions from Eq.\ref{eq:GDF_approx}, the sound speed relation $c_s=\vkep(h/r)$, and the disk size from Eq.~\ref{eq:disk_in}. This suggests that a higher SMBH mass leads to a slightly lower disk BH number density, reducing the binary-single interaction rate and binary formation rate, ultimately decreasing the overall merger rate. This effect becomes significant when the disk BH population is insufficient for frequent hierarchical mergers, as seen in the isotropic distribution. However, for the Gaussian distribution, even with a factor of $10^{-1/4}\approx 1/3$ decrease in number density compared to fiducial mass, the number of BHs remains sufficient to sustain enough hierarchical mergers.

A key factor explaining the trend difference is the critical number density for hierarchical mergers discussed in Sec.~\ref{sec:merger_feature}. In the Gaussian distribution, this density threshold persists at early times but shifts closer to the SMBH as $M_{\SMBH}$ increases, facilitating more hierarchical mergers and leading to a higher maximum merger mass. In contrast, for the isotropic distribution, this critical number density fails to emerge, particularly in environments with a high SMBH mass, making it difficult to sustain high-generation mergers.

\subsubsection{Implications}

Combining the results from Sec.~\ref{sec:merger_feature} with our previous discussion, we identify several key factors influencing both the merger rate and the highest merger mass. If a disk parameter modifies the type I/II migration rate on the order of $\mathcal{O}(1)$, it typically affects the merger mass by facilitating the migration of binaries into high-density gas regions, thereby enhancing binary hardening. Additionally, if a disk parameter variation moves the location of a migration trap (due to discontinuities in disk properties) at an $\mathcal{O}(1)$ scale, it can slow down inward migration at small radii, increasing the likelihood of hierarchical mergers and contributing to a higher maximum merger mass.

Parameters that directly increase the number of disk BHs also tend to raise the merger rate, as binary-single interactions play a crucial role in the binary hardening process. If the increase in disk BH numbers is substantial, it can further elevate the maximum merger mass, similar to the effect of modifying a migration trap. This effect is particularly pronounced in the isotropic model, where the number of disk BHs is initially insufficient to sustain frequent binary-single interactions and binary formations.

A sufficiently long AGN lifetime would smooth out the structure of merger positions and times, as the merger rate and the rate of BHs migrate into $r<r_{\rmin}$ both approach a steady-state value. While migration speed remains a key factor influencing the merger rate, its impact on the maximum merger mass diminishes over time. This is evidenced by the fact that $f_{\mig}$ becomes irrelevant to the highest merger mass of isotropic distribution.

\section{Discussion} \label{sec:discussion}
\subsection{Convergence test}
\label{sec:convergence}
We examined the impact of numerical resolution on the merger rate and the highest merger mass by varying the number of radial cells  ($N_{\cell}$), the number of mass cell ($N_{\rm{mass}}$) and the time step parameter $\eta_t$. When adjusting the number of radial cells to $N_{\cell}=80,160$, the merger rate $\Gamma_{\rm{merger}}$ changed by only $\sim$2\% for both Gaussian models, while the top 1\% mass $M_{\rm{top}}$ changes 15\% and 8\% for Gaussian models. In contrast, for the isotropic model, both the merger rate and top 1\% mass both changed by less than 4\%.  The slower convergence of the Gaussian model likely stems from its higher merger rate, where 10,000 mergers may not be sufficient for full statistical convergence.

When varying the number of mass cells to $N_{\rm{mass}}=80,120$, the merger rate $\Gamma_{\rm{merger}}$ and the top 1\% mass $M_{\rm{top}}$ exhibited only minor fluctuations of 1\%$-$5\% for both isotropic and Gaussian models. Additionally, we tested the time step parameter at $\eta_t=0.05$ and $\eta_t=0.2$. For both Gaussian and isotropic model, the merger rate and highest mass changes by $\sim$15\% for $\eta_t=0.05$ and $\eta_t=0.2$. However, with the $\eta_t=0.05$ applied, the time cost becomes incredibly large (increase by more than 5 times from $\eta_t=0.1$).  This larger variation arises because the weak scattering diffusion term scales as $\sqrt{\Delta t}$, as indicated in Eq.~\ref{eq:WS,diff}. Although convergence is slower for the Gaussian distribution, our main conclusions regarding the trends of parameter variations remain unaffected by numerical resolution. In the next section, we provide potential methods to solve this issue. 

\subsection{Limitations and neglected effects}
\label{sec:limitation}

In this work, we adopt a highly simplified model at the statistical background level. We assume a collisionless BH disk with scale height $h_{\DBH}$, neglecting both binary-single interactions and binary formation processes. While this approximation may hold in noncollisional or weakly collisional systems, it introduces several limitations and potential inaccuracies in the context of AGN disks, where dynamical interactions are expected to be significant.

First, the collisionless assumption breaks down in the inner regions of the disk, where the binary-single interaction rate becomes comparable to the gas interaction rate (as shown in Fig.~\ref{fig:timecale}). These interactions play a crucial role in slowing BH migration and enriching the BH population. Neglecting them leads to an underestimated migration timescale in these regions. Moreover, binary-single interactions can eject BHs from the disk, temporarily reducing the local number density. Although ejected BHs may return to the disk on low-inclination orbits, this recycling process introduces a time delay in the buildup of the BH population. This effect is likely less significant in AGNs with shorter lifetimes, where such interactions are confined to smaller radii, and in isotropic models, where the overall number density of disk BHs is relatively low.

Second, neglecting interactions leads to an overestimation of the local BH number density and thus an inflated binary formation rate. Although we apply a correction at the Monte Carlo stage using Eq.~\ref{eq:DBH_modification} to account for number reduction due to binary formation, this only considers binaries formed within the simulated BH sample. Background BHs that are not explicitly tracked can also form binaries and deplete the local population—an effect not included in our current model. As a result, the background BH number density remains systematically overestimated.

Third, by ignoring binary formation in the background, we implicitly assume that all binary partners are drawn from the first-generation BH population. In other words, only 1g+Ng hierarchical mergers are included, while Mg+Ng mergers involving two higher-generation BHs are omitted. This constraint affects the merger weighting in Eqs.~\ref{eq:weight_rate} and \ref{eq:weight_mass}, leading to an overestimated merger rate and potentially underestimating the maximum BH mass attainable in the system.

A potential solution to address all three issues is to explicitly track single and binary BH populations at the background level, while incorporating higher-generation mergers. However, this requires accurately modeling binary-single interactions, including ejection and reentry processes, without significantly increasing computational cost—a challenge we aim to tackle in future work.

On addition to the modeling assumptions described above, our simulations neglect several physical processes. Notably, we do not account for the spin evolution of BHs. High-spin BHs can result in significant gravitational recoil velocities following mergers \citep{rezzolla2008final}. However, since the majority of mergers occur in close proximity to the central SMBH (as shown in Fig.~\ref{fig:merger_feature}), we do not expect the recoiling remnants to become unbound from the SMBH. While spin-induced differences in recoil velocity could slightly alter the return timescale of remnant BHs to the disk, such variations are likely minor due to the high ambient gas density. These effects are expected to have limited impact, especially considering that we focus primarily on the fastest mergers in the simulation.

Another major omission is the feedback from BHs. Radiative and kinetic feedback from accreting BHs can heat the surrounding gas or drive outflows, suppressing further accretion at late times \citep{skadowski2015global, Jiang2014global}. Furthermore, our assumption of artificially high BH number densities may lead to unphysical scenarios in which the total BH mass approaches or exceeds the local AGN disk mass. In such cases, dynamical feedback could significantly alter the disk’s structure and stability. These effects are highly model-dependent, emphasizing the need for more realistic treatments of BH–gas interactions and feedback in AGN disks. A promising direction would be to adopt time-dependent AGN disk models, such as that proposed by \citet{epstein2025time}. 

Another potentially important effect is the Kozai–Lidov (KL) mechanism, which can induce large-amplitude oscillations in eccentricity and inclination in hierarchical triple systems \citep{kozai1962, lidov1962}. In dense, gas-rich AGN environments, KL oscillations can drive BH triples to high eccentricities, enhancing merger rates \citep{tagawa2020formation, Dittmann2023sha, Mockler2023ttb} and possibly contributing to the formation of the most massive BHs in the system.

We also cap BH accretion rates slightly above the Eddington limit. However, recent radiation-hydrodynamic simulations indicate that BHs embedded in dense, optically thick environments can accrete at super-Eddington rates—up to $\sim10^3$ times the Eddington limit—due to photon trapping and anisotropic radiation \citep{Jiang2014global, ogawa2017radiation, toyouchi2024radiation}. Ignoring such extreme accretion regimes may lead to significant underestimation of BH mass growth from gas accretion, especially in regimes where gas accretion dominates over mergers (see Sec. ~\ref{sec:gas_int_impact}).

Finally, our results are highly sensitive to the assumed AGN disk model. Variations of treatment in turbulence and magnetic fields in the AGN disk can lead to different disk structures and local properties. These differences in turn affect BH migration rates \citep{secunda2019orbital} and binary accretion processes \citep{chen2023chaotic}, ultimately influencing both the merger rate and the maximum BH mass attainable.

\subsection{Migration traps and extreme mass-ratio inspirals}
\label{sec:mig_trap}
\begin{figure*}
    \centering
    \includegraphics[width=1.0\linewidth]{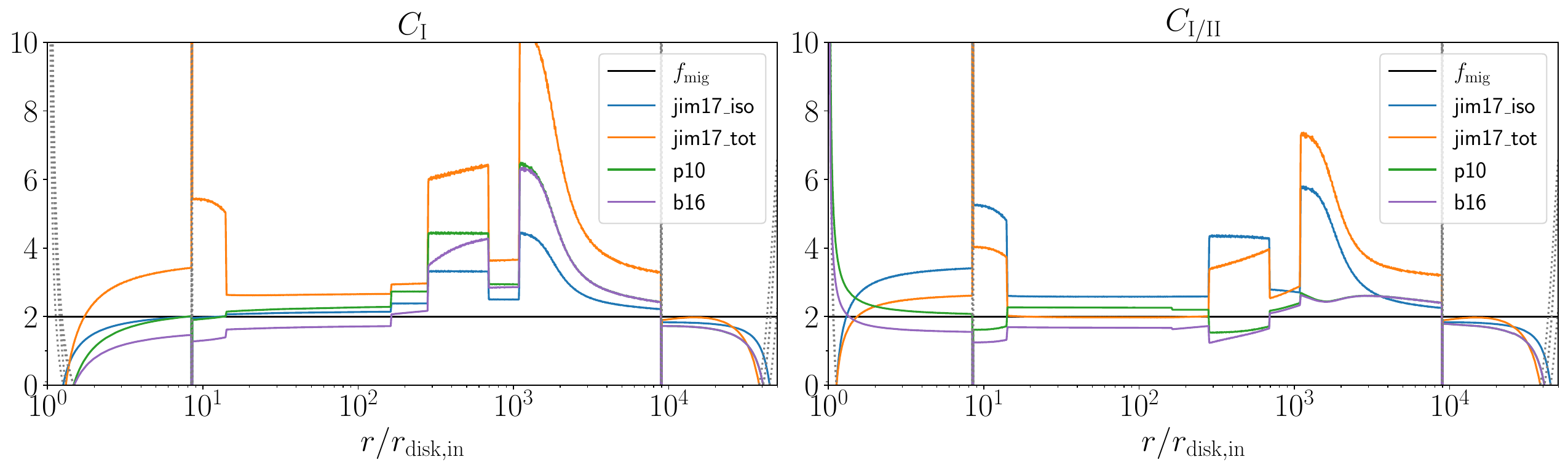}
    \caption{  The migration normalized factor $C_{\rm{I}}$ and $C_{\rm{I/II}}$ for type I and type I/II migration. The black line represent our simplified model $f_{\mig}=2$ used in the simulation. Additionally, we plot torque models from \cite{jimenez2017improved} (\textit{jim17}) for both isothermal (\textit{jim17\_iso}, blue) and non-isothermal cases (\textit{jim17\_tot}, orange), \cite{paardekooper2010torque} (\textit{p10}, green) for the isothermal case, and \cite{bellovary2016migration} (\textit{b16}, purple), which adopts the formula from \cite{paardekooper2010torque}. The gray dotted line indicates regions where the torque results in outward migration. }
    \label{fig:mig_trap}
\end{figure*}

As discussed in Appendix~\ref{sec:mig}, a BH orbiting within a gas disk exchanges angular momentum with its surroundings, leading to net inward radial migration. \cite{paardekooper2006halting} demonstrated that, under certain conditions, migration can proceed outward, resulting in the formation of a migration trap where outwardly migrating objects meet inwardly migrating ones. Such migration traps have been predicted in protoplanetary disks \citep{lyra2010orbital} and in AGN disks, as suggested by \cite{bellovary2016migration} using the torque formula from \cite{paardekooper2010torque}. Numerous studies highlight their significance in contributing to hierarchical mergers \citep{mckernan2012intermediate,yang2019agn,tagawa2020formation,santini2023black,gilbaum2022feedback,vaccaro2024impact}.

However, recent studies have challenged the existence of migration traps in AGN disks. \cite{peng2021last} found that the outward migration region vanishes when considering the torque exerted by mini disks surrounding black holes due to gas drag. \cite{pan2021formation} reached a similar conclusion, accounting for the influence of density wave generation and headwind effects. \cite{grishin2024effect}, employing an updated torque formula from \cite{jimenez2017improved}, also observed the disappearance of the outward migration region. However, \cite{grishin2024effect} introduced the role of thermal torque, generated by the thermal response of the AGN to black holes accreting via their own mini disks, which in turn facilitates the formation of migration traps. \cite{gangardt2024pagn} then applied this mechanism to the AGN disk model from \cite{thompson2005radiation} and \cite{sirko2003spectral}, successfully recovering migration traps at various radii. More recently, \citet{gilbaum2025escape} investigated various mechanisms by which BHs might escape from these migration traps and assessed their implications for BH mass growth.

The aforementioned studies consider only type I migration torque, which is the direct summation of corotation torque and Lindblad torque. In AGN disks, the gap-opening condition is typically satisfied when $K\gtrsim20$ \citep{kanagawa2018radial}, where $K$ is defined in Eq.~\ref{eq:K_factor}. Given that our disk model is very thin $h/r\sim10^{-3}$, this condition is easily met at small radii. We thus normalize the type I and type I/II migration interaction rates as $\Gamma_{\rm{I}}=C_{\rm{I}}\Gamma_0$ and $\Gamma_{\rm{I/II}}=C_{\rm{I/II}}\Gamma_0/(1+0.04K)$, where
\begin{align}
    \Gamma_0&=2\left(\frac{M_{\BH}}{M_{\SMBH}}\right)
\left(\frac{2\rho r^2\vkep}{M_{\SMBH}}\right)\left(\frac{h}{r}\right)^{-1},
\end{align}
and adopt the formulation from \cite{kanagawa2018radial} to account for the gap opening:
\begin{equation}
    C_{\rm{I/II}}=C_{\rm{L}}+C_{\rm{C}}\exp(-K/K_t),\;\;\;K_t=20.
\end{equation}
Here, the subscripts $\rm{L}$ and $\rm{C}$ represent the Lindblad and corotation torques, respectively, with the latter often simplified to a horseshoe drag in the nonlinear regime. We explore the coefficients derived in \cite{paardekooper2010torque} (\textit{p10}) and \cite{jimenez2017improved} (\textit{jim17}):
\begin{equation}
    C_{\rm{L}}=\begin{cases}
       -2.5+0.1\nabla_\Sigma+0.5\nabla_{T} &\textit{p10\_iso}\\
       (-2.5+0.1\nabla_\Sigma-1.7\nabla_{T})/\gamma_{\gas} &\textit{p10\_ad}\\
       -2.34+0.1\nabla_\Sigma-1.5\nabla_T&\textit{jim17\_iso}\\
       \left(-2.34+0.1\nabla_\Sigma-1.5\nabla_T\right)f_{\gamma}\left(\frac{\chi}{\chi_c}\right)&\textit{jim17\_tot}\\
    \end{cases}
\end{equation}
 and 
\begin{equation}
    C_{\rm{C}}=\begin{cases}
       1.65-1.1\nabla_\Sigma-1.4\nabla_T &\textit{p10\_iso}\\
       (1.65-1.1\nabla_\Sigma +7.9\nabla_S/\gamma_{\gas})/\gamma_{\gas} &\textit{p10\_ad}\\
       0.976-0.64\nabla_{\Sigma}+\nabla_T &\textit{jim17\_iso}\\
       (0.46-0.96\nabla_\Sigma+1.8\nabla_T)/\gamma_{\gas} &\textit{jim17\_tot}\\
    \end{cases}
\end{equation}
where $\nabla_\Sigma=-d\ln\Sigma/{d\ln r}$, $\nabla_T=-d\ln T/d\ln r$, and the entropy gradient can be written as $\nabla_S=-d\ln S/d\ln r=\nabla_T-(\gamma_{\gas}-1)\nabla_\Sigma$. We adopt $\gamma_{\gas}=5/3$ as the fiducial adiabatic index. The function $f_{\gamma}\left(\chi/\chi_c\right)$ is defined as $f_{\gamma}(x)=(\sqrt{x/2}+1/\gamma_{\gas})/(\sqrt{x/2}+1)$, with
\begin{equation}
    \chi=\frac{16\gamma_{\gas}(\gamma_{\gas}-1)\sigma_{\rm{SB}}T^4}{3\rho^2c_s^2\kappa},\;\chi_c=\frac{c_s^2}{\Omega}
\end{equation}
The \cite{bellovary2016migration} (\textit{b16}) combines the adiabatic and isothermal torque in \cite{paardekooper2010torque} (\textit{p10}), expressed as
\begin{align}
    C_{\rm{L,C}}&=\frac{C_{\rm{L,C}}^{\mathit{ad}}\Theta^2+C_{\rm{L,C}}^{\mathit{iso}} } {(1+\Theta)^2}, &\textit{b16}
\end{align}
where the dimensionless factor $\Theta$ is the ratio of the radiative and dynamical timescales, defined in \cite{lyra2010orbital} as:
\begin{equation}
    \Theta=\frac{k_B (\vkep/r)\tau_{\eff}}{12\pi m_{\gas}(\gamma_{\gas}-1)\sigma_{\rm{SB}}T^3},
\end{equation}
and effective optical depth is given by:
\begin{equation}
    \tau_{\eff}=\frac{3\tau}{8}+\frac{\sqrt{3}}{4}+\frac{1}{4\tau}.
\end{equation}

Fig.~\ref{fig:mig_trap} illustrates how the migration factors $C_{I}$ and $C_{I/II}$ for a $10\Msun$ BH  vary with radius in our fiducial SMBH and AGN disk model. The gray dotted line marks regions experiencing outward migration torque, while the different colored lines correspond to different torque models. We observe the formation of a narrow migration trap at $r\sim r_{\disk,\rmin}$ and $r\sim 10^4r_{\disk,\rmin}$, arising from discontinuities in surface density and temperature. \cite{pan2021formation} demonstrated that these migration traps are caused by opacity gaps and can be nullified by incorporating a more accurate radiation pressure equation. In our simulation, we account for this effect by directly adjusting the migration speed at the discontinuity, as depicted in Figure~\ref{fig:mig}. This effectively assumes that all BHs traverse the narrow discontinuity trap while experiencing a slowdown in migration. A potential "pile-up" region may emerge near $r_{\disk,\rmin}$, except when using the type I/II migration factor from the \cite{bellovary2016migration} model. The existence of such a trap region depends on a correction factor $(1-\sqrt{r_{\disk,\rmin}/r})$ for the accretion rate $\dot{M}$ at inner region. Although this factor generates persistent outward torque, our chosen $r_{\disk,\rmin}$ is typically too large due to model limitations. If a more refined model were applied with $r_{\disk,\rmin}\sim 6R_g=6GM_{\SMBH}/c^2$, this region might vanish or become dominated by the gravitational radiation from extreme mass ratio inspirals (EMRIs) \citep{grishin2024effect}. While \cite{grishin2024effect} introduced thermal torque as a mechanism to reinstate migration traps, its behavior in gap-opening regions remains unstudied and may not be directly applicable to this AGN disk model. Additionally, an outward migration region may emerge at $r\sim r_{\disk,\out}$, potentially preventing stars and BHs formed at very large radii from migrating inward. Since BHs from star formation have a negligible influence on the overall process, this effect can be considered insignificant. However, our results indicate that variations in migration speed can substantially impact both the merger rate and the maximum merger mass. Given this, and in light of Figure~\ref{fig:mig_trap}, a more precise treatment of migration speed is necessary for accurate predictions. 

We can estimate the number of EMRI events in our simulation by tracking the number of BHs that migrate into $r<r_{\rmin}$. However, due to computational constraints, our fiducial AGN lifetime of $t_{\AGN}=10$ Myr is insufficient to observe such BHs in the Gaussian models. In contrast, for the isotropic models, we find an average of 10.35 BHs with a total mass of $132\Msun$ migrate into this region. This aligns with the reduced binary-single interactions in the isotropic model. Consequently, even if a migration trap forms at $r\sim r_{\rmin}$, it would only accumulate a total mass of $132\Msun$ within 10 Myr, leading to a slight increase in the maximum merger mass in the isotropic model without significantly altering the merger rate.

\section{Conclusion} \label{sec:conclusion}

We carried out a set of AGN simulations to model AGN-assisted black hole mergers. In particular, we aimed to determined the maximum black hole mass expected to be observed from AGNs as a function of binary and AGN parameters (see Table \ref{table:par}). We examined, through multiple simulations, the dependence of the maximum mass on each of our model parameters, while keeping all other parameters at its fiducial value to be able to explore the 24-dimensional parameters space (see Fig. \ref{fig:merger_mass} and \ref{fig:t_AGN}). Our conclusions are the following.
\begin{enumerate}

\item We found the most substantive change in the maximum mass is related to the lifetime of the AGN disk, $t_{\mathrm{AGN}}$. The maximum attainable mass appears to exponentially grow with $t_{\mathrm{AGN}}$.
\item We found that a black hole mass of $\gtrsim200$\,M$_\odot$ is only attainable for AGN lifetimes $\gtrsim 40$\,Myr, independently of the other parameters. Therefore, the possible future detection of such a high-mass black hole in a merger, and the identification of the host AGN through electromagnetic followup, would give a unique measurement of the minimum lifetime of an AGN disk.
\item Several other parameters affect the maximum mass more significantly, in particular the SMBH mass ($M_{\rm SMBH}$), the angular momentum transport rate ($m_{\rm AM}$), the disk viscosity parameter ($\alpha_{\mathrm{SS}}$), conversion efficiency ($\epsilon$), and the AGN disk size ($r_{\disk,\rmin}$, $r_{\disk,\out}$). 
\item The remaining model parameters have limited / marginal effect on the expected maximum black hole mass. This substantially simplifies the task of constraining the remaining model parameters with observations.
\item The production of a $\sim100$\,M$_\odot$ black hole, like the one in GW190521, requires one of the above parameters to deviate from our fiducial model.
\item We similarly find the dependence of the AGN-assisted merger rate as a function of each of our model parameters (see Figs. \ref{fig:merger_rate} and \ref{fig:t_AGN}), which can also be used to constrain the AGN scenario once it becomes sufficiently clear what subset of the observed black hole mergers are of AGN origin.
\end{enumerate}

\begin{acknowledgments}
The authors are thankful for Yang Yang and Yue Yu for valuable discussions and assistance with the programming process. H.T. was supported by the National Key R\&D Program of China (Grant No. 2021YFC2203002). Z.H. is grateful for support from NASA under Grants No. 80NSSC22K0822 and No. 80NSSC24K0440. I.B. acknowledges support from the National Science Foundation under Grant No. PHY-2309024. 
\end{acknowledgments}

\section*{DATA AVAILABILITY}
                          
The data that support the findings of this article are not publicly available. The data are available from the authors upon reasonable request.

\appendix

\section{preexisting stellar and BH populations}
\label{sec:pre-exist}
 In the fiducial model, the preexisting single BH mass is drawn from the initial BH mass function, which we define as \citep{chabrier2003galactic,salpeter1955luminosity}:
\begin{equation}
\frac{dN}{dM_{\BH}}=\mathcal{F}_{\preexist}(M_{\BH})= \frac{(1-\beta_{\IMF})M_{\BH}^{-\beta_{\IMF}}}{M_{\mathrm{BH,ini,max}}^{1-\beta_{\IMF}}-M_{\mathrm{BH,ini,min}}^{1-\beta_{\IMF}}},\label{eq:IMF}
\end{equation}
within the range of maximum and minimum BH masses,  $M_{\mathrm{BH,ini,max}}$ and $M_{\mathrm{BH,ini,min}}$ respectively. 

Observational studies of the Galactic center suggest that the initial mass function (IMF) of stars in this region may be top-heavy \citep{bartko2009extremely, lu2013stellar}. Motivated by this, we explore a range of BH mass functions centered around the fiducial power-law index $\beta_{\IMF} = 2.35$, which corresponds to both the standard Salpeter IMF and some theoretical predictions for BH birth mass distributions \citep{o2009gravitational,woosley2020birth}.

We use the stellar distribution as a background parameter that remains constant at the same radial distance over time. In our fiducial model, we assume an average stellar mass $\bar{m}_*=1\Msun$ as the main sequence stellar mass, and adopt the stellar number density from \cite{o2009gravitational}:
\begin{equation}
n_*=1.38\times10^5\,\mathrm{pc}\left(\frac{10^6\Msun}{M_{\SMBH}}\right)^{1/2}\left(\frac{r}{r_{i}}\right)^{-1.4},\label{eq:n_s}
\end{equation}
where $r_{i}$ is assumed to be the  nuclear star cluster (NSC) size, determined by SMBH mass $M_{\SMBH}$ and velocity dispersion $\sigma_*$:
\begin{equation}
    r_{i}=\frac{GM_{\SMBH}}{\sigma_*^2},\label{eq:r_BH,out}.
\end{equation}
To avoid the singularity in our simulation and match up with the simulation result in \cite{o2009gravitational}, we assume that the stellar density outside the NSC ($r>r_i$) is 
\begin{equation}
    n_*=1.38\times10^5\,\mathrm{pc}\left(\frac{10^6\Msun}{M_{\SMBH}}\right)^{1/2}\exp\left(-\frac{r}{r_i}\right).
\end{equation}
Although many recent studies have reported variations in the index of the $M-\sigma$ relation of NSC \citep{mcconnell2011two,kormendy2013coevolution,davis2017updating}, we adopt the conventional form \citep{merritt1999black}, where the stellar velocity dispersion of host SMBH mass $M_{\SMBH}$ is:
\begin{equation}
M_{\SMBH}=3.1\times10^8\;M_{\odot}\left(\frac{\sigma_*}{200\;\mathrm{km/s}}\right)^{4}.\label{eq:M_sigma}
\end{equation}
This choice ensures that the dependence of the preexisting BH maximum radius on the SMBH mass coincides with that of the AGN disk size (Eq. \ref{eq:disk_out}). The Keplerian speed can then be calculated by 
\begin{equation}
\vkep=\sqrt{\frac{G(M_{\SMBH}+M_{<})}{r}},\label{eq:kep}
\end{equation}
where the enclosed mass $M_{<}$, is defined as:
\begin{equation}
M_{<}(r)=4\pi\int_{0}^{r}dr\,r^2 \bar{m}_* n_*(r).
\end{equation}

In our fiducial model, the BH number density is given by:
\begin{equation}
    n_{\BH}(r)=\frac{N_{\BH,\ini} (\gamma_\rho+1)}{4\pi}\left(\frac{r^{\gamma_{\rho}-2}}{r_{\BH,\out}^{\gamma_{\rho}+1}-r_{\rmin}^{\gamma_{\rho}+1}}\right),\label{eq:n_BH_gamma_rho}
\end{equation}
where the maximum BH distance to center $r_{\BH,\out}$ is set to be the same as the galactic nucleus size $r_{i}$ (Eq. \ref{eq:r_BH,out}), and the total number of preexisting BHs scales with SMBH mass fiducially as \citep{miralda2000cluster}:
\begin{equation}
    N_{\BH,\ini}=20000\times\frac{M_{\SMBH}}{4\times10^6M_\odot}.\label{eq:N_BH_ini}
\end{equation}

We select $M_{\SMBH}=4\times10^{6}M_{\odot}$ as the default SMBH mass. As a comparison with \cite{tagawa2020formation}, we also set a default stellar population for an SMBH mass of $4\times10^6\,\Msun$,  where the stellar number density in Eq. \ref{eq:n_s} is replaced by:
\begin{equation}
        n_*(r)=\left(\frac{M_{*,3pc}}{4.3 \;\bar{m}_*}\;\mathrm{pc}^{-3}\right)
       \left(\frac{r}{0.3\;\mathrm{pc}}\right)^{-0.5}\left[1+\left(\frac{r}{0.3\;\mathrm{pc}}\right)^{4}\right]^{-0.325}\label{eq:n_s_fiducial}
\end{equation}
where $M_{*,3pc}=10^7 M_\odot$ is the total stellar mass in 3 pc. The average stellar mass $\bar{m}_*$ is calculated using initial mass function in Eq. \ref{eq:IMF} with index $\beta_{\IMF}$, for the stellar mass $m_{*}$ between $0.1\Msun$ and $20\Msun$.

We set the fraction $f_{\preexist}$ as the ratio of preexisting binaries to the total number of BH systems (single number + binary number), and adopt a fiducial value of $f_{\preexist}=0.15$ following \cite{tagawa2020formation}. The masses of the binary components, denoted as $M_1$ and $M_2$, are generated independently through Eq. \ref{eq:IMF} and $M_{\BH}=M_1+M_2$ becomes the total mass of the binary. In our simulations, we treat a BH binary of mass $M_1$ and $M_2$ as a single BH of mass $M_{\BH}=M_1+M_2$. The binary separation $s$ of preexisting BH follows a log-flat distribution, ranging from $s_{\min}=R_{*,1}+R_{*,2}$ to $s_{\max}$, where $R_{*,1}$ and $R_{*,2}$ are the stellar radii of components 1 and 2, determined by their progenitor stellar masses \citep{torres2010accurate}:
\begin{equation}
R_*=R_{\odot}\left(\frac{m_{s}}{\Msun}\right)^{1/2}.\label{eq:R_s}
\end{equation}
We apply the relationship between BH mass and progenitor stellar mass in \cite{tagawa2020formation} as:
\begin{equation}
M_{\BH}=
\begin{cases}
\frac{m_{*}}{4}&20\le m_{*}<40\\
10&40\le m_{*}<55\\
\frac{m_{*}}{13}+5.77&55\le m_{*}<120\\
15&120\le m_{*}<140\\
\end{cases}.\label{eq:ms_mbh}
\end{equation}
The masses of preexisting single BHs are initialized in the range $M_{\BH,\ini,\min}=5\Msun$ and $M_{\BH,\ini,\max}=15\Msun$, consistent with stellar evolution predictions \citep{belczynski2010maximum}. When varying $M_{\BH,\ini,\max}$, we correspondingly derive the progenitor stellar mass using the inverse of Eq.~\ref{eq:ms_mbh}, and generalized it as $M_{*}=13(M_{BH}-15\Msun)+140\Msun$ for $M_{\BH}>15\Msun$.

%\begin{equation}
%m_s=
%\begin{cases}
%4M_{BH}
%&M_{BH}<10\\
%13(M_{BH}-15)+120
%&10\le M_{BH}<15\\
%13(M_{BH}-15)+140
%&M_{BH}>15
%\end{cases}.\label{eq:ms_mbh}
%\end{equation}
%To compare with (ref Tagawa), when SMBH mass is chosen to be ``Default'' mode, mass mode=1 and $r$ mode=0 (only applies to model named ``Tagawa'') , we use $s_{\max}=R_{\max}$ and delete the soft binaries, which effectively reduce the binary fraction to around 7\%. In other cases we directly add the soft hard boundary in the $s_{\max}=\max(R_{\max},s_{bound})$ and adjust the binary number by changing $f_{pre}$. 
The maximum preexisting BH separation $s_{\max}=\max(R_{\max},s_{\mathrm{bound}})$ is set to be the maximum value of $R_{\max}=10^5R_{\odot}$ \citep{belczynski2008compact,kinugawa2014possible} and the soft-hard boundary at distance $r$ derived in Eq. \ref{eq:soft_hard}, in case the soft binaries are usually breaks into singles due to the soft binary single interaction in Appendix  \ref{sec:BS_int} \citep{tagawa2020formation}. 

The simulation also tracks the relative velocity with respect to the AGN disk's rotational (Keplerian) velocity as they cross the disk. The inclination angle $i$ to the AGN disk is approximated with the $z$ components of relative speed, $\frac{v_z}{\vkep}=\sin i$. The $x$ and $y$ velocity relative to the $v_{x}$ and $v_y$ are generated in the same manner as $v_{z}$ to save the computational cost. We investigate two basic inclination/dispersion models: (i) an isotropic model, where the BH position vector's direction and velocity's direction are both randomly distributed on a 2D sphere \citep{o2009gravitational}, and (ii) a highly anisotropic Gaussian distribution of BH as in \cite{tagawa2020formation}.  In the isotropic model the inclination angle distribution follows:
\begin{equation}
\frac{dN}{d i}=\frac{i}2\sin i.\label{eq:incl_uniform}
\end{equation}
In the anisotropic model  $v_{x,y,z}/\vkep$  follow a normal distribution with a mean of 0 and dispersion $\beta_v$, as suggested by \cite{tagawa2020formation}:
\begin{equation}
\frac{dN}{d\sin i}=\frac{1}{\sqrt{2\pi\left(\frac{\beta_v}{\sqrt{3}}\right)^2}}\exp\left[-\frac{\sin^2 i}{2\left(\frac{\beta_v}{\sqrt{3}}\right)^2}\right].\label{eq:incl_Gaussian}
\end{equation}
The rare occurrence of conditions of $\sin i>1$ or $\sin i<-1$ indicates that the BHs are free and unbound from the SMBH. Additionally, a BH is also considered free if its relative velocity $v=\sqrt{v_x^2+v_y^2+v_z^2}$ exceeds $\sqrt{2}\vkep$. In such cases, the simulation for that BH is terminated. 

For all subsequent discussions involving the inclination angle $i$, its negative counterpart $-i$ will automatically included, since the AGN disk structure and interactions are symmetric with respect to the disk ($xy$) plane. The fraction of the BHs with inclination $0\le i\le i_0$ (also including $-i_0\le i\le 0$) relative to the total number of BHs at radius $r$ is then given by:
\begin{equation}
F(i_0)=\begin{cases}\erf\left(\frac{\sqrt{3}\sin i_0}{\sqrt{2}\beta_v}\right)&\text{Gaussian}\\
\sin i_0-i_0\cos i_0&\text{isotropic}
\end{cases}\label{eq:incl_frac}.
\end{equation}

\begin{figure}[!t]
    \centering
    \includegraphics[width=0.5\linewidth]{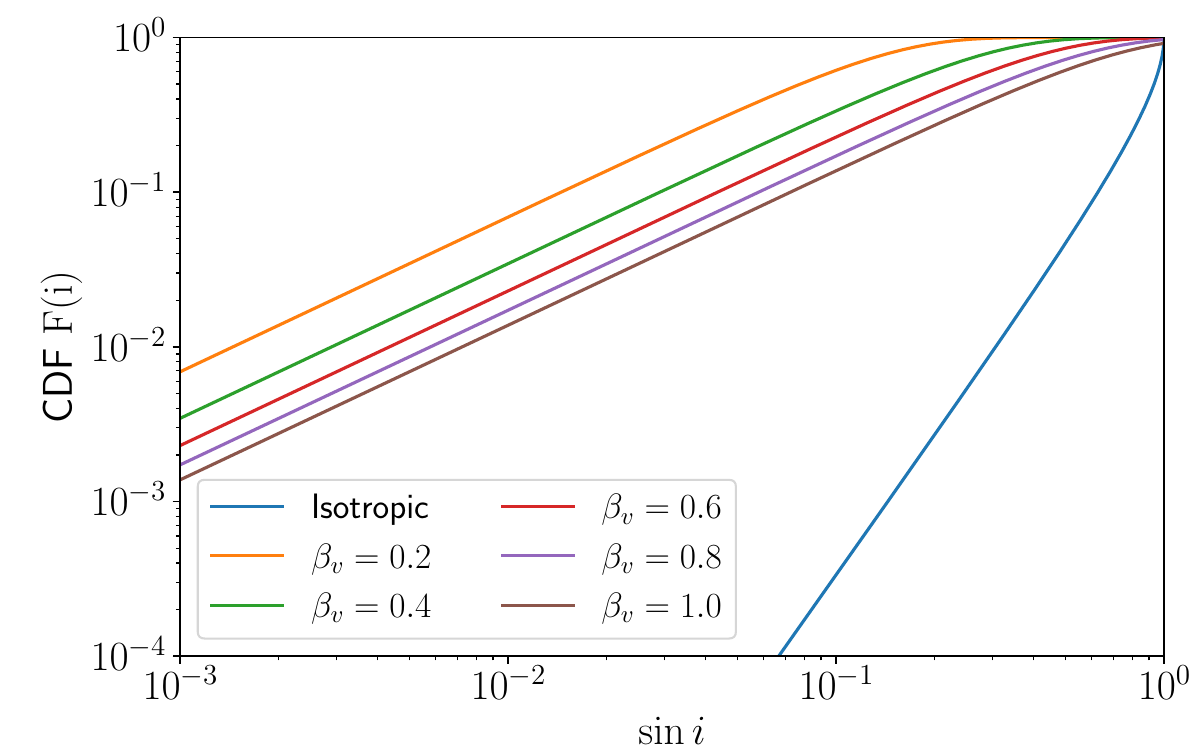}
    \caption{The cumulative density function (CDF) of the inclination distribution $F(i_0)$, as defined in Eq. \ref{eq:incl_frac}. Compared to the isotropic model, the Gaussian distribution yields a significantly larger population of low-inclination preexisting BHs, leading to a greater number of BHs falling into the disk gradually over time.}
    \label{fig:inclination_dis}
\end{figure}

Fig.~\ref{fig:inclination_dis} presents the cumulative density function of the inclination distribution $F(i_0)$ from Eq. \ref{eq:incl_frac}. The Gaussian distribution, regardless of $\beta_v$, exhibits a similar growth rate on a log scale, whereas the isotropic model starts with a low fraction at small inclinations before rising sharply. As a result, the number density of BHs captured by the AGN disk remains comparable across Gaussian models but is significantly lower for the isotropic model, which lead to a reduced merger rate. A higher disk BH density in the Gaussian model also increases the frequency of binary-single interactions and binary formation via gas capture, as discussed in Sec.~\ref{sec:merger_feature}.

\begin{figure*}[!t]
    \centering
    \includegraphics[width=1.0\linewidth]{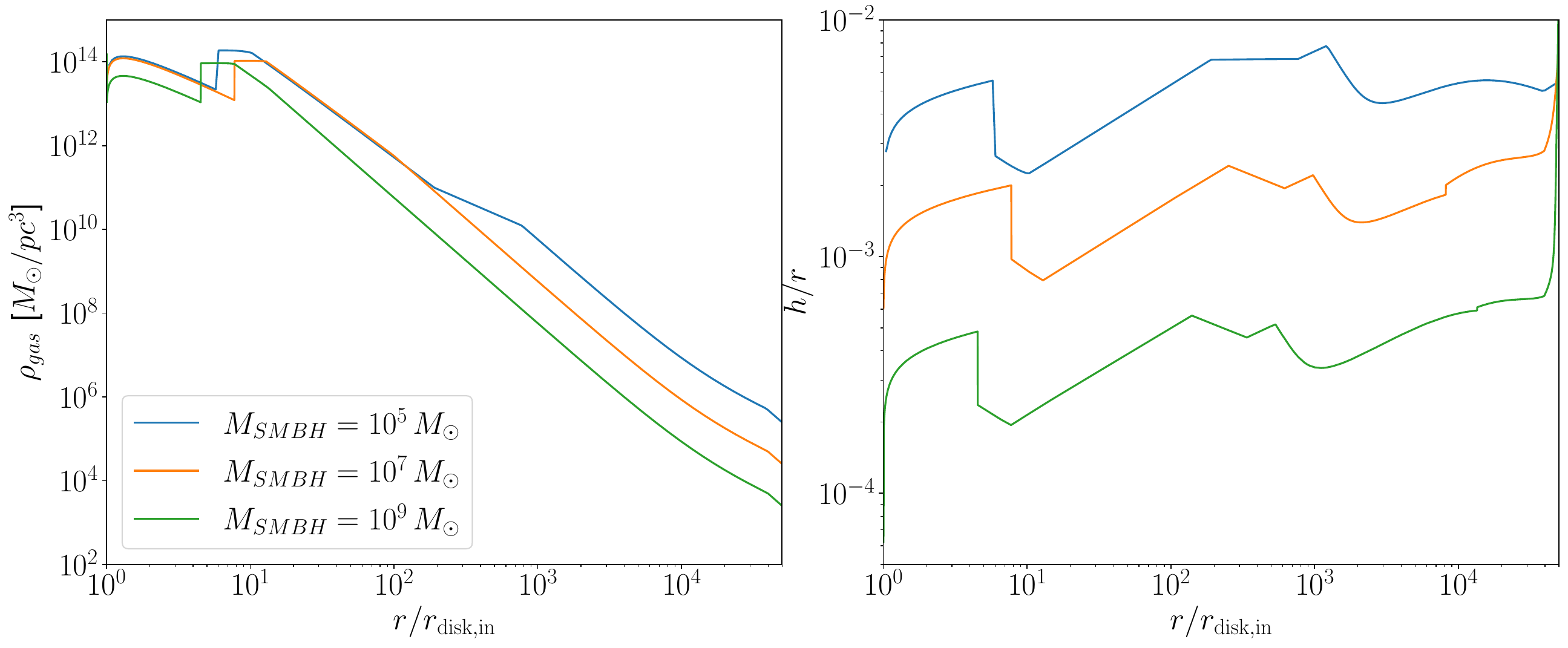}
    \caption{Example of gas density $\rho$ (in $\Msun/\mathrm{pc}^{3}$) and aspect ratio $h/r$ as functions of radius $r$ in our fiducial model ($n_s$ with Eq. \ref{eq:n_s}). The blue, orange, and green curves correspond to the gas density or aspect ratio for host SMBH mass $10^{5}\Msun$, $10^{7}\Msun$, and $10^{9}\Msun$. The radius $r$ is normalized by the inner radius of the disk (Eq. \ref{eq:disk_in}).}
    \label{fig:disk_pro}
\end{figure*}

\section{AGN disk model}
\label{sec:AGN}

We adopt the AGN disk model proposed by \cite{thompson2005radiation} and the Keplerian speed is modified by Eq. \ref{eq:kep} as suggested by \cite{tagawa2020formation}. We assume that the AGN disk rotates with Keplerian speed in the $xy$ plane. Given the mass of the SMBH, $M_{\SMBH}$, and assuming the disk velocity at radius $r$ is the Kepler velocity $\vkep(r)$, the gas properties of an AGN disk at radius $r$ are: gas density $\rho$, aspect ratio $h/r$, gas temperature $T$, effective temperature of radiation $T_{\mathrm{eff}}$, the accretion rate $\dot{M}$, opacity $\kappa(\rho,T)$, the vertical optical depth $\tau_V$, and  the star formation rate $\dot{\Sigma}_{*}$. The Kepler speed determined by the stellar density (Eq. \ref{eq:kep}) will replace equation C1 in \cite{thompson2005radiation}. Equations C2-C13 in \cite{thompson2005radiation} still apply. The equations to solve the properties are listed below:
\begin{equation}
    \rho\frac{k_B}{m_p}T+\epsilon\dot{\Sigma}_{*}c\left(\frac12\tau_V+\xi\right)=\rho\left(\frac{h}{r}\right)^2\vkep^2
    \label{eq:disk_1}
\end{equation}
\begin{equation}
     T^4=\frac34 T_{\mathrm{eff}}^4\left(\tau_V+\frac{4}{3}+\frac{2}{3\tau_V}\right)
     \label{eq:disk_2}
\end{equation}
\begin{equation}
    \tau_V=\rho h\kappa(\rho,T)
    \label{eq:disk_3}
\end{equation}
\begin{equation}
    \dot{M}=\dot{M}_{\out}-\int_{r}^{r_{\disk,\out}}\;2\pi r\dot{\Sigma}_{*}\,dr
    \label{eq:disk_4}
\end{equation}
\begin{equation}
    \sigma_{\mathrm{SB}}T_{\mathrm{eff}}^4=\frac12 \epsilon\dot{\Sigma}_{*}c^2+\frac{3}{8\pi}\dot{M}\left(1-\sqrt{\frac{r_{\disk,\rmin}}{r}}\right)\left(\frac{\vkep}{r}\right)^2,
    \label{eq:disk_5}
\end{equation}
where $c$ is the speed of light, $\sigma_{\mathrm{SB}}$ is the Stefan-Boltzmann constant, $k_B$ is the Boltzmann constant, $m_p$ is the proton mass, and    $r_{\rmin}$  and $r_{\out}$ are the inner and outer boundary of the AGN disk.

We adopt the analytical Rosseland mean opacity model as described in Appendix A of \cite{bell1993using}. The accretion rate at the outer boundary, $\dot{M}_{\out}$ is selected to be a fraction of the Eddington accretion rate, with its fiducial value given by:
\begin{equation}
\frac{\dot{M}_{\out}}{0.1\dot{M}_{\Edd}}=1\label{eq:disk_6},
\end{equation}
where $\dot{M}_{\Edd}=L_{\Edd}(M_{\SMBH})/(0.1c^2)$ is the Eddington accretion limit with radiation efficiency $\eta=0.1$ \citep{tagawa2020formation}.

In the outer region, the accretion heat is insufficient to maintain the Toomre parameter $Q=1$ and thus the star formation rate is nonzero. The Toomre parameter condition $Q=1$ is thus equivalent to:
\begin{equation}
    \rho=\frac{\vkep^2/r^2 }{\sqrt{2}\pi G}.\label{eq:Toomre}
\end{equation}
Angular momentum transport is primarily driven by global torques, represented by a constant fraction $m_{AM}$ :
\begin{equation}
    \dot{M}=4\pi r^2\left(\frac{h}{r}\right)^2\rho\, \vkep m_{AM}.
    \label{eq:AM_transport}
\end{equation}
The parameters in the outer disk region are solved based on Eqs. \ref{eq:disk_1}--\ref{eq:disk_6}, subject to the conditions outlined in Eqs. \ref{eq:Toomre}
and \ref{eq:AM_transport}. 

In the middle region, the accretion heat is sufficient to maintain $Q=1$, resulting in the cessation of star formation. In this region, the degree of angular momentum transport $m_{AM}$ is adjusted to ensure that the Toomre parameter remains $Q=1$. The parameters in the middle region are then determined based on Eqs. \ref{eq:disk_1}--\ref{eq:disk_6}, subject to conditions $\dot{\Sigma}_{*}=0$ and Eq. \ref{eq:Toomre}

In the inner region where $Q>1$, the accretion is driven by a local viscosity, represented by
\begin{equation}
    \dot{M}\left(1-\sqrt{\frac{r_{\disk,\rmin}}{r}}\right)=4\pi \alpha_{\mathrm{ss}}\rho r^3\left(\frac{h}{r}\right)^3\frac{\vkep}{r}\left|\frac{d\ln\Omega}{d\ln r}\right|\label{eq:viscosity}.
\end{equation}
where $\alpha_{\mathrm{ss}}$ is the $\alpha$ parameter of the Shakura-Sunyaev model, $\Omega=\vkep/r$ is the Keplerian angular velocity, and the factor $\left|\frac{d\ln\Omega}{d\ln r}\right|$ is given based on stellar density (Eq.  \ref{eq:n_s}),
\begin{equation}
    \left|\frac{d\ln\Omega}{d\ln r}\right|=\left|\frac32-\frac{G\bar{m}_* n_{*} r^2}{\vkep^2}\right|.\label{eq:dln_Omega}
\end{equation}
The parameters in the inner region are then determined based on Eqs. \ref{eq:disk_1}--\ref{eq:disk_6}, with condition $\dot{\Sigma}_{*}=0$ and Eq. \ref{eq:viscosity}. For simplicity we assume that, when $r=r_{\rmin}$, all disk properties are 0. We adopt a fiducial value of $\alpha_{\rm{SS}} = 0.1$, consistent with previous studies \citep{bai2013local,king2007accretion}, and explore a broad range from 0.005 to 0.3 to encompass a wide variety of possible disk conditions.

For stars and BHs formed in the AGN disk, we adopt the model of \cite{tagawa2020formation}, where the stars are formed following initial mass function with index $\beta_{\cre}$ between $m_{\mathrm{s,min}}=0.1\Msun$ and $m_{\mathrm{s,max}}=140\Msun$:
 \begin{equation}
\frac{dN}{dm_{*}}=\mathcal{F}_{\cre}(m_{*},m_{\mathrm{*,min}},m_{*,\max})=\frac{(1-\beta_{\cre})m_{*}^{-\beta_{\cre}}}{m_{*,\max}^{1-\beta_{\cre}}-m_{*,\min}^{1-\beta_{\cre}}}.\label{eq:IMF_star}
\end{equation}
Stars of mass between $m_{*,\mathrm{change}}=20\Msun$ and $m_{*,\max}=140\Msun$ will become BHs immediately after they are formed. The mass of BHs follow the initial mass function of their progenitors and the relation in Eq. \ref{eq:ms_mbh}. The power-law index of the stellar IMF for stars formed in the disk, denoted here by $\beta_{\cre}$, is varied between $-1.7$ and $-2.35$, based on the discussion in Appendix~\ref{sec:pre-exist}.

 The conversion efficiency $\epsilon$ in Eqs. \ref{eq:disk_1} and \ref{eq:disk_5} represents the efficiency of converting the star formation rate into radiation. \cite{thompson2005radiation} assumes $\epsilon=10^{-3}$ for a Salpeter IMF (Eq. \ref{eq:IMF_star}) in the range $1-100\Msun$, while \cite{tagawa2020formation} and \cite{epstein2025time} suggests that $\epsilon$ may vary with AGN lifetime, stellar lifetime, and the star formation IMF index $\beta_{\cre}$. Using the method outlined in Appendix A of \cite{tagawa2020formation}, we estimate $\epsilon$ to range from $10^{-4}$ to $10^{-3}$ within our simulation parameters ($10\,\mathrm{Myr}\le t_{\AGN}\le100\,\mathrm{Myr}$, $1.7\le\beta_{\cre}\le2.35$). In this paper, we treat $\epsilon$ as a free parameter independent of $\beta_{\cre}$ and $t_{\AGN}$, setting a default value of $10^{-4}$ for $t_{\AGN}=10\,\mathrm{Myr}$ and $\beta_{\cre}=2.35$. 
 
 The AGN lifetime itself is highly uncertain, with estimates ranging from an order of 1 to 100 Myr \citep{ho2004coevolution,haiman2001constraining,martini2001quasar,marconi2004local,schawinski2009moderate}. For computational efficiency, we adopt $t_{\AGN} = 10,\mathrm{Myr}$ as our baseline and limit high-lifetime runs to a few representative cases. Since feedback processes are neglected in our model, the AGN lifetime effectively serves as the total simulation time, starting from the formation of the AGN disk.

We then denote following parameters: the average stellar mass $\bar{m}_{*,\cre}$ for stars that are not BH progenitors, the average stellar mass $\bar{m}_{*,\change}$ of BH progenitors, and the average BH mass $\bar{m}_{\BH,\cre}$ from star formation. The mass fraction of BH progenitor stars, relative to the total stellar mass, is denoted as $f_{\BH}$. The star and BH formation rates (number per area per time) can be expressed as follows:
\begin{align}
    &\left(\frac{dN_{\BH}}{dA\,dt}\right)_{\cre}=\frac{f_{\BH}\dot{\Sigma}_* }{\bar{m}_{\mathrm{\change}}},\label{eq:BH_cre}\\
    &\left(\frac{dN_{\mathrm{*}}}{dA\,dt}\right)_{\cre}=\frac{(1-f_{\BH})\dot{\Sigma}_* }{\bar{m}_{*}}.
\end{align}
We compute the average number of BHs formed within the AGN lifetime, assuming their formation times are uniformly distributed between 0 and $t_{\AGN}$. The masses of the newly formed BHs are randomly generated according to the initial mass function in Eqs. \ref{eq:IMF_star} and  \ref{eq:ms_mbh}. The velocities relative to AGN disk are assumed to be the sound speed, given by $\vec{v}=c_s\hat{n}$ where $\hat{n}$ is a random 3D unit vector. For simplicity, we assume that the binary fraction of star formation is consistent with that of the background preexisting BHs.

In a disk where the outer boundary $r_{\disk,\out}$ exceeds our simulation range $r_{\out}$, we can effectively consider an AGN disk of $r_{\disk,\out}=r_{\out}$ with a much lower accretion rate than normal outer boundary accretion rate as $\dot{M}_{\mathrm{5\,pc}}/(0.1\dot{M}_{\Edd})<1$. Changes in the disk size can thus be accounted for by adjustments in the accretion rate at $r_{out}$, which may result in significant orders of magnitude changes $\dot{M}_{\mathrm{5\,pc}}/(0.1\dot{M}_{\Edd})\ll1.0$. In our default settings, the size of the disk is scaled similarly to the AGN size, as given by Eq. \ref{eq:r_BH,out}, in relation to the SMBH mass. Specifically, we define the outer boundaries of the disk proportional to $\sqrt{M_{\SMBH}}$, motivated by mid-IR observations \citep{burtscher2013diversity}. For simplicity, we assume that the inner boundaries scale the same way as the outer boundaries:
\begin{align}
&r_{\disk,\rmin}=10^{-4}\,\mathrm{pc}\left(\frac{M_{\SMBH}}{4\times10^{6}M_{\odot}}\right)^{1/2},\label{eq:disk_in}\\
&r_{\disk,\out}=5\,\mathrm{pc}\left(\frac{M_{\SMBH}}{4\times10^{6}M_{\odot}}\right)^{1/2}.\label{eq:disk_out}
\end{align}
In reality, the inner radius of the AGN disk is expected to be smaller than the value given by Eq. \ref{eq:disk_in}. In the model by \cite{thompson2005radiation}, the equations governing the inner region (Eqs. \ref{eq:disk_1}-\ref{eq:disk_6}, $\dot{\Sigma}_{*}=0$ and Eq. \ref{eq:viscosity}) fail to yield solutions near the radius $r_{\disk,\rmin}$ (e.g. $r_{\disk,\rmin}=6GM_{\SMBH}/c^2=r_{\rm{ISCO}}$). To avoid this issue and effectively analyze the impact of migration traps in the AGN disk, we adopt a larger $r_{\disk,\rmin}$ in Eq. \ref{eq:disk_in}, which ensures that the impact of the migration trap becomes more pronounced, as discussed in Appendix  \ref{sec:mig} below.

In Eq. \ref{eq:disk_4}, the AGN accretion rate decreases rapidly, and thus the number of cells needs to be increased to avoid a negative accretion rate. In this way, the number of (log scale) radius steps is thus determined by Eq. \ref{eq:disk_4}:
\begin{align}
    &\Delta(\ln r)_{\max}=\frac{0.01\dot{M}_{\out}}{2\pi r_{\disk,out}^2 \dot{\Sigma}_{*}(r_{\disk,\out})},\label{eq:dlog_r_disk}\\
    &N_{\disk}=\left[\ln\left(\frac{r_{\disk,\out}}{r_{\disk,\rmin}}\right)\Big/\Delta(\ln r)_{\max}\right]+1.\label{eq:N_disk}
\end{align}
The disk properties are converted into values at standard grid boundary and cell center (determined by $N_{\cell}$) in the next step.

In Fig.~\ref{fig:disk_pro}, we illustrate how the gas density $\rho$ and  the aspect ratio $h/r$ changes with the radius. The gas density $\rho$ decreases slightly as the SMBH mass increases, with radii rescaled to $r_{\disk,\rmin}$ (left pannel), while $h/r$ decreases correspondingly (right pannel). Notably, the aspect ratio $h/r$ remains approximately constant for a given SMBH mass, which implies that the initial inclination of objects entering the disk is largely independent of radius. Both the gas density and disk thickness are critical factors in determining the number of BHs within the disk, as they directly affect the disk's capacity to capture and retain BHs within its structure. The variations in gas density and aspect ratio in the AGN disk, influenced by different host SMBH masses, lead to complex effects on the physical processes discussed in later sections.

\section{Gravitational interaction} 
\label{sec:GW}
\subsection{Gravitational wave radiation}
\label{sec:GW_radiatio}
Gravitational radiation gradually hardens binary systems through energy loss. Using the gravitational-wave hardening rate for a binary with zero eccentricity from \cite{peters1964gravitational}, we express this rate as:
\begin{equation}
\Gamma_{GW}=-\frac{(ds/dt)_{\mathrm{GW}}}{s}=\frac{64}{5}\frac{G^3 M_1M_2(M_1+M_2)}{c^5s^4},
\end{equation}
where $M_{1}$ and $M_2$ are the masses of the primary and secondary BHs in the binary system. 

\subsection{Merger prescription}
\label{sec:Merger}
We assume that the BH binary merges and forms a new remnant single BH when the separation of the BH binary reaches the innermost stable circular orbit $r_{\mathrm{ISCO}}=6GM_{BH}/c^2$. The newly formed BH experiences a recoil due to anisotropic gravitational-wave emission and undergoes mass loss from GW radiation. We adopt the model used in \cite{tagawa2020formation} and neglect the effect of spin and mass ratio, the single mass \citep{tichy2008final} and velocity after merger is then described by:
\begin{align}
&M_{\BH,\mathrm{single}}=M_{\BH}\left[1-\frac{q}{5(1+q)^2}\right],\\
&\vec{v}_{\mathrm{single}}=\vec{v}_{\mathrm{bin}}+v_{GW}\left[\frac{q^2(1-q)}{(1+q)^5}\right]\hat{n},
\end{align}
where $\hat{n}$ is a random 3D vector and $v_{GW}=8830\,\mathrm{km/s}$ \citep{baker2007modeling}.

\section{Gas interaction}
\label{sec:Gas}
Each BH crosses the AGN disk twice per orbital period, interacting with the gas on both occasions. Although the relative velocity of the BH varies during each crossing, we use the same relative velocity in simulations to reduce computational costs. This is justified since the eccentricity of BH orbits is typically close to zero. A BH is considered to be fully embedded in the disk when its orbital height is smaller than the disk’s half-thickness $h_{z}=r\frac{|v_z|}{\vkep}<h$. The ratio of time the BH spends within the disk per orbit period, denoted as $p_{\disk}$ is given by:
\begin{equation}
p_{\disk}=\begin{cases}
1&\text{BH resides in the disk}\\
\frac{2}{\pi}\arcsin\left[\left(\frac{h}{r}\right)\Big/\left(\frac{|v_z|}{\vkep}\right)\right]&\text{BH is outside the disk}
\end{cases}.\label{eq:p_disk}
\end{equation}

When a BH is inside the disk, the angular momentum flux from the BH locally dominates the viscous flux due to a large mass difference from stellar mass BH and SMBH. As a consequence, the torque it exerts depletes the surrounding gas density, forming a gap 
 at its radius, in which the surface density is smaller than its unperturbed value. 
When the BH is outside the disk, we neglect the angular momentum flux injected into the disk as it is small compared to the transport of angular momentum due to the disk. As a result, the gas density remains approximately unperturbed. We approximate the gas density around the BH suggested by \cite{kanagawa2018radial} as:
\begin{equation}
\rho_{\gas}=\begin{cases}
\rho/(1+0.04K)&\text{BH resides in the disk}\\
\rho&\text{BH is outside the disk}
\end{cases},\label{eq:rho_gas}
\end{equation}
where the dimensionless factor $K$ considering the local disk angular momentum transport is given by:
\begin{equation}
K=\left(\frac{M_{\BH}}{M_{\SMBH}}\right)^2\left(\frac{h}{r}\right)^{-5}\alpha_{\mathrm{eff}}^{-1}.\label{eq:K_factor}
\end{equation}
The effective $\alpha$ parameter $\alpha_{\mathrm{eff}}$, is calculated throughout the simulation range by replacing $\alpha_{\mathrm{SS}}$ with $\alpha_{\mathrm{eff}}$ in Eq. \ref{eq:viscosity}.  In the inner region of AGN disk $\alpha_{\mathrm{eff}}=\alpha_{\mathrm{SS}}$ while in other regions $\alpha_{\mathrm{eff}}>\alpha_{\mathrm{ss}}$.

\subsection{Accretion}
\label{sec:acc}
We assume that the gas within the Bondi-Hoyle-Lyttleton (BHL) radius is interacting with the BH. The BHL radius is defined as:
\begin{equation}
r_{\mathrm{BHL}}=\frac{GM_{\BH}}{(c_s^2+v^2)}.
\end{equation} 
where $c_s$ is the sound speed of the gas and $v$ is the relative velocity between the BH and the AGN disk. The rescaled Bondi-Hoyle-Lyttleton accretion rate is
\begin{equation}
\Gamma_{\acc}=\frac{\dot{m}_{BHL}}{M_{BH}}=\frac{4\pi r_w r_h\rho_{\gas}(c_s^2+v^2)^{1/2}}{M_{BH}}\label{eq:acc}
\end{equation}
where
\begin{align}
&\rHill=r\left(\frac{M}{3M_{\SMBH}}\right)^{1/3},\\
&r_{w}=\min\left(r_{\mathrm{BHL}},\rHill,r_{\mathrm{shear}}\right),\\
&r_{h}=\min(r_w,h),\\
&r_{\mathrm{shear}}=\frac{GM_{\BH}}{\left(\rHill\vkep/r\right)^2}.
\end{align}
The relative velocity of the BH to the gas decreases at a rate:
\begin{equation}
\frac{d\vec{v}}{dt}=-\Gamma_{\acc}p_{\disk}\vec{v},
\end{equation}
where $p_{\disk}$ is the ratio of time the BH spends in the disk every orbit period.

The mass of the BH increases but is limited by the Eddington rate. The default Eddington limit is given by $\dot{M}_{\Edd}(M_{\BH})=L_{\Edd}(M_{\BH})/(0.1 c^2)$, where $L_{\Edd}$ is the Eddington luminosity and the fiducial radiation efficiency is set to $\eta_c=0.1$. The BH mass increases as:
\begin{equation}
\frac{dM_{\BH}}{dt}=\min\left(M_{\BH}\Gamma_{\acc}p_{\disk},\frac{\Gamma_{\Edd}}{\eta_c/0.1}\dot{M}_{\Edd}(M_{\BH})\right).\label{eq:acc_mass}
\end{equation}
Here $\Gamma_{\Edd}=1$ is the fiducial Eddington ratio of stellar mass BHs. To investigate the impact of accretion, we vary $\Gamma_{\Edd}/(\eta_c/0.1)$, which effectively alters both the accretion limit and the radiation efficiency $\eta_c$.

For binary BHs, the total mass $M_{\BH}=M_1+M_2$ is used to compute the mass increase (also applicable to gas dynamical friction and migration). The component masses should change as follows:
\begin{align}
\frac{dM_2}{dt}=\min\left(\frac{\lambda}{1+\lambda}\Gamma_{\acc}M_{\BH}p_{\disk}, \frac{\Gamma_{\Edd}}{\eta_c/0.1}\dot{M}_{\Edd}(M_{2})\right)\\
\frac{dM_1}{dt}=\min\left( M_{\BH}\Gamma_{\acc}p_{\disk}-\frac{dM_2}{dt}, \frac{\Gamma_{\Edd}}{\eta_c/0.1}\dot{M}_{\Edd}(M_{1})\right)\label{eq:acc_bin}
\end{align}
where $\lambda$ is a dimensionless function of mass ratio $q$, given by Eq. (1) in \cite{kelley2019massive}.
\begin{figure}
\centering
\includegraphics[width=0.5\linewidth]{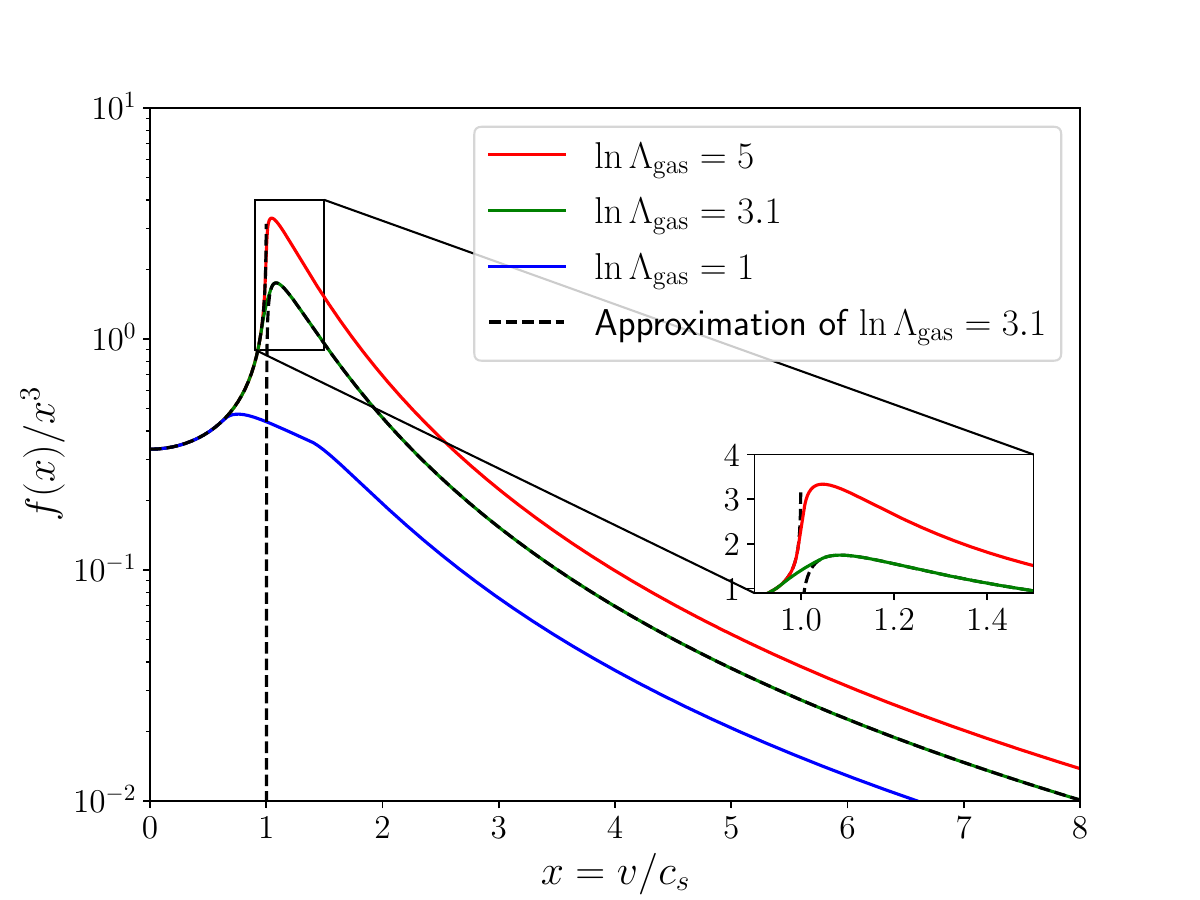}
\caption{The gas dynamical friction factor $f(x)/x^3$ at different $x=v/c_s$ for $\ln\Lambda_{\gas}=$ 1 (blue), 3.1 (fiducial value, green), and 5 (red). The singularity of the approximation (dashed black) is removed at $x=1$, and the peak of this function is at $x\sim1.1$ that varies slightly due to $\ln\Lambda_{\gas}$.} 
\label{fig:GDF}
\end{figure}

\subsection{Gas dynamical friction}
\label{sec:GDF}
We adopt the formulation of \cite{ostriker1999dynamical} for the deceleration of a BH with a non    zero relative velocity to the gaseous disk. The deceleration rate due to gas dynamical friction (GDF) is given by
\begin{equation}
\Gamma_{\GDF}=-\frac{(dv/dt)_{\GDF}}{v}=\frac{4\pi G^2M_{\BH}\rho_{\gas}}{c_s^3}\frac{f(x)}{x^3},\label{eq:GDF}
\end{equation}
where $x=v/c_s$, and we can analytically integrate $f(x)$ in \cite{ostriker1999dynamical} and assume the $\ln\Lambda_{\gas}=-\ln x_m=\ln\frac{c_st}{r_{\min}}$ in the formula, then
\begin{equation}
f(x)=\begin{cases}
\frac12\ln\left(\frac{1+x}{1-x}\right)-x&x<1-x_m\\
\frac12\ln\left(\frac{1+x}{x_m}\right)+\frac{(x-x_m)^2-1}{4x_m}&1-x_m\le x<1+x_m\\
\frac{1}{2}\ln(x^2-1)+\ln\Lambda_{\gas}&x\ge1+x_m
\end{cases}.
\end{equation}
We adopted a Coulomb logarithm for gas $\ln\Lambda_{\gas}=3.1$ as the fiducial value suggested by \cite{chapon2013hydrodynamics}. Compared to the approximation used in \cite{tagawa2020formation}, we remove the singularity at $v=c_s$, which occurs when the BH orbits near the AGN disk, as shown in the Fig. \ref{fig:GDF}. When a BH is outside but near the disk $i\sim h/r$, the gas dynamical friction reaches the maximum and it helps the BH rapidly fall into the disk. When a BH has a large inclination, its relative velocity is very large compared to the sound speed, leading to a low dynamical friction rate $\Gamma_{\GDF}\sim\ln x/x^3$. Additionally, we approximate the formula for very low velocities $v\ll c_s$:
\begin{equation}
\Gamma_{\GDF}=\frac{4\pi G^2M_{\BH}\rho_{\gas}}{3c_s^3}+\mathcal{O}(x).\label{eq:GDF_approx}
\end{equation}

The gas dynamical friction disappears when the Bondi-Hoyle-Lyttleton radius is smaller than the size of the H II sphere \citep{park2017gaseous}. The condition can be rewritten as \citep{tagawa2020formation}:
\begin{equation}
    \left(\frac{\rho_{\gas}}{2m_{\gas}\times10^{14}\,\mathrm{m^{-3}}}\right)\left(\frac{M_{\BH}}{10\Msun}\right)\left(\frac{v}{10\;\mathrm{km/s}}\right)^{-3}>1,\label{eq:feedback}
\end{equation}
where the average atomic mass of gas is $m_{\gas}=(1-Y_{\rm{He}})m_{\rm{H}}+Y_{\rm{He}}m_{\rm{He}}$ and we adopt the default Helium mass fraction $Y_{\rm{He}}=0.24$ as suggested by \cite{dors2022chemical}.
The gas dynamical friction recovers when $v>50$ km/s as suggested by \cite{park2017gaseous} and \cite{park2013accretion}.

\begin{figure}[!t]
    \centering
    \includegraphics[width=0.5\linewidth]{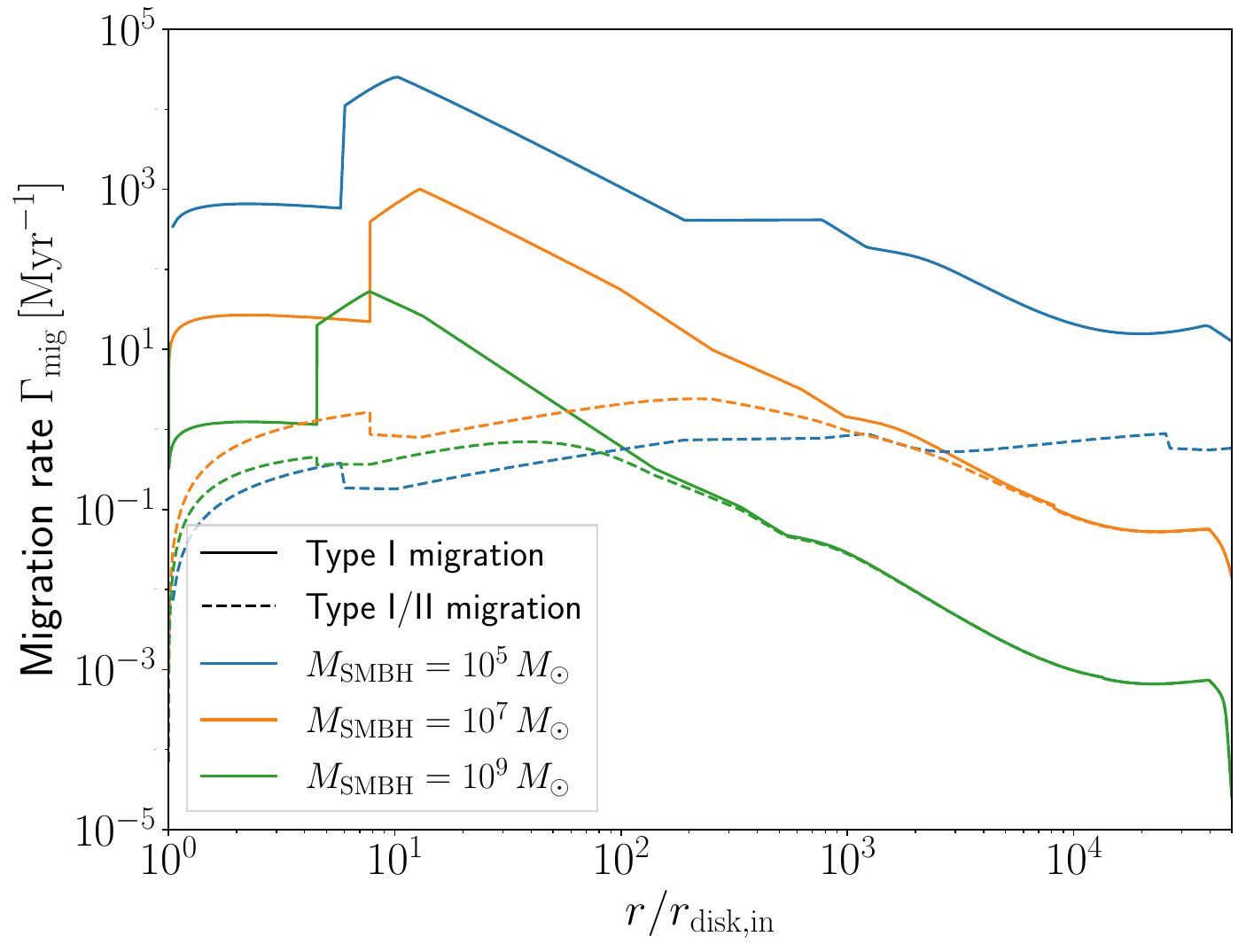}
    \caption{Migration rate of a $10\Msun$ test BH as a function of radius for different SMBH masses. The radius is normalized by the inner disk radius $r_{\disk,\rmin}$ (Eq. \ref{eq:disk_in}). The solid lines represent the type I migration rate, while the dashed lines show the type I/II migration rate, accounting for gap opening in the disk.}
    \label{fig:mig}
\end{figure}
\subsection{Migration}
\label{sec:mig}

In the studies of protoplanetary disks \citep{tanaka2002three,ward1997protoplanet,goldreich1979excitation}, the Lindblad resonances between a low-mass planet and the surrounding gas lead to a loss of angular momentum. This process results in radial migration, known as type I migration. When a planet becomes sufficiently massive, it repels the surrounding gas due to nonlinear torques, creating a density gap along its orbital path, which is known as type II migration. Similar migration behavior also applies to the interaction between BHs or stars and an AGN disk, as discussed by \cite{bellovary2016migration}. In particular, the conditions for type II migration is not hard to be satisfied ($K\gtrsim 25$) in our AGN disk models when BHs reside in the disk.

For simplicity, we adopt the same formula in \cite{tagawa2020formation}, accounting for both type I and II migration with the reduced gas in the gap \citep{kanagawa2018radial}:
%both  (for BHs outside the disk) and type II migration (for BHs within the disk):
\begin{equation}
\Gamma_{\mig}=2f_{\mig}\left(\frac{M_{\BH}}{M_{\SMBH}}\right)
\left(\frac{2\rho_{\gas}r^2\vkep}{M_{\SMBH}}\right)\left(\frac{h}{r}\right)^{-1},\label{eq:mig}
\end{equation}
where $\rho_{\gas}$ is the perturbed gas density profile discussed in Eq. \ref{eq:rho_gas}, and $f_{\mig}$ is a dimensionless parameter that quantifies the strength of the migration effect. In particular, the dimensionless factor $f_{\mig}$ is approximately $2.0$ but generally varies depending on the local disk properties, \citep{tanaka2002three,paardekooper2010torque,baruteau2011rapid}. We treat $f_{\mig}$ as a free parameter to assess its impact on merger rate and merger mass.

Fig. \ref{fig:mig} illustrates the dependence of the migration rate on the SMBH mass. The type I migration rate, which applies to BHs outside the AGN disk, follows a similar trend as the gas density profile and approximately scales with $M/M_{\SMBH}$ as in Eq. \ref{eq:mig}. For BHs embedded in the AGN disk, the combined type I/II migration is suppressed in the inner region due to gap opening (Eq. \ref{eq:rho_gas}), maintaining a similar order of magnitude for all SMBH mass. In the outer region, where BHs are generally too light to open gaps (except for very low SMBH masses, e.g., $M_{\SMBH}=10^5\Msun$), the migration rate converges with type I migration and remains lower. A deceleration in migration velocity occurs around $r\sim10r_{\disk,\rmin}$ due to a discontinuity in the gas density, effectively forming a discontinuous migration trap \citep{bellovary2016migration}. The details of migration traps were  discussed in Sec.  \ref{sec:mig_trap}.

\subsection{Binary hardening due to gas interaction}
\label{sec:gas_hard}
The analytical formulation of gas-driven binary hardening remains uncertain. In our model, a 2D circum-binary disk may form around the binary, exhibiting a higher gas density than its surroundings (Fig. \ref{fig:Interactions}). Therefore, both gas dynamical friction and the torque exerted by the circum-binary disk contribute to the binary hardening process:
\begin{equation}
    \Gamma_{\gas,s}=-\frac{(ds/dt)_{\gas}}{s}=\Gamma_{\GDF,s}+\Gamma_{\CBD,s}\label{eq:gas_hard}
\end{equation}
Many simulation studies \citep{baruteau2010binaries,li2021orbital,li2022long,rowan2023black} confirm that gravitational drag dominates over gas accretion effects, leading to merger timescales much shorter than the AGN disk lifetime, as illustrated in Fig. \ref{fig:timecale}. For simplicity, we do not distinguish between prograde and retrograde binaries, despite their hardening rate differences \citep{li2022hydrodynamical}.

We adopt gas dynamical friction for the gas hardening timescale following Eq. \eqref{eq:GDF}, with the feedback condition given by Eq. \ref{eq:feedback} or for $v>50$ km/s:
\begin{equation}
    \Gamma_{\GDF,s}=\frac{4\pi G^2M_{\BH}\rho_{\gas}}{c_s^3}\frac{f(x)}{x^3},\;\;\;x=\frac{\sqrt{GM_{\BH}/s}}{c_s}.
\end{equation}
Here, the relative speed is approximated as the binary's orbital velocity $\sqrt{GM_{\BH}/s}$, the gas density is taken as the gap density $\rho_{\gas}$, and the sound speed $c_s$ corresponds to that of the AGN disk. At large radii, where the gas density is lower, feedback effect generally stops gas dynamical friction when the binary separation is large. A In contrast, in the dense inner regions of the disk, the critical velocity defined by the feedback condition Eq.~\ref{eq:feedback} can exceeds 50 km/s, and gas dynamical friction remains effective throughout the hardening process.

We adopt the gas hardening formalism by \cite{ishibashi2020evolution} for prograde binaries to evaluate the viscous interaction with circum-binary disk, combined with accretion torque using Eq.~23 in \cite{ishibashi2024gravitational}
, assuming eccentricity $e=0$. The circum-binary disk hardening rate is then:
\begin{equation}
    \Gamma_{\CBD,s}=\frac{24\pi (1+q)^2 \alpha_{\CBD} c_{s,\CBD}^2\Sigma_{\gas,\CBD}}{q M \omega },\label{eq:hard_CBD}
    %(12\Sigma_{\gas,\CBD}-\frac{dM_{\BH}}{dt}\frac{1}{\omega s^2})
\end{equation}
where $\omega=\sqrt{GM_{\BH}/s^3}$ is the Keplerian angular velocity of the binary, %$\frac{dM_{\BH}}{dt}$ is the binary total accretion rate in Eq. \ref{eq:acc}, 
and the subscript $\CBD$ denotes the value of circum-binary disk.

 The properties of the circum-binary disk are calculated using Eqs. 5-14 from \cite{haiman2009population}. We assume a gas-dominated disk in the inner region, with the viscous $\alpha$ parameter for circum-binary disk set to be 
 $\alpha_{\CBD}=0.1$ as the fiducial value. The reduced accretion rate factor $\frac{\dot{m}}{\eta_c/0.1}$ is then given by Eq. \ref{eq:acc_mass}:
 \begin{equation}
     \frac{\dot{m}}{\eta_c/0.1}=\frac{dM_{\BH}/dt}{\dot{M}_{\Edd}(M_{\BH})}=\min\left(\Gamma_{\acc}\frac{M_{\BH}}{\dot{M}_{\Edd}},\frac{\Gamma_{\Edd}}{\eta_c/0.1}\right).
 \end{equation} 
 The surface gas density of the circum-binary disk, $\Sigma_{\CBD}$, is then given by
\begin{align}
\Sigma_{\CBD}&= 3.464\times10^8{\mathrm{g/cm^2}}\notag\\
     &\left( \frac{\dot{m}}{\eta_c/0.1}\right)^{3/5}\left(\frac{\alpha_{\CBD}}{0.3}\right)^{-4/5}\left(\frac{M_{\BH}}{\Msun}\right)^{1/5}\left(\frac{s}{10^{3}R_S}\right)^{-3/5}\notag\\
     &\quad\quad\quad\quad\quad\quad\quad\quad\quad\quad\quad\quad\text{if  }s<s_{\mathrm{es/ff}}\quad,\label{eq:Sigma_CBD_1}\\
\Sigma_{\CBD}&= 2.388\times10^6{\mathrm{g/cm^2}}\notag\\
     &\left( \frac{\dot{m}}{\eta_c/0.1}\right)^{7/10}\left(\frac{\alpha_{\CBD}}{3/10}\right)^{-4/5}\left(\frac{M_{\BH}}{\Msun}\right)^{1/5}\left(\frac{s}{10^{3}R_S}\right)^{-3/4}\notag\\
     &\quad\quad\quad\quad\quad\quad\quad\quad\quad\quad\quad\quad\quad\text{if  }s>s_{\mathrm{es/ff}}\quad,\label{eq:Sigma_CBD_2}
\end{align}
and the aspect ratio of circum-binary disk $(h/r)_{\CBD}=h_{\CBD}/s$ is given by    
\begin{align}
 \left(\frac{h}{r}\right)_{\CBD}&= 7.5\times10^{-3}\dot{m}\left(\frac{s}{10^3R_S}\right)^{-1}\quad\quad\text{if  }s<s_{\mathrm{gas/rad}}\quad,\label{eq:h_r_CBD_1}\\
\left(\frac{h}{r}\right)_{\CBD}&= 1.019\times10^{-2}\notag\\
     &\left( \frac{\dot{m}}{\eta_c/0.1}\right)^{1/5}\left(\frac{\alpha_{\CBD}}{0.3}\right)^{-1/10}\left(\frac{M_{\BH}}{\Msun}\right)^{-1/10}\left(\frac{s}{10^{3}R_S}\right)^{1/20}\notag\\
     &\quad\quad\quad\quad\quad\quad\quad\quad\text{if  }s_{\mathrm{gas/rad}}<s<s_{\mathrm{es/ff}}\quad,\label{eq:h_r_CBD_2}\\
\left(\frac{h}{r}\right)_{\CBD}&= 6.121\times10^{-3}\notag\\
     &\left( \frac{\dot{m}}{\eta_c/0.1}\right)^{3/20}\left(\frac{\alpha_{\CBD}}{0.3}\right)^{-1/10}\left(\frac{M_{\BH}}{\Msun}\right)^{-1/10}\left(\frac{s}{10^{3}R_S}\right)^{1/8}\notag\\
     &\quad\quad\quad\quad\quad\quad\quad\quad\quad\quad\quad\quad\text{if  }s>s_{\mathrm{es/ff}}\quad.\label{eq:h_r_CBD_3}
\end{align}

The separation of transition between inner region and middle region $s_{\mathrm{gas/rad}}$ and between middle region and outer region $s_{\mathrm{es/ff}}$ are given by
\begin{align}
    \frac{s_{\mathrm{gas/rad}}}{R_S}&=0.104\left( \frac{\dot{m}}{\eta_c/0.1}\right)^{16/21}\left(\frac{\alpha_{\CBD}}{0.3}\right)^{2/21}\left(\frac{M_{\BH}}{\Msun}\right)^{2/21}\\
    \frac{s_{\mathrm{es/ff}}}{R_S}&=4.10\times10^3\left( \frac{\dot{m}}{\eta_c/0.1}\right)^{2/3}
\end{align}

In practice, if the circum-binary disk gas density $\rho_{\CBD}=\Sigma_{\CBD}/(2h_{\CBD})$ is lower than the gap gas density $\rho_{\gas}$, we assume that the circum-binary disk does not form, where the $\Sigma_{\gas,\CBD}$ in Eq. \ref{eq:hard_CBD} is assumed to be 0.

\section{Interaction between stars and BHs}
\label{sec:int}
BHs interact with objects in their surroundings, which alters their velocity, and can result in binary formation or the change of binary separation. We consider the interaction between the main BH (or BH binary) and three types of surrounding components (denoted with subscript $c$): spherical background stars (back), BHs in the AGN disk (DBH), and stars in the AGN disk (Ds).  For each component type, we calculate four key parameters: the average mass $m_c$, the average local number density $n_c$, the average velocity dispersion relative to disk velocity $\sigma_c$, and the average scale height $h_c$. Note that $\sigma_c=0$ indicates that the component corotates with the disk. We assume that the fraction of time a BH or BH binary interacts with a given component, $p_{c}$ corresponds to the fraction of time it remains within the scale height of that component.

We treat the spherical background stellar component (denoted  with subscript ${\back}$) as a static background that does not evolve over time. Any BH will always interact with the spherical stars. The spherical background stellar properties are then:
\begin{align}
    &(m_{\back},n_{\back},\sigma_c,h_{\back})=\left(\bar{m}_*,n_*(r),\frac{\vkep(r)}{\sqrt{3}},\frac{r}{\sqrt{2}}\right),\\
    &p_{\back}=1.
\end{align}

The number density and average mass of disk components $(m_{\DBH}, n_{\DBH}, m_{\Ds}, n_{\Ds})$ are precalculated at the background level in Appendix  \ref{sec:disk_com}.  We assume that the velocity dispersion and the scale height of the disk components are identical, i.e., $\sigma_{\DBH}=\sigma_{\Ds}$, $h_{\DBH}=h_{\Ds}$. These values are derived in Appendix  \ref{sec:v_disp}, and recalculated during each loop of the individual sample simulations. The ratio of time that a BH or BH binary interacts with disk components $p_c$ is
\begin{equation}
p_c=\begin{cases}
1&\frac{h_c}{r}\ge\frac{|v_z|}{\vkep}\\
\frac{2}{\pi}\arcsin\left[\left	(\frac{h_c}{r}\right)/\left(\frac{|v_z|}{\vkep}\right)\right]& \frac{h_c}{r}<\frac{|v_z|}{\vkep}
\end{cases}.
\end{equation}

In this scenario, we assume that all interactions between an individual sample (a single BH or BH binary) and the surrounding components occur within the Hill radius $\rHill$. At this distance, the relative speed between the binary and the surrounding components, $\vrel$, is then calculated as:
\begin{equation}
\vrel=\max(\sqrt{3}\sigma_c,v,v_{\mathrm{rel,mig}},v_{\mathrm{shear}}),\label{eq:v_rel}
\end{equation}
where the relative migration velocity is $v_{\mathrm{rel,mig}}=(\Gamma_{\mig}-\Gamma_{\mig,c})r$. The migration rate of spherical background stars is set to be $\Gamma_{\mig,\back}=0$, and the migration rate of the disk components is computed using Eq. \ref{eq:mig} with average mass $m_c$. The shear velocity between the BH sample and the third object from the   component is
\begin{equation}
    v_{\mathrm{shear}}=p_{\uni}\rHill\frac{\vkep}{r},\label{eq:shear}
\end{equation}
where $p_{\uni}$ is uniform between 0 and 1.

When analyzing the interaction between individual samples and disk BH components, we account for the fact that binary formation depletes the population of single BHs at approximately the same radii. To reflect this, we adjust the number density of disk BHs as follows:
\begin{equation}
    n_{\DBH,\mathrm{int}}=\max\left(n_{\DBH}-\frac{N_{\mathrm{form}}}{V_{\DBH}},0\right),\label{eq:DBH_modification}
\end{equation}
where $n_{\DBH}$ is calculated in Appendix  \ref{sec:disk_com} as an evolving background, $N_{\mathrm{form}}$ represents the number of binaries formed within the cell (which resets to zero once the BH migrates to a different cell), and $V_{\DBH}$ is the volume of the disk BH population within the cell, given by:
\begin{equation}
    V_{\DBH}=\pi (r_{\mathrm{right}}^2-r_{\mathrm{left}}^2) h_{\DBH},
\end{equation}
where $r_{\mathrm{left}}$ and $r_{\mathrm{right}}$ is the right-left bound radii of the radial grid.

As a result, the modified number density, $n_{\DBH,\mathrm{int}}$, which is applied to interactions with disk BH components, effectively suppresses binary formation and binary-single interactions at small radii. This mechanism naturally limits the formation of new binaries as the available population of single BHs is depleted.

We define the impact parameter for a $90^\circ$ deflection angle due to gravity as $b_{90}$:
\begin{equation}
b_{90}=\frac{G(M_{BH}+m_{c})}{\vrel^2},\label{eq:90}
\end{equation} 
where $\vrel$ is the relative velocity between the BH and the surrounding object.

When discussing the interactions between other components (denoted by subscript $c$), the scale height is modified as   
\begin{equation}
    h_{\eff}=\max\left(\frac{|v_z|}{\vkep}\cdot r,h_c\right).
\end{equation}

\subsection{Weak scattering}
\label{sec:WS}
A weak interaction refers to the velocity exchange resulting from encounters between BH and its surrounding objects. According to the Fokker-Planck approximation in \cite{binney2011galactic}, the change in velocity due to weak scattering, $\Delta \vec{v}_{\WS}$, consists of a dynamical friction term $\Delta \vec{v}_{\DF}$ and a diffusion term $\Delta \vec{v}_{\diff}$. The change of the velocity in time $\Delta t$ using Monte Carlo method is :
\begin{align}
&\Delta v_{\DF}=p_{c}D[\Delta v_{\parallel}]\Delta t,\label{eq:WS,DF}\\
&\Delta v_{\diff}=p_{c}\sqrt{\left(D[\Delta v_{\perp}^2]+D[\Delta v_{\parallel}^2]\right)\Delta t}\label{eq:WS,diff},\\
&\Delta \vec{v}_{\WS}=\Delta v_{\mathrm{DF}}\hat{v}+\Delta v_{\mathrm{diffusion}}\hat{n},\label{eq:WS_tot}
\end{align}
where $\hat{v}$ is the unit vector of velocity direction, $\hat{n}$ is a random 3D vector.

When the impact parameter is smaller than the scale height of the surrounding components, $b_{90}< h_c$, we apply the weak scattering approximation for an infinite homogeneous medium (Equation 7.92 in \cite{binney2011galactic}). The Coulomb logarithm $\Lambda_{\WS}$ is set to be $\Lambda_{\WS}=h_c/b_{90}$, suggested by \cite{papaloizou2000orbital}, and the mass density of the surrounding component is $\rho_c=n_c m_c$. The dynamical friction and diffusion coefficients of velocity change are then given by
\begin{align}
D[\Delta v_{\parallel}]&=-\frac{4\pi G^2 (M_{BH}+m_c)m_cn_c\ln\Lambda_{\WS}}{\sigma_c^2}G(X),\label{eq:WS_1}\\
D[\Delta v_{\parallel}^2]&=\frac{4\pi \sqrt{2}G^2m_cn_c\ln\Lambda_{\WS}}{\sigma_c}\frac{G(X)}{X},\label{eq:WS_2}\\
D[\Delta v_{\perp}^2]&=\frac{4\pi \sqrt{2}G^2m_cn_c\ln\Lambda_{\WS}}{\sigma_c}\left[\frac{\erf(X)-G(X)}{X}\right].\label{eq:WS_3}
\end{align}
The quantity $X=|\vec{v}-\vec{v}_c|/(\sqrt{2}\sigma_c)$, where $v_c=0$ for disk components since they corotate with disk with low dispersion, and $v_c=\vkep$ for spherical component. The function $G(x)$ is given by
\begin{equation}
G(X)=\frac{1}{2X^2}\left[\erf(X)-\frac{2X}{\sqrt{\pi}}e^{-X^2}\right].
\end{equation}

When the impact parameter is larger than the scale height of the surrounding components, $b_{90}\ge h_c$,  the scattering angle is less than 90 degrees, and the dynamical friction and diffusion terms are confined to two dimensions. The modified diffusion terms are (Appendix B of \cite{tagawa2020formation}):
\begin{align}
&D[\Delta v_{\parallel}]=\begin{cases}
0&\text{c=DBH \& Ds}\\
-9.765\,Gm_{c}n_{c}h_{c}&\text{c=back}
\end{cases},\\
&D[\Delta v_{\parallel}^2]=D[\Delta v_{\perp}^2]=\begin{cases}
(2\pi)^{3/2}Gm_{c}n_{c}h_{c}\frac{\sigma_c m_c}{M_{BH}+m_c}&\text{c=DBH \& Ds}\\
12.7\,Gm_{c}n_{c}h_{c}\frac{\sigma_c m_c}{M_{BH}+m_c}&\text{c=back}
\end{cases}.
\end{align}
where the  $\hat{n}$ in Eq. \ref{eq:WS_tot} becomes a random vector in the $xy$ plane.

\subsection{Binary single interaction}
\label{sec:BS_int}
Binary BHs sometimes undergo significant changes in their separation when interacting with surrounding objects, e.g. \cite{binney2011galactic}. These interactions are characterized by close gravitational interactions between the binary and a third object. The cross section for these encounters is approximated as $\sigma_{\rm{coll}}=b_{xy}b_z$, where the effective impact parameters in the $xy$ plane (disk plane) and z direction $b_{xy}$ and $b_z$ are given by
\begin{align}
b_{xy}&=\min\left(s\sqrt{1+\frac{2b_{90}}{s}},\rHill\right),\label{eq:xy_coll}\\
b_{z}&=\min\left(b_{xy},h_{\eff}\right).
\end{align}
The rate of binary-single interactions for the component $c$, denoted as $\Gamma_{\BS,c}$, is given by:
\begin{equation}
\Gamma_{\BS,c}=p_c n_c\sigma_{\mathrm{coll}}\vrel.
\end{equation}
The probability of a BH undergoing a binary-single interaction within time $\Delta  t$ is $\Gamma_{BS,c}\Delta t$.

Binary-single interactions can be classified into two regimes based on the comparison between the binary's binding energy and the kinetic energy of the surrounding objects. When the binary's binding energy $E_b=GM_1M_2/(2s)$ is smaller than the kinetic energy of the surrounding objects $E_c=\frac32 m_{c}\sigma_{c}^2$, the binary is ``soft" and experiences a softening interaction, primarily relevant for encounters with background stars. The soft-hard boundary separation is given by

\begin{equation}
    s_{\mathrm{soft,hard}}=\frac{Gm_1m_2}{3m_{\back}\sigma_{\back}^2}=\frac{q}{(1+q)^2}\frac{GM_{\BH}^2}{\bar{m}_{s}\vkep^2},\label{eq:soft_hard}
\end{equation}
Initially, preexisting binary separations must be smaller than the soft-hard boundary, as wide binaries are destroyed before AGN disk formation. The softening rate applied to the BH binary is derived in \cite{gould1991binaries}:
\begin{equation}
\left(\frac{ds}{dt}\right)_{\BS,\back}=\frac{16}{3}\frac{Gn_{\back}m_{\back}s^2}{M_{\BH}\sigma_{\back}^3}(E_b-E_c)\ln\left(\frac{GM_{\BH}}{s\sigma_{\back}^2}\right).\label{eq:soft_rate}
\end{equation}

Conversely, "hard" binaries, where the binding energy exceeds the kinetic energy of the surrounding objects, undergo hardening \citep{leigh2018rate}. This process is common in interactions with disk components due to their low velocity dispersion, resulting from significant dynamical friction. The total energy of the system, consisting of the binary BH and the surrounding object, is determined by
\begin{equation}
E_0=\frac12\frac{m_cM_{\BH}}{m_c+M_{\BH}}\vrel^2-\frac{GM_{\BH}m_c}{\rHill}-E_b.
\end{equation}

We choose the escape velocity from \cite{leigh2018rate} to estimate the velocity change of the main object $\Delta v_{BH}$, since it is typically smaller than the binary velocity $v_{bin}=(GM_1M_2/s)^{1/2}$. The recoil velocity after the interaction is described by
\begin{equation}
\Delta \vec{v}_{\BH}=\vec{v}_{\rm{recoil}}=\alpha\left[\frac{ M_{\BH}}{m_c(m_c+M_{\BH})}|E_0|\right]^{1/2}\hat{n},
\end{equation}
and the separation hardening rate is
\begin{align}
    &E_{b,f}=E_b+\frac{\alpha^2}{2}|E_0|.\\
    &s_f=\frac{GM_1M_2}{2E_{b,f}}\\
&\left(\frac{ds}{dt}\right)_{\BS}=\frac{s_f-s_i}{\Delta t}.
\end{align}

\subsection{Binary formation due to three-body interaction}
\label{sec:BBF}
In dense or low-velocity dispersion regions, a close encounter of three objects can result in binary formation (e.g., \citealt{binney2011galactic}). The encounter impact parameter is constrained by the gravitational interaction and is selected as:
\begin{equation}
b_i=\min(b_{90},\rHill).
\end{equation}
The vertical ($z$direction) extent of the disk, combined with the component scale height, further limits the impact parameters, such that the effective vertical impact parameter becomes $b_{z,\eff}=\min(b_i,h_{\eff})$. The mean rate of a two-body encounter is given by $\Gamma_2=p_cn_cb_ib_{z,\eff}\vrel$, where the third object must be located within a volume of $b_i^2b_{z}$, with $b_z=\min(b_i,h_c)$. This leads to the three-body encounter rate \citep{tagawa2020formation}:
\begin{equation}
\Gamma_{3bbf}=p_{\DBH}n_{\DBH}\left(\frac12n_{\DBH}+n_{\Ds}\right)b_i^3b_{z,\eff}b_z\vrel.\label{eq:BBF}
\end{equation}
The factor $1/2$ on disk BH number density accounts for the double counting effect, and the third object could potentially be the disk stars instead of BHs. The probability of a three-body encounter resulting in binary formation within time $\Delta t$ is $P=\Gamma_{3bbf}\Delta t$. The binary that forms from such an interaction will have the following properties:
\begin{align}
&M_{\mathrm{bin}}=M_{\BH}+m_{\BH,2}\\
&\vec{v}_{\mathrm{bin}}=\vec{v}_{\mathrm{kick}}+\vec{v}_{\mathrm{cen}}\\
&s_{\mathrm{bin}}=b_i
\end{align}
where the kick velocity is
\begin{equation}
    \vec{v}_{\mathrm{kick}}=\frac{m_c}{M_{\BH}+m_{\BH,2}+m_{c}}\sqrt{\frac{GM_{\BH}m_{\BH,2}}{b_i}},
\end{equation}
and the center-of-mass velocity is
\begin{equation}
\vec{v}_{\mathrm{cen}}=\frac{M_{\BH}\vec{v}_{\BH}+m_{\BH,2}\vec{v}_2+m_{c}\sigma_{\DBH}\hat{n}_2}{M_{\BH}m_{\BH,2}+m_{c}}.
\end{equation}
The partner BH of this binary formation is chosen directly from the preexisting BH at the same cell. The mass $m_{\BH,2}$ and velocity $\vec{v}_2$ are generated from Eq. \ref{eq:IMF} and inclination in Eq. \ref{eq:incl_Gaussian}.

\subsection{Binary formation via gas capture}
\label{sec:gas_cap}

In AGN disks, a binary may form if sufficient energy is dissipated during the interaction of two BHs. Gas dynamical friction is the primary mechanism responsible for dissipating energy in such encounters near the disk \citep{delaurentiis2023gas}. \cite{qian2024dynamical} also considers a gravitational wave-assisted formation channel for close encounters with an impact parameter of $\sim2\times10^{-5}\rHill$. More recent studies \citep{rowan2024black_2,rowan2024black} provide refined predictions for separation, eccentricity, and encounter rates. To optimize computational efficiency, we adopt the formalism from \cite{goldreich2002formation} and \cite{tagawa2020formation}, where the fraction of binding energy lost during a passage through the mutual Hill sphere is proportional to the fraction of time that two bodies remain within this Hill radius. The timescale for crossing the Hill radius is
\begin{equation}
t_{\mathrm{pass}}=\frac{\rHill}{v_{\rel,c}},
\end{equation}

The gas dynamical friction rate, $\Gamma_{\GDF}(v_{\rel,c})$, reduces the relative velocity between the two bodies during this interaction following Eq. \ref{eq:GDF}. The probability of binary formation due to gas capture during a single encounter is
\begin{equation}
P_{\mathrm{cap}}=\min(1,\Gamma_{\GDF}t_{\mathrm{pass}}).
\end{equation}

The rate of BH encounters with disk BH components is \citep{tagawa2020formation}:
\begin{equation}
\Gamma_{\mathrm{enc}}=n_{\DBH}\rHill z_{\mathrm{Hill}}v_{\rel} p_{\DBH},
\end{equation}
where
\begin{equation}
    z_{\mathrm{Hill}}=\min(\rHill,h_{\DBH}).
\end{equation}
The gas capture rate of a single BH is then
\begin{equation}
\Gamma_{\mathrm{cap}}=\Gamma_{\mathrm{enc}}P_{\capt}.\label{eq:cap}
\end{equation}
Thus, the probability of gas capture during a time step $\Delta t$ is $P=\Gamma_{\mathrm{cap}}\Delta t$. When a gas capture event occurs, it is assumed to happen within the AGN disk, where gas dynamical friction remains effective. The BH binary properties are given by
\begin{align}
    &M_{\mathrm{bin}}=M_{\BH}+m_{\BH,2},\\
    &\vec{v}_{\mathrm{bin}}=\vec{v}_{\BH},\\
    &s_{\mathrm{bin}}=\rHill,
\end{align}
where in our simulation, the mass of the partner BH is assumed to be $m_{\BH,2}=m_{\DBH}(r,t)$.

\section{Statistical Background}
\label{sec:disk_com}
\subsection{Number density and average mass}
\label{sec:nd&md}
\begin{figure*}[!t]
    \centering
    \includegraphics[width=1.0\linewidth]{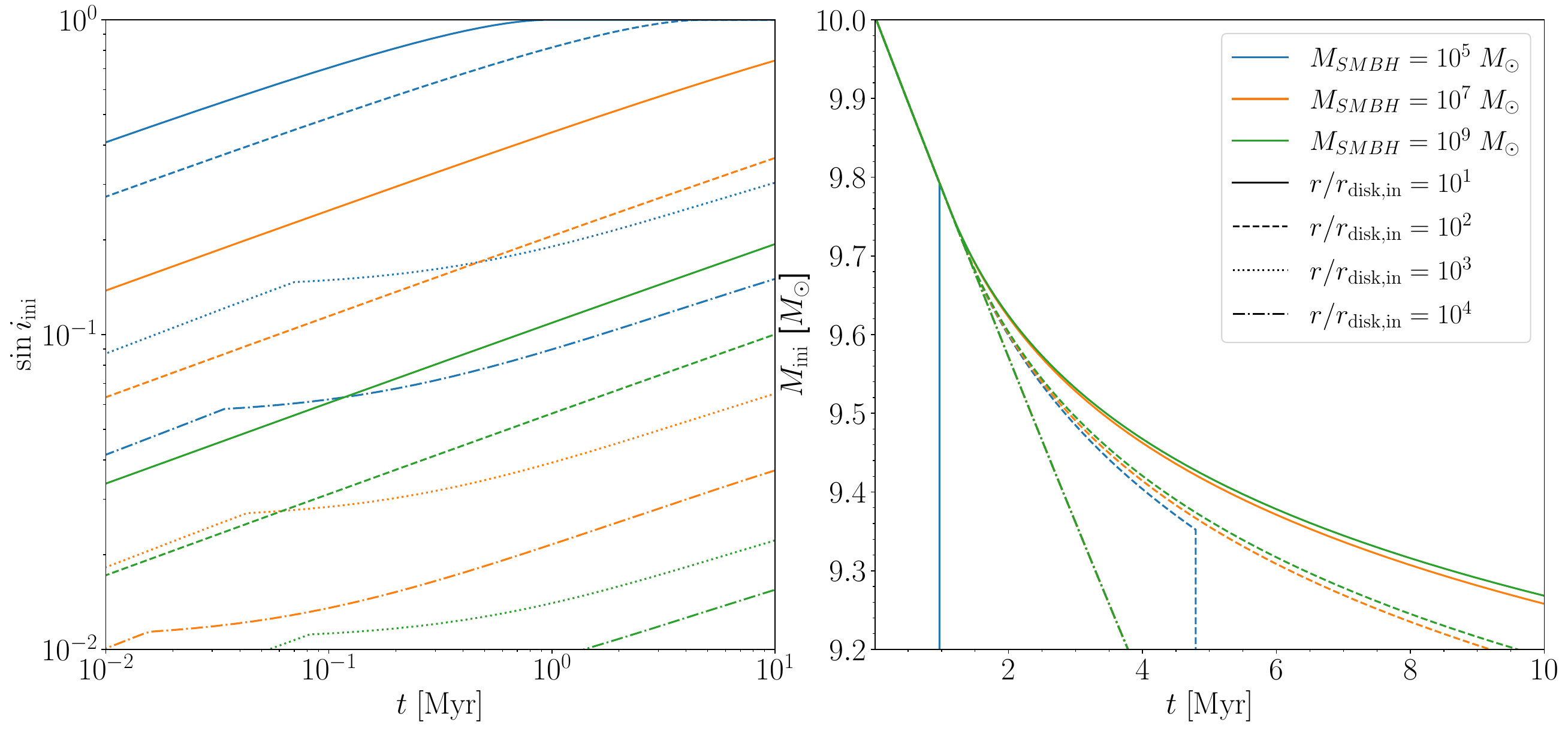}
    \caption{Initial inclination (left) and initial mass (right) as functions of falling time $t$ of a BH mass $10\Msun$ at different radii. The solid, dashed, dotted, and dash-dotted lines refer to the BHs falling into disk at radius $r/r_{\disk,\rmin}=10^{1},10^{2},10^{3},10^{4}$. The red, blue, and green lines correspond to the host SMBH mass of $10^5$, $10^7$, and $10^9$ $\Msun$.}
    \label{fig:ini_inclination}
\end{figure*}
The surface distribution of disk BHs is characterized as
\begin{equation}
  \frac{dN_{\DBH}}{(2\pi rdr)\,dM}=  \Sigma(r,M).
\end{equation}
We then obtain the maximum disk BH distribution, assuming that all BHs at $r$ fall into the disk without radial migration:
\begin{equation}
\Sigma_{\max}(r,M)=2rn_{\BH}(r) \mathcal{F}_{\preexist}(M),
\end{equation}
where the $\mathcal{F}_{\preexist}(M)$ is the Salpeter initial mass function in Eq. \ref{eq:IMF}. The initial surface distribution of disk BHs is only a fraction of the maximum surface density, scaled by the function $F(i_{\disk})$ (Eq. \ref{eq:incl_frac}). The initial inclination at the disk surface, $i_{\disk}$—which also corresponds to the inclination at time $t=0$ when AGN disk is formed—is defined as:
\begin{equation}
    i_{\disk}=i_{\ini}(t=0)=\arcsin\left(\frac{h}{r}\right)
\end{equation}
Accordingly, the initial surface density distribution is expressed as:
\begin{equation}
\Sigma(t=0,r,M)=\Sigma_{\max}(r,M)F\left(i_{\disk}\right).
\end{equation}

The distribution of disk BHs evolves according to the following equation:
\begin{align}
%\begin{split}
&\frac{1}{r}\left[\frac{\partial (r\Sigma(t,r,M))}{\partial t}+\frac{\partial(r \Sigma(t,r,M))}{\partial r}\frac{dr}{dt}+\frac{\partial(r \Sigma(t,r,M))}{\partial M}\frac{dM}{dt}\right]\notag\\
&=\frac{dN_{\mathrm{fall}}}{dAdt\,dM}+\left(\frac{dN_{\BH}}{dA\,dt}\right)_{\cre}\mathcal{F}(M).\label{eq:Back_evo}
%\end{split}
\end{align}
Here $\frac{dr}{dt}=-\Gamma_{\mig}r$ and $\frac{dM}{dt}=\min(\Gamma_{\acc}M,\frac{\Gamma_{\Edd}}{\eta_c/0.1}\dot{M}_{\Edd}(M))$ represent the migration and accretion effect in the disk, with rate $\Gamma_{\mig}$ and $\Gamma_{\acc}$ defined in Eqs. \ref{eq:mig} and \ref{eq:acc}. The term $\frac{dN_{\mathrm{fall}}}{dA\,dt}$ represents the rate at which BHs fall from outside the disk, while $\left(\frac{dN_{\BH}}{dA\, dt}\right)_{\cre}$ denotes the BH formation rate within the disk in Eq. \ref{eq:BH_cre}.  The function $\mathcal{F}(M)$ is the normalized IMF for BHs formed due to star formation governed by $\beta_{\cre}$, stellar creation function $\mathcal{F}_{\cre}$ and star-BH relation Eq. \ref{eq:ms_mbh}. In our timescale calculations, we assume that the velocity of the BH relative to the AGN disk is zero, as it is typically small compared to the sound speed, $c_s$. Additionally, we do not consider the formation of binary systems from single BHs in disk component calculation. The effect of these assumptions was   be discussed in Sec. \ref{sec:limitation}.

The number of time steps $\Delta t=t_{\AGN}/N_{\mathrm{time}}$ syncretizing this equation is then given by  
\begin{equation}
N_{\mathrm{time}}=\left[t_{\AGN}/\left(0.01/\max\left(\Gamma_{\mig},\Gamma_{\acc}\right)\right)\right], \label{eq:time_step_back}
\end{equation}

To evaluate the rate of BHs that fall into the AGN disk, we consider a BH of mass $M_{\BH}$ that enters the disk (inclination angle decrease to $i_{\disk}$) at time $t_0$. The BH initially has mass $M_{\ini}$ and inclination angle $i_{\ini}$. The BH relative velocity to disk should evolve as 
\begin{equation}
     \frac{d\vec{v}}{dt}=-\vec{v}(\Gamma_{\acc}+\Gamma_{\GDF}) p_{\disk}(i). 
\end{equation}
 Assuming a circular orbit, the relative velocity to the disk is given by $v=2\vkep\sin\frac{i}{2}$, and the evolution of the BH’s mass and inclination then follows:
\begin{align}
    \frac{dM_{\BH}}{dt}&= M_{\BH}\cdot\max\left(\Gamma_{\acc}p_{\disk},\frac{\Gamma_{\Edd}L_{\Edd}(M_{\BH})}{M_{\BH}\eta_c c^2}\right),\label{eq:M_ini}\\
     \frac{d i}{dt}&=-2\tan \frac{i}{2}\, (\Gamma_{\acc}+\Gamma_{\GDF}) p_{\disk}(i).\label{eq:psi_ini}
 \end{align}
The gas dynamical friction rate $\Gamma_{\GDF}$ and accretion rate $\Gamma_{\acc}$ are calculated by Eqs. \ref{eq:GDF} and \ref{eq:acc}, and  the time fraction $p_{\disk}(i)$ is calculated  by Eq. \ref{eq:p_disk}. Note that if the BH with initial mass  $M_{\ini}$ and initial inclination $i_{\ini}$ enters the disk at time $t$, BH with same mass $M_{\ini}$ and inclination in range $-i_{\ini}\le i\le i_{\ini}$ should be all in the disk. The rate at which BHs fall into the disk is given by
\begin{equation}
\begin{split}
        r\frac{dN_{\mathrm{fall}}}{dAdt\,dM}&=r\frac{\Sigma_{\max}(r,M_{\ini}(t+\Delta t))F(i_{\ini}(t+\Delta t))}{\Delta t}\\
        &-r\frac{\Sigma_{\max}(r,M_{\ini}(t))F(i_{\ini}(t))}{\Delta t}\\
        &=r\frac{d[\Sigma_{\max}(r,M_{\ini})F(i_{\ini})]}{dt}
        .\label{eq:fall_rate}
\end{split}
\end{equation}
The time steps in Eq. \ref{eq:fall_rate} is set to be $\Delta t=t_{\AGN}/N_{\mathrm{time}}$, consistent with Eq. \ref{eq:Back_evo}. The values $i_{\ini}(t+\Delta t)$ and $M_{\ini}(t+\Delta t)$ can be computed directly from Eqs. \ref{eq:M_ini} and \ref{eq:psi_ini} using initial conditions $i_{\ini}(t+\Delta t)$ and $M_{\ini}(t+\Delta t)$ and evolving over time $\Delta t$.

Fig.~\ref{fig:ini_inclination} illustrates the evolution of the initial inclination $i_{\ini}$and initial mass $M_{\ini}$ as functions of the BH infall time $t$. On a log-log scale, the growth of $\sin i_{\ini}$ appears nearly linear, slowing only when gas dynamical friction becomes ineffective in specific velocity regimes due to radiation feedback (Eq. \ref{eq:feedback}). The initial inclination angle $i_{\ini}(t=0)$ is comparable to the disk aspect ratio $h/r$. It then increases rapidly over a short timescale (e.g. $\lesssim10^{-2}$ Myr) due to the strong dynamical friction near the disk.

BHs around more massive SMBHs start at lower initial inclinations but increase at a nearly identical rate on a logarithmic scale (e.g., $M_{\SMBH}=10^9\Msun$, shown by the green line). Due to the lower gas density at larger radii, most disk BHs originate with small inclination angles [i.e. $\sin i_{\ini}\sim\mathcal{O}(10^{-2})-\mathcal{O}(10^{-1})$] and remain near the disk for an extended time. As a result, the mass gained through accretion becomes more significant at larger radii, as shown in the right panel Fig.~\ref{fig:ini_inclination}. The vertical line in the right panel indicates the point at which mass accretion ceases, as all BHs have reached the disk within the given time frame. The initial mass shows minimal variation across different SMBH masses, attributed to their similar gas density profiles.

We neglect the migration of BHs outside the disk in this step to reduce computational cost. The surface number density and average mass of disk BHs at time $t$ and radial distance $r$ is then
\begin{align}
\Sigma_{\DBH}&=\int_{M_{\min}}^{M_{\max}}dM\,\Sigma(t,r,M),\\
m_{\DBH}&=\frac{\int_{M_{\min}}^{M_{\max}}dM\,M\Sigma(t,r,M)}{\int_{M_{\min}}^{M_{\max}}dM\,\Sigma(t,r,M)}.
\end{align}
The required disk BH number density $n_{\DBH}$ should be calculated by  
\begin{equation}
    n_{\DBH}=\Sigma_{\DBH}/(2h_{\DBH}),\label{eq:DBH_density}
\end{equation}
where $h_{\DBH}$ is the scale height of disk BH in Eq.\ref{eq:scale_height} , determined by weak scattering process, as discussed in the Appendix  \ref{sec:v_disp}.

Using the discretization method from Sec.  \ref{sec:Method}, we estimate the average mass and number density of disk BHs at any time within the AGN lifetime. The disk star average mass is set to be $m_{\Ds}=\bar{m}_{*}$, while the number density of disk stars is calculated similarly. We do not account for the dynamical friction of stars, and the accretion of stars is assumed to proceed at the maximum Bondi-Hoyle-Lyttleton rate. We adopt Equation 11 of \cite{2017ApJ...835..165B} as the timescale of the accretion rate of stars:
\begin{align}
&r_{\mathrm{BHL},*}(i)=\frac{2G\bar{m}_*}{\vkep^2\left[4\sin^2\frac{i}{2}+\left(\frac{h}{r}\right)^2\right]},\\
&t_{*}(i)=\frac{\pi r}{\vkep}\frac{\bar{m}_*\cos\frac{i}{2}}{\rho\pi r\left(\frac{h}{r}\right)r_{\mathrm{BHL},*}^2}.
\end{align}
The maximum initial inclination of disk stars at time $t$ is solved via $t=t_*(i_{*,\ini})$, and the number density of the disk stars is then 
\begin{equation}
n_{Ds}(t)=\left[n_*(1-\cos i_{*,\ini})+\frac{1}{r}\left(\frac{dN_*}{dAdt}\right)_{\cre}\right]/\left(2\frac{h_{\Ds}}{r}\right).
\end{equation}

\subsection{Velocity dispersion and scale height}
\label{sec:v_disp}
We assume that the disk star component has the same velocity dispersion and scale height as the disk BH component. As an approximation, the scale height of the disk component is estimated using the velocity dispersion $\sigma_{\DBH}$:
\begin{equation}
h_{\Ds}=h_{\DBH}=\frac{\sigma_{\DBH}}{\vkep}r.\label{eq:scale_height}
\end{equation}

The velocity dispersion of the disk component is taken to be the average velocity in the $x,y$ and $z$ components of the BH's relative velocity in the disk, with $v\sim\sqrt{3}\sigma_{\DBH}$. We assume that the velocity dispersion represents the root mean square velocity in each component for a disk BH with mass $m_{\DBH}$. 
When a BH falls into the disk, the $z$ component of velocity $v_z=\left(\frac{h}{r}\right)\vkep=c_s$. At this point, the gas dynamical friction is near its maximum since  $v\gtrsim c_s$. The dynamical friction rapidly reduces the velocity, resulting in $\sigma_{\DBH}\ll c_s$. This effect continues to reduce the velocity until equilibrium is reached between the dominant gas dynamical friction, which drives the velocity decrease, and weak scattering diffusion. To capture this process, we use the leading-order term of gas dynamical friction rate in Eq. \ref{eq:GDF_approx}  and set the velocity change rate equal to the diffusion velocity as described in Eq. \ref{eq:WS,diff},
\begin{equation}
\sqrt{\left(D[\Delta v_{\perp}^2]+D[\Delta v_{\parallel}^2]\right)_c\Delta t}=\Gamma_{\GDF}\sigma_{\DBH}\Delta t .
\end{equation}
The relative velocity of a disk BH relative to background stellar component is determined by $\sqrt{3}\sigma_{\back}=\vkep$ as the other terms are too small. Thus the impact parameter for 90-degree deflection is $b_{90}=G(m_{\back}+m_{\DBH})/\vkep^2$. Consequently, the Coulomb logarithm for weak scattering is given by:
\begin{equation}
\Lambda_{\WS}=\left[\frac{\sqrt{2}G(m_{ss}+m_{DBH})}{r\vkep^2}\right]^{-1}\sim\frac{M_{\SMBH}}{m_{ss}+m_{DBH}}\gg1.
\end{equation}
$X$ is taken to be $\sqrt{3/2}$ in Eqs. \ref{eq:WS_2} and \ref{eq:WS_3} due to the low velocity of disk BHs, and the time interval is chosen to be $1/\Gamma_{\GDF}$, the velocity dispersion is thus:
\begin{equation}
    \sigma_{\DBH}=4.93\sqrt{\frac{G^2m_{\back}n_{\back}\ln\Lambda_{\WS}}{\vkep \Gamma_{\GDF}}}.\label{eq:v_dispersion}
\end{equation}
The disk BH and star components share the same velocity dispersion, and the relative velocity between two disk BHs is small, leading to $\Lambda_{\WS}<1$. The diffusion velocity from disk components is thus only on the x-y plane and does not affect the scale height. We neglect its effect on velocity dispersion and thus the velocity dispersion is slightly underestimated. The velocity dispersion of the disk component is normally $\sim 10^{-3}c_s$ at the outer region of the AGN disk and $\sim 10^{-2}c_s$ at the inner region of AGN disk ($r\lesssim10r_{\disk,\rmin}$).

\section {Time step and evolution in individual BH simulation}
\label{sec:time_step}
 
In simulations of individual BHs, the number density and mass density of disk BHs is precalculated at the background level (Appendix  \ref{sec:nd&md}) with time step $t_{\mathrm{AGN}}/N_{\mathrm{time}}$ (Eq. \ref{eq:time_step_back}). We calculate the disk components' velocity dispersion $\sigma_{\DBH}$ and scale height $h_{\DBH}$ in each loop (Appendix  \ref{sec:v_disp}), and then all the timescales of interactions with gas and stellar components. The time step of an individual BH simulation is then determined by
\begin{equation}
    \Delta t=\eta_t/\max(\Gamma_{\mig}p_{\disk},\Gamma_{\acc}p_{\disk},\Gamma_{\gas,s},\\
    \Gamma_{GW},\Gamma_{BS,c}p_{c}),
\label{eq:time_step}
\end{equation}
where $\eta_t=0.1$ is the default value and subscript $c$ denotes the spherical background, disk BH, and disk star components. The gas dynamical friction rate is very large near or in the disk, as shown in Fig.~\ref{fig:GDF}. We thus neglect the gas dynamical friction term in the time step to reduce the computational cost. 

The properties are updated in each time step as
\begin{align}
    r_{\new}=&r(1-\Delta t\cdot\Gamma_{\mig}p_{\disk}),\label{eq:update_r}\\
    s_{\new}=&s(1-\Delta t\cdot\Gamma_{\gas,s})(1-\Delta t\cdot\Gamma_{GW})\notag\\
    &+\Delta t\cdot\sum_{c}\left(\frac{ds}{dt}\right)_{\BS},\label{eq:update_s}\\
    \vec{v}_{\new}=&\vec{v}\cdot\exp(-\Gamma_{\GDF}\Delta t\,p_{\disk})(1-\Gamma_{\acc}\Delta t\,p_{\disk})\notag\\
    &+\sum_{c}\Delta\vec{v}_{\WS}+\sum_{c}(\vec{v}_{rec})_{\BS},\label{eq:update_v}\\
    M_{\new}=&M+\left(\frac{dM}{dt}\right)_{\acc}.\label{eq:update_M}
\end{align}
To ensure consistency with the velocity dispersion calculation in Appendix ~\ref{sec:v_disp}, we set the time interval $\Delta t_{\WS}$ in Eqs.~\ref{eq:WS,DF} and \ref{eq:WS,diff} as $\Delta t_{\WS} = \min(\Delta t, 1/\Gamma_{\GDF}/p_{\disk})$.

\bibliography{BH_in_AGN}{}
\bibliographystyle{aasjournal}

\end{document}